\documentclass[traditabstract,10pt]{aa}

\usepackage[T1]{fontenc}
\usepackage[stable]{footmisc}
\usepackage{amsmath, amssymb, latexsym}
\usepackage{epstopdf}
\usepackage[breaklinks, colorlinks, citecolor=blue]{hyperref} 
\usepackage{ifthen}
\usepackage[usenames,dvipsnames,table]{xcolor}
\usepackage{placeins}
\usepackage{overpic}
\usepackage{txfonts}

\usepackage{graphicx}
\graphicspath{{Figures/}}
\usepackage{float}

\usepackage[nonamebreak]{natbib}
\bibpunct{(}{)}{;}{a}{}{,}
\providecommand{\sorthelp}[1]{}

\def\farcm{\ifmmode {^{\scriptstyle\prime}} \else $^{\scriptstyle\prime}$\fi}
\newcommand{\draft}{false}

\newcommand{\lcdm}{{$\rm{\Lambda CDM}$}}
\newcommand{\As}{A_{\rm s}}
\newcommand{\Alens}{A_{\rm L}}
\def\GHz{\ifmmode $\,GHz$\else \,GHz\fi}
\newcommand{\alm}{\ifmmode {\vec{a}_{\ell m}} \else $\vec{a}_{\ell m}$\fi}

\newcommand{\W}[4]{{W^{\scriptscriptstyle #1,#2}_{\scriptscriptstyle #3#4}}}
\newcommand{\Cl}[2]{C_{#1}^{#2}}
\def\VEV#1{{ \left\langle #1 \right\rangle }}
\newcommand{\lm}{{\ell m}}
\newcommand{\lmprime}{{\ell'm'}}
\newcommand{\lmun}{{\ell_1m_1}}
\newcommand{\lmdeux}{{\ell_2m_2}}
\newcommand{\lmtrois}{{\ell_3m_3}}
\newcommand{\wigner}[6]
{{
\left( 
\begin{array}{ccc} #1 & #2 & #3 \\#4 & #5 & #6 \end{array}
\right) 
}}

\newenvironment{eqs}
{\begin{subequations}\begin{eqnarray}}
{\end{eqnarray}\end{subequations}}

\newcommand{\xpol}{\emph{Xpol}}
\newcommand{\hillipop}{{HiLLiPOP}}
\newcommand{\hlp}{\emph{hlp}}

\def\planck{\textsc{Planck}}

\begin{document}


\title{Cosmology with the CMB temperature-polarization correlation}
\author{F.~Couchot\inst{1}, S.~Henrot-Versill\'e\inst{1}, O.~Perdereau\inst{1}, S.~Plaszczynski\inst{1},\\ B.~Rouill\'e d'Orfeuil\inst{1}, M.~Spinelli\inst{1,2}, and M.~Tristram\inst{1}\thanks{Corresponding author: \href{mailto:tristram@lal.in2p3.fr}{tristram@lal.in2p3.fr}}}
\authorrunning{F. Couchot et al.}

\institute{Laboratoire de l'Acc\'el\'erateur Lin\'eaire, Univ. Paris-Sud, CNRS/IN2P3, Universit\'e Paris-Saclay, Orsay, France
\and 
Department of Physics and Astronomy, University of the Western Cape, Robert Sobukwe Road, Bellville 7535, South Africa}

\abstract{
We demonstrate that the  cosmic microwave background (CMB) temperature-polarization cross-correlation provides accurate and robust constraints on cosmological parameters.
We compare them with the results from temperature or polarization and investigate the impact of foregrounds, cosmic variance, and instrumental noise.
This analysis makes use of the \planck\ high-$\ell$ \hillipop\ likelihood based on angular power spectra, which takes into account systematics from the instrument and foreground residuals directly modelled using \planck\ measurements. 
The temperature-polarization correlation ($TE$) spectrum is less contaminated by astrophysical emissions than the temperature power spectrum ($TT$), allowing  constraints that are less sensitive to foreground uncertainties to be derived. 
For \lcdm\ parameters, $TE$ gives very competitive results compared to $TT$. 
For basic \lcdm\ model extensions (such as $A_{\rm L}$, $\sum$$m_\nu$, or $N_{\rm eff}$), it is still limited by the instrumental noise level in the polarization maps.
}

\keywords{cosmology: observations -- cosmic background radiation -- surveys -- methods: data analysis}

\date{\today}

\maketitle

\section{Introduction}

The results from the \planck\ satellite have recently demonstrated the consistency between the temperature and the polarization data \citep{planck2014-a15}. 
Adding  the information coming from the velocity gradients of the photon--baryon fluid through the polarization power spectra to the measurement of the temperature fluctuations improves the constraints on cosmological parameters and helps break some degeneracies.
One of the best examples is the measurement of the reionization optical depth using the large-scale signature that reionization leaves in the $EE$ polarization power spectrum  \citep{planck2014-a25}.
Moreover, as suggested in~\citet{galli:2014}, for a cosmic variance limited experiment, polarization power spectra alone can provide tighter constraints on cosmological parameters than the temperature power spectrum, while for an experiment with \planck-like noise, constraints should be comparable.

In this paper, we discuss  in greater detail the constraints on cosmological parameters obtained with the \planck\ 2015 polarization data (including foregrounds and systematic residuals). We find that the level of instrumental noise allows for an accurate reconstruction of cosmological parameters using temperature-polarization cross-correlation $C^{TE}_\ell$ only.
Constraints from \planck\ $EE$ polarization spectrum are dominated by instrumental noise.
In addition, we investigate the robustness of the cosmological interpretation with respect to astrophysical residuals. 

In the \planck\ analysis \citep{planck2013-p08,planck2014-a13}, the foreground contamination is mitigated using masks which are adapted to each frequency, reducing the sky fraction to the region where the foreground emission is low. The residuals of diffuse foreground emission are then taken into account using models at the spectrum level in the likelihood. Most of the results presented in \citet{planck2014-a15} are based on $TT$ angular power spectra which present the higher signal-to-noise ratio.
However, foreground residuals in temperature combine several different components which are difficult to model in the power spectra domain as they are both non-homogeneous and non-Gaussian. Any mismatch between the foreground model and the data can thus result in a bias on the estimated cosmological parameters and, in all cases, will increase their posterior width.
On the contrary, in polarization, even though the signal-to-noise ratio is lower, the only foreground that affects the \planck\ data is the polarized emission of the Galactic dust.
As we  show, this allows for a precise reconstruction of the cosmological parameters (especially with $TE$ spectra) with less impact from foreground uncertainties.

The cosmological parameters reconstructed with $TT$ spectra are compared to those obtained independently with $TE$ and $EE$. In each case, we detail the foreground modelling and the propagation of its uncertainties.
We use the High-$\ell$ Likelihood on Polarized Power spectra (\hillipop ) likelihood  which is based on the \planck\ data in temperature and polarization. \hillipop\ is one of the four high-$\ell$ likelihoods developed within the \planck\ consortium for the 2015 release and is briefly presented and compared to others in \citet{planck2014-a15}. It is a full temperature+polarization likelihood based on cross-spectra from \planck\ maps at 100, 143, and 217\GHz. It is based on a Gaussian approximation of the $C_\ell$ likelihood which is well suited for multipoles above $\ell=30$.
In contrary to the \planck\ public likelihood \citep{planck2014-a15}, the foreground description in \hillipop\ directly relies on the \planck\ astrophysical measurements.
For the \lcdm\ cosmology, using a $\tau$ prior, it gives results very compatible with the \planck\ public likelihood,  except for the $(\tau,\As)$ pair which is more consistent with the low-$\ell$ data. Consequently, it also shows a better lensing amplitude $\Alens$ \citep[see the discussion in][]{couchot:2015}.

The paper is organized as follows. In Sect.~\ref{sec:data}, we describe the power spectra used in this analysis. We discuss the \planck\ maps and the sky region for the power spectra estimation. Section~\ref{sec:lik} presents the likelihood functions both in temperature and in polarization, and details the model of each associated foreground emission. We then present in Sect.~\ref{sec:results} the results for the \lcdm\ cosmological model and check the impact of priors on the astrophysical parameters. Section~\ref{sec:lcdm+} gives the results on the $A_{\rm L}$ parameter considered as an internal cross-check of the CMB likelihoods. Finally, in Sect.~\ref{sec:systematics}, we demonstrate the impact of the foreground parameters for the temperature likelihood and the $TE$ likelihood in terms of both the bias and the precision of the cosmological parameters.

\section{Data set}
\label{sec:data}

\subsection{Maps and masks}
\label{sec:data:maps}

\begin{figure}[!ht]
        \includegraphics[draft=\draft,width=\columnwidth]{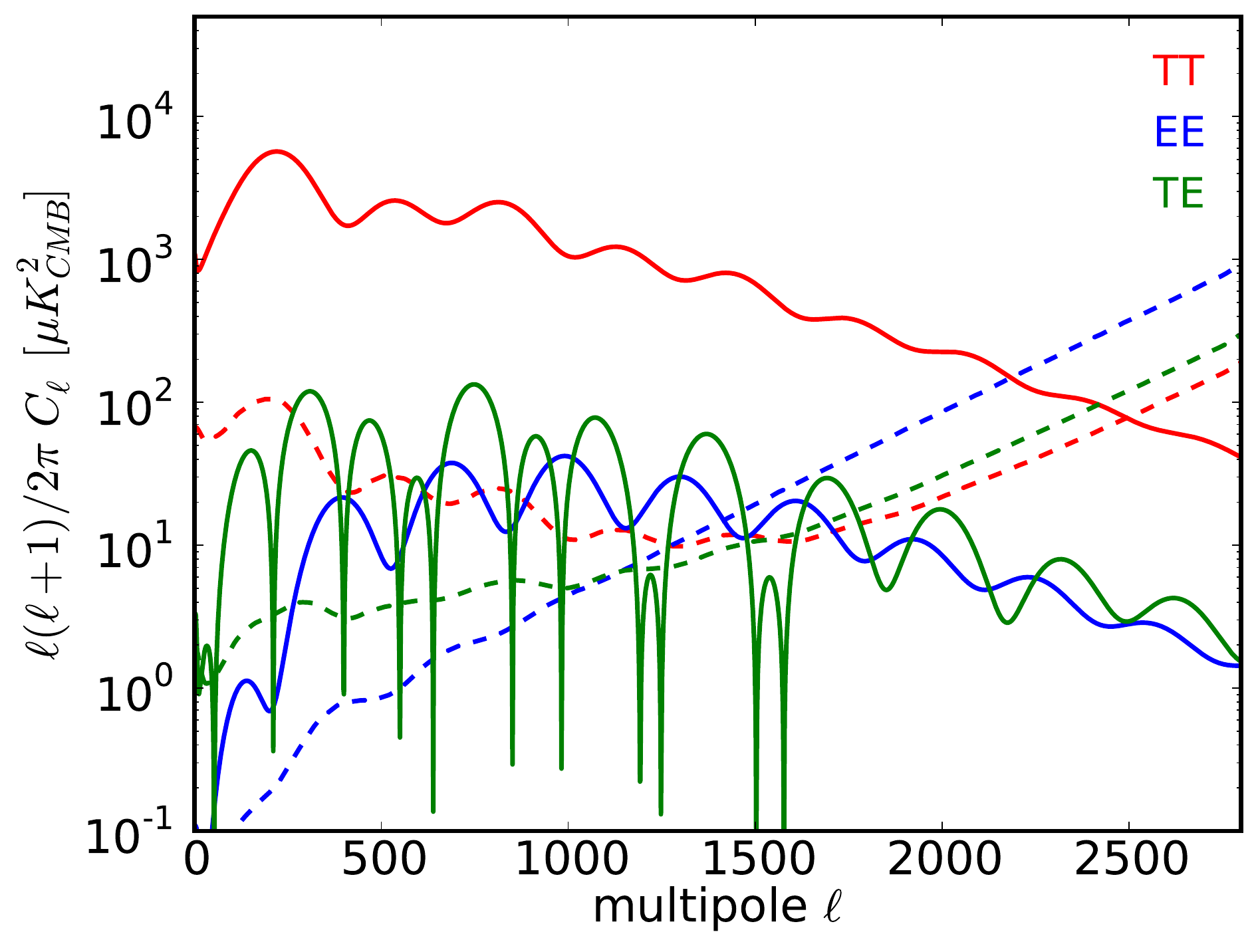}
        \caption{Signal ({\it solid line}) versus noise ({\it dashed line}) for the \planck\ cross-spectra for each mode $TT$, $EE$, and $TE$ (in {\it red}, {\it blue}, and {\it green}, respectively).}
        \label{fig:signal_vs_noise}
\end{figure}

The maps used in this analysis are taken from the \planck\ 2015 data release\footnote{Planck PLA: \url{http://pla.esac.esa.int}} and described in detail in \citet{planck2014-a09}. We use two maps per frequency ($A$ and $B$, one for each \emph{half-mission}) at 100, 143, and 217\GHz.
The beam associated with each map is provided by the \planck\ collaboration \citep{planck2014-a08}.
Figure~\ref{fig:signal_vs_noise} compares the signal with the noise of the \planck\ maps for each mode $TT$, $EE$, and $TE$.

Frequency-dependent apodized masks are applied to these maps in order to limit the foreground contamination in the power spectra.
We use the same masks in temperature and polarization. 
The masks are constructed first by thresholding the total intensity maps of diffuse Galactic dust to exclude strong dust emission. In addition, we also remove regions with strong Galactic CO emission, nearby galaxies, and extragalactic point sources. 

Diffuse Galactic dust emission is the main contaminant for CMB measurements in both temperature and polarization at frequencies above 100\GHz.
We build Galactic masks using the \planck\ 353\GHz\ map as a tracer of the thermal dust emission in intensity. 
In practice, we smoothed the \planck\ 353\GHz\ map to increase the signal-to-noise ratio before applying a threshold which depends on the frequency considered. 
Masks are then apodized using a $8^\circ$ Gaussian taper for power spectra estimation.
For polarization, \planck\ dust maps show that the diffuse emission is strongly related to the Galactic magnetic field at large scales \citep{planck2014-XIX}. However, at the smaller scales which matter here ($\ell > 50$), the orientation of dust grains is driven by local turbulent magnetic fields which produce a polarization intensity proportional to the total intensity dust map. We thus use the same Galactic mask for polarization as for temperature.

Molecular lines from CO produce diffuse emission on star forming region. Two major CO lines at 115\GHz\ and 230\GHz\ enter the \planck\ bandwidths at 100 and 217\GHz,\ respectively \citep{planck2013-p03a}.
We smoothed the \planck\ reconstructed CO map to 30~arcmin before applying a threshold at 2~K.km/s. The resulting masks are then apodized at 15~arcmin.
In practice, the CO masks are almost completely included in the Galactic masks, decreasing the accepted sky fraction only by a few percentage points.

For point sources, the \planck\ 2013 and 2015 analyses mask the sources detected with a signal-to-noise ratio above 5 in the \planck\ point-source catalogue \citep{planck2014-a35} at each frequency \citep{planck2013-p11,planck2014-a13}.
On the contrary, the masks used in our analysis rely on a more refined procedure that preserves Galactic compact structures and ensures the completeness level at each frequency, but with a higher flux cut (340, 250, and 200 mJy at 100, 143, and 217\GHz, respectively). The consequence is that these masks leave a slightly greater number of  unmasked extragalactic sources, but preserve the power spectra of the dust emission \citep[as described in][]{planck2014-XXX}.
For each frequency, we mask a circular area around each source using a radius of three times the effective Gaussian beam width ($\sigma = FWHM/\sqrt{\ln8}$) at that frequency. We apodize these masks with a Gaussian taper of FWHM = 15~arcmin.

Finally, we also mask strong extragalactic objects including both point sources and nearby extended galaxies.
The masked galaxies include the LMC and SMC and also M31, M33, M81, M82, M101, M51, and CenA.

\begin{figure}[!ht]
        \includegraphics[draft=\draft,width=\columnwidth]{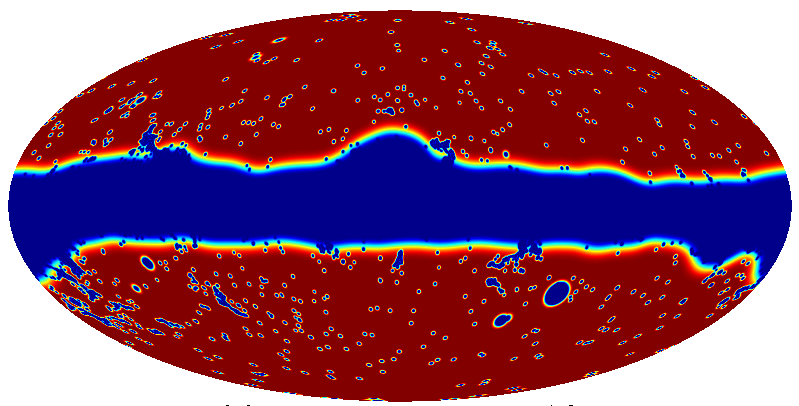}
        \includegraphics[draft=\draft,width=\columnwidth]{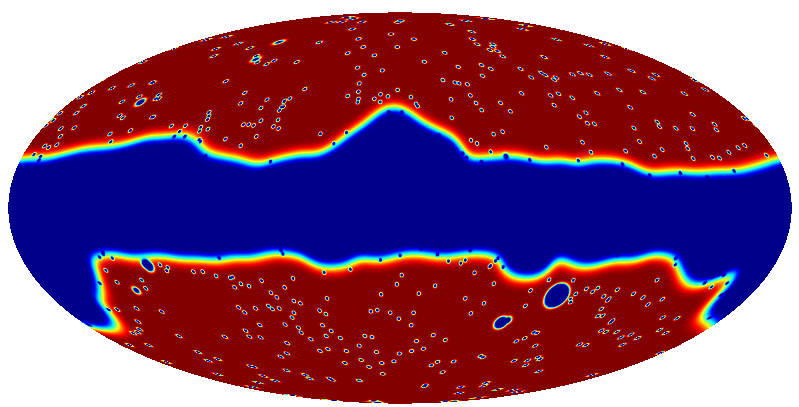}
        \includegraphics[draft=\draft,width=\columnwidth]{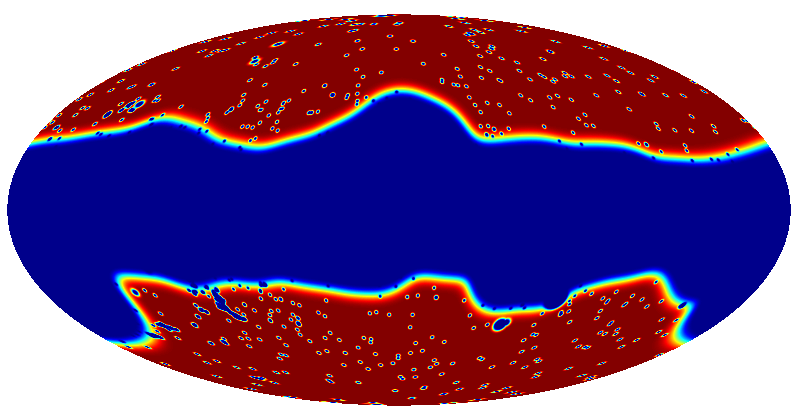}
        \caption{ M80, M70, and M55 masks. A combination of an apodized Galactic mask and a compact object mask is used at each frequency (see text for details).}
        \label{fig:masks}
\end{figure}

The combined masks used are named M80, M70, and M55 (corresponding to effective $f_{\mathrm{sky}}=72\%,62\%,48\%$), associated with the 100, 143, and 217~GHz channels, respectively (Fig.~\ref{fig:masks}).
Tests have been carried out using more conservative Galactic masks (with $f_{\rm sky}$ = 65\%, 55\%, and 40\% for 100, 143, and 217~GHz, respectively) showing perfectly compatible results   with those of the smaller masks.
Compared to the masks used in the \planck\ 2015 analysis, the retained sky fraction is almost identical. Indeed, the Galactic masks used in \citet{planck2014-a13} retain 70\%, 60\%, and 50\% respectively.

\subsection{Power spectra}
We use \xpol\ \citep[an extension to polarization of][]{tristram:2005} to compute the cross-power spectra in temperature and polarization ($TT$, $EE$, and $TE$). \xpol\ is a pseudo-$C_{\ell}$ method which also computes an analytical approximation of the $C_\ell$ covariance matrix directly from data. 
Using the six maps presented in Sect.~\ref{sec:data:maps}, we derive the 15 cross-power spectra for each CMB mode: one each for 100$\times$100, 143$\times$143, and 217$\times$217; four each for 100$\times$143, 100$\times$217, and 143$\times$217 as outlined below.

From the coefficients of the spherical harmonic decomposition of the ($I$,$Q$,$U$) masked maps $\vec{\tilde a}_{\ell m}^X = \{\tilde a^T_{\ell m},\tilde a^E_{\ell m},\tilde a^B_{\ell m}\}$, we form the pseudo cross-power spectra between map $i$ and map $j$,
\begin{equation}
        \tilde{\vec C}_\ell^{ij} = \frac{1}{2\ell+1} \sum_{m} \vec{\tilde a}^{i*}_{\ell m} \vec{\tilde a}^{j}_{\ell m} \, ,
\end{equation}
where the vector $\vec{\tilde C}_\ell$ includes the four modes $\{\tilde C^{TT}_\ell,\tilde C^{EE}_\ell,\tilde C^{TE}_\ell,\tilde C^{ET}_\ell\}$.
 We note that the $TE$ and $ET$ cross-power spectra do not carry exactly the same information since computing T from  map $i$ and E from  map $j$ is not the same as computing E from  map $j$ and T from $i$. They are computed independently and averaged afterwards using their relative weights for each cross-frequency.
The pseudo-spectra are then corrected from beam and sky fraction using
\begin{equation}
        \tilde{\vec C}_\ell^{ij} = (2\ell'+1) \tens{M}^{ij}_{\ell \ell'} \vec{C}^{ij}_{\ell'}
        \label{eq:master}
,\end{equation}
where the coupling matrix $\tens{M}$ depends on the masks used for each set of maps \citep{peebles:1973} and includes beam transfer functions usually extracted from Monte Carlo simulations~\citep{hivon2002}.

The multipole ranges used in the likelihood analysis have been chosen to limit the contamination of the Galactic dust emission at low-$\ell$ and the noise at high-$\ell$. Table~\ref{tab:multipoles} gives the multipole ranges, $[\ell_{\rm min},\ell_{\rm max}]$, considered for each of the six cross-frequencies in TT, TE, and EE.
The spectra are cosmic-variance limited up to $\ell \simeq 1500$ in $TT$ and $\ell \simeq 700$ in $TE$ (outside  the troughs of the CMB signal). The $EE$ mode is dominated by instrumental noise.

\begin{table}[!ht]
        \begin{center}
        \begin{tabular}{l|ccc}
        \hline
        \hline
        &       TT &    EE &    TE\\
        \hline
        100$\times$100  & [~50,1200]    & [100,1000]    & [100,1200]    \\
        100$\times$143  & [~50,1500]    & [100,1250]    & [100,1500]    \\
        100$\times$217  & [500,1500]    & [400,1250]    & [200,1500]    \\
        143$\times$143  & [~50,2000]    & [100,1500]    & [100,1750]    \\
        143$\times$217  & [500,2500]    & [400,1750]    & [200,1750]    \\
        217$\times$217  & [500,2500]    & [400,2000]    & [200,2000]    \\
        $n_\ell$                & $9\,556$      & $7\,256$      & $8\,806$      \\
        \hline
        \end{tabular}
        \caption{Multipole ranges used in the analysis and corresponding number of multipoles available ($n_{\ell}=\ell_{\rm max}-\ell_{\rm min}+1$). The total number of multipoles is $25\,618$.}
        \label{tab:multipoles}
        \end{center}
\end{table}

\begin{figure*}[!ht]
        \center
        \includegraphics[width=0.9\textwidth]{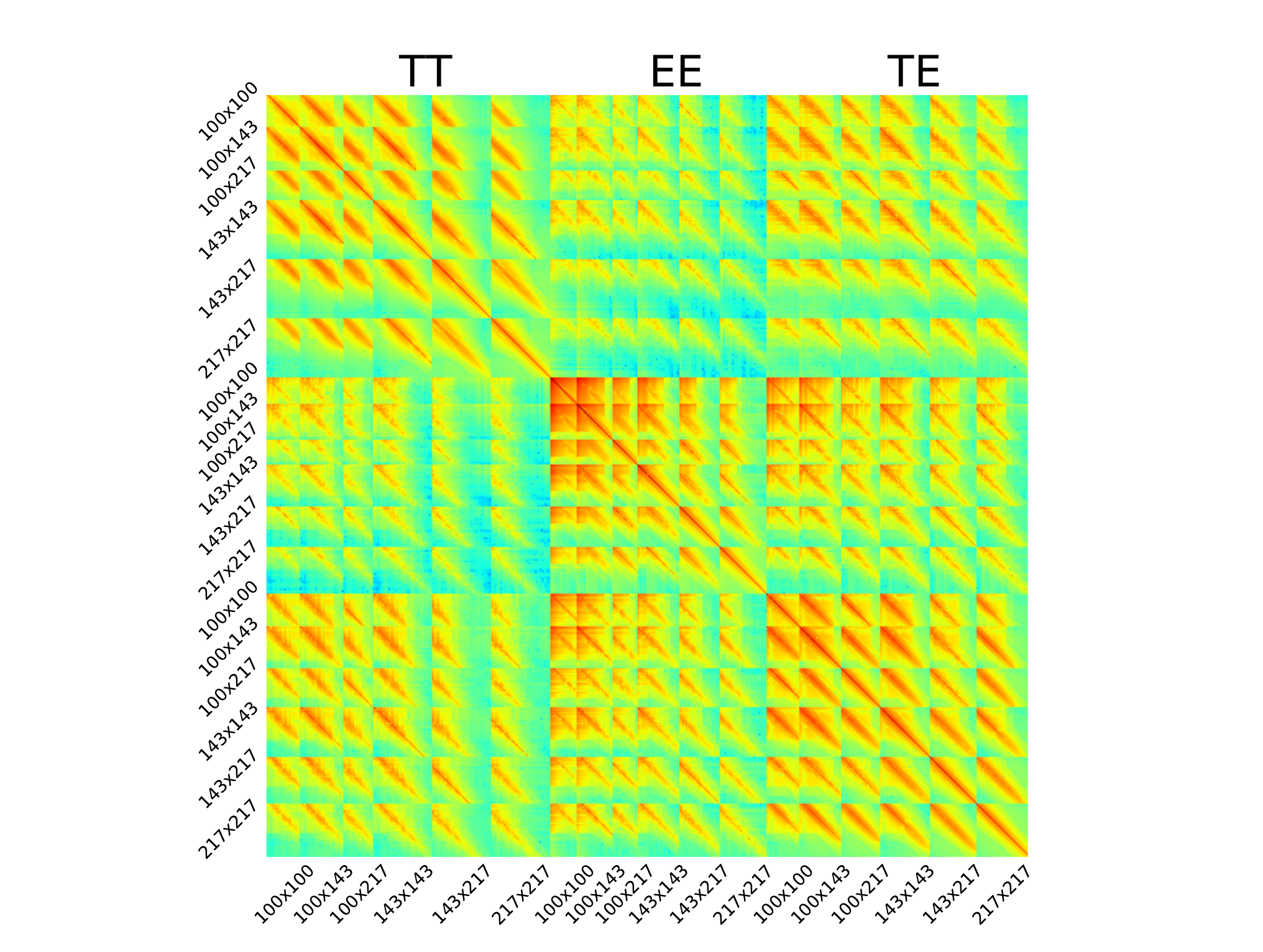}
        \caption{ Full \hillipop\ covariance matrix including all correlations in multipoles between cross-frequencies and power spectra.}
        \label{fig:CovMat}
\end{figure*}

\section{ Likelihood function}
\label{sec:lik}
On the full-sky, the distribution of auto-spectra is a scaled-$\chi^2$ with $2\ell+1$ degrees of freedom. The distribution of the cross-spectra is slightly different \citep[see Appendix A in][]{mangilli:2015}; however, above $\ell \geqslant 50$ the number of modes is large enough so that we can safely assume that the $C_\ell$ are Gaussian distributed.
When considering only a part of the sky, the values of $C_\ell$ are correlated so that for high multipoles, the resulting distribution can be approximated by a multi-variate Gaussian taking into account $\ell$-by-$\ell$ correlations
\begin{equation}
        \label{eq:likelihood}
        -2 \ln \mathcal{L} = \sum_{\substack{i \leqslant j \\ i' \leqslant j'}} \sum_{\ell\ell'} 
                               \vec{R}_\ell^{ij} \, 
                               \left[\tens{\Sigma}^{-1}\right]_{\ell\ell^\prime}^{ij,{i'}{j'}}  \, 
                               \vec{R}_{\ell^\prime}^{{i'}{j'}}
                               + \ln | \tens{\Sigma} |
,\end{equation}
where $\vec{R}^{ij}_\ell = \vec{C}^{ij}_\ell - \vec{\hat C}^{ij}_\ell$ denotes the residual of the estimated cross-power spectrum $\vec{C}_\ell$ with respect to the model $\vec{\hat C}_\ell$ for each polarization mode considered ($TT$, $EE$, $TE$) and each frequency ($\{i,j\} \in [100,143,217]$). The matrix $\tens{\Sigma} = \left< \vec{R} \vec{R}^T \right>$ is the full covariance matrix which includes the instrumental variance from the data as well as the cosmic variance from the model. The latter is directly proportional to the model so that the matrix $\tens{\Sigma}$ should, in principle, depend on the model. 
In practice, given our current knowledge of the cosmological parameters, the theoretical power spectra typically differ from each other at each $\ell$ by less than they differ from the observed $C_\ell$ so that we can expand $\tens{\Sigma}$ around a reasonable fiducial model. As described in~\citet{planck2013-p08}, the additional terms in the expansion are small if the fiducial model is accurate and its absence does not bias the likelihood. Using a fixed covariance matrix $\tens{\Sigma}$, we can drop the constant term $\ln|\tens{\Sigma}|$. 
We therefore expect the likelihood to be $\chi^2$-distributed with a mean equal to the number of degrees of freedom $n_{\rm dof} = n_\ell - n_{\rm p}$ ($n_\ell$ is given in Table~\ref{tab:multipoles} and $n_{\rm p}$ is the number of fitted parameters) and a variance equal to $2n_{\rm dof}$. 

We define several likelihood functions based on the information used: \hlp T for TT cross-spectra, \hlp E for EE cross-spectra, \hlp X for TE cross-spectra, and \hlp TXE for the combination of all cross-spectra. The \hlp X likelihood combines information from TE and ET cross-spectra.

The next two sections describe the computation of the covariance matrix and the building of the model, focusing on the differences with the \planck\ public likelihood.

\subsection{Semi-analytical covariance matrix}

We use a semi-analytical estimation of the $C_\ell$ covariance matrix computed using \xpol. The matrix encloses the $\ell$-by-$\ell$ correlations between all the power spectra involved in the analysis. The computation relies directly on data estimates. It follows that contributions from noise (correlated and uncorrelated), sky emission (from astrophysical and cosmological origin),  and the cosmic variance are implicitly taken into account in this computation without relying on any model or simulations.

The covariance matrix $\tens\Sigma$ of cross-power spectra is directly related to the covariance $\tens{\tilde\Sigma}$ of the pseudo cross-power spectra through the coupling matrices:
\begin{eqnarray}
        \Sigma_{\ell_1\ell_2}^{ab,cd} 
        \equiv \left<\Delta C_{\ell}^{ab}\Delta C_{\ell^\prime}^{cd*}\right>
        = \left(M_{\ell\ell_1}^{ab}\right)^{-1} \tilde\Sigma_{\ell_1\ell_2}^{ab,cd} \left(M_{\ell^\prime\ell_2}^{cd*}\right)^{-1}
\end{eqnarray}
with $(a,b,c,d) \in \{T,E\}$ for each map $A,B,C,D$.

We compute $\tens{\tilde\Sigma}$ for each cross-spectra block independently that includes $\ell$-by-$\ell$ correlation and four-spectra mode correlation $\{TT,EE,TE,ET\}$.
The TE and ET blocks are both computed individually and finally averaged.
The matrix $\tens{\tilde\Sigma}$, which gives the correlations between the pseudo cross-power spectra ($ab$) and ($cd$), is an N-by-N matrix (where $N=n^{TT}_\ell+n^{EE}_\ell+n^{TE}_\ell+n^{ET}_\ell$) and reads
\begin{eqs}
\label{eq:correlation}
        \tilde\Sigma_{\ell\ell^\prime}^{ab,cd} &\equiv& \left<\Delta\tilde{C}_{\ell}^{ab}\Delta\tilde{C}_{\ell^\prime}^{cd*}\right>
        =  \left<\tilde{C}_{\ell}^{ab}\tilde{C}_{\ell^\prime}^{cd*}\right>-\tilde{C}_{\ell}^{ab}\tilde{C}_{\ell^\prime}^{cd*}  \nonumber \\
        &=& \sum_{mm^\prime} \frac{
\left<\tilde{a}_{\ell m}^{a}\tilde{a}_{\ell^\prime m^\prime}^{c*}\right>\left<\tilde{a}_{\ell m}^{b*}\tilde{a}_{\ell^\prime m^\prime}^{d}\right>+
\left<\tilde{a}_{\ell m}^{a}\tilde{a}_{\ell^\prime m^\prime}^{d*}\right>\left<\tilde{a}_{\ell m}^{b*}\tilde{a}_{\ell^\prime m^\prime}^{c}\right>
                }{(2\ell+1)(2\ell^\prime+1)}  \nonumber
\end{eqs}
by expanding the four-point Gaussian correlation using  Isserlis' formula (or Wick's theorem).

Each two-point correlation of pseudo-$\alm$ can be expressed as the convolution of $\vec C_\ell$ with a kernel which depends on the polarization mode considered
\begin{eqnarray*}
        \VEV{ \tilde a^{T_a*}_{\lm}\tilde a^{T_b}_{\lmprime}} &=& \sum_{\lmun} \Cl{\ell_1}{T_aT_b} \W{0}{T_a}{\lm}{\lmun} \W{0}{T_b*}{\lmprime}{\lmun}
        \\
        \VEV{ \tilde a^{E_a*}_{\lm} \tilde a^{E_b}_{\lmprime}} &=& \frac{1}{4} \sum_\lmun
        \left\{
                \Cl{\ell_1}{E_aE_b} \W{+}{E_a*}{\lm}{\lmun} \W{+}{E_b}{\lmprime}{\lmun}
                + \Cl{\ell_1}{B_aB_b} \W{-}{E_a*}{\lm}{\lmun} \W{-}{E_b}{\lmprime}{\lmun}
        \right\}
        \\
        \VEV{\tilde a^{T_a*}_{\lm} \tilde a^{E_b}_{\lmprime}} &=& \frac{1}{2} \sum_\lmun
                \Cl{\ell_1}{T_a E_b} \W{0}{T_a*}{\lm}{\lmun} \W{+}{E_b}{\lmprime}{\lmun}
\end{eqnarray*}
where the kernels $W^{0}$, $W^{+}$, and $W^{-}$ are defined as linear combination of products of $Y_{\ell m}$ of spin 0 and $\pm 2$ (see Appendix~\ref{ann:xpol_covariance}).
As suggested in \citet{efstathiou:2006}, neglecting the gradients of the window function and applying the completeness relation for spherical harmonics \citep{varshalovich:1988}, we can reduce the products of four $W$ into kernels similar to the coupling matrix $\tens{M}$ defined in Eq.~\ref{eq:master}.
In the end, the blocks of $\tens{\Sigma}$ matrices reads
\begin{eqnarray*}
        \Sigma^{T_aT_b,T_cT_d}
        &\simeq
        \Cl{\ell\ell'}{T_aT_c}\Cl{\ell\ell'}{T_bT_d} \tens{M}_{TT,TT} &+\ \Cl{\ell\ell'}{T_aT_d}\Cl{\ell\ell'}{T_bT_c} \tens{M}_{TT,TT}
        \\
        \Sigma^{E_aE_b,E_cE_d}
        &\simeq
        \Cl{\ell\ell'}{E_aE_c}\Cl{\ell\ell'}{E_bE_d} \tens{M}_{EE,EE} &+\ \Cl{\ell\ell'}{E_aE_d}\Cl{\ell\ell'}{E_bE_c} \tens{M}_{EE,EE}
        \\
        \Sigma^{T_aE_b,T_cE_d}
        &\simeq
        \Cl{\ell\ell'}{T_aT_c}\Cl{\ell\ell'}{E_bE_d} \tens{M}_{TE,TE} &+\ \Cl{\ell\ell'}{T_aE_d}\Cl{\ell\ell'}{E_bT_c} \tens{M}_{TT,TT}
        \\
        \Sigma^{T_aT_b,T_cE_d}
        &\simeq
        \Cl{\ell\ell'}{T_aT_c}\Cl{\ell\ell'}{T_bE_d} \tens{M}_{TT,TT} &+\ \Cl{\ell\ell'}{T_aE_d}\Cl{\ell\ell'}{T_bT_c} \tens{M}_{TT,TT}
        \\
        \Sigma^{T_aT_b,E_cE_d}
        &\simeq
        \Cl{\ell\ell'}{T_aE_c}\Cl{\ell\ell'}{T_bE_d} \tens{M}_{TT,TT} &+\ \Cl{\ell\ell'}{T_aE_d}\Cl{\ell\ell'}{T_bE_c} \tens{M}_{TT,TT}
        \\
        \Sigma^{E_aE_b,T_cE_d}
        &\simeq
        \Cl{\ell\ell'}{E_aT_c}\Cl{\ell\ell'}{E_bE_d} \tens{M}_{TE,TE} &+\ \Cl{\ell\ell'}{E_aE_d}\Cl{\ell\ell'}{E_bT_c} \tens{M}_{TE,TE}
\end{eqnarray*}
which are thus directly related to the measured auto- and cross-power spectra (see Appendix~\ref{ann:xpol_covariance} for details). In practice, to avoid any correlation between $C_\ell$ estimates and their covariance, we use a smoothed version of each measured power spectrum (using a Gaussian filter with $\sigma_\ell=5$) to estimate the covariance matrix. 

The analytical full covariance matrix (Fig.~\ref{fig:CovMat}) has $25\,618\times25\,618$ elements, is symmetric and positive definite. Its condition number is $\sim 10^8$.

This semi-analytical estimation has been tested against Monte Carlo simulations. In particular, we tested how accurate the approximations are in the case of a non-ideal Gaussian signal (due to the presence of small foregrounds residuals), Planck realistic (low) level of pixel-pixel correlated noise, and apodization length used for the mask.
We have found no deviation to the sample covariance estimated from the 1000 realizations of the full focal plane Planck simulations \citep[FFP8, see][]{planck2014-a14} including anisotropic correlated noise and foreground residuals. To go further and to check the detailed impact from the sky mask (including the choice of the apodization length), we simulated CMB maps from the \planck\ 2015 best-fit $\Lambda$CDM angular power spectrum, to which we added realistic anisotropic Gaussian noise (but without correlation) corresponding to each of the six data set maps. We then computed their cross-power spectra using the same foreground masks as for the data. A total of $15\,000$ sets of cross-power spectra have been produced. 

\begin{figure}[!ht]
        \centering
        \includegraphics[draft=\draft,width=\columnwidth]{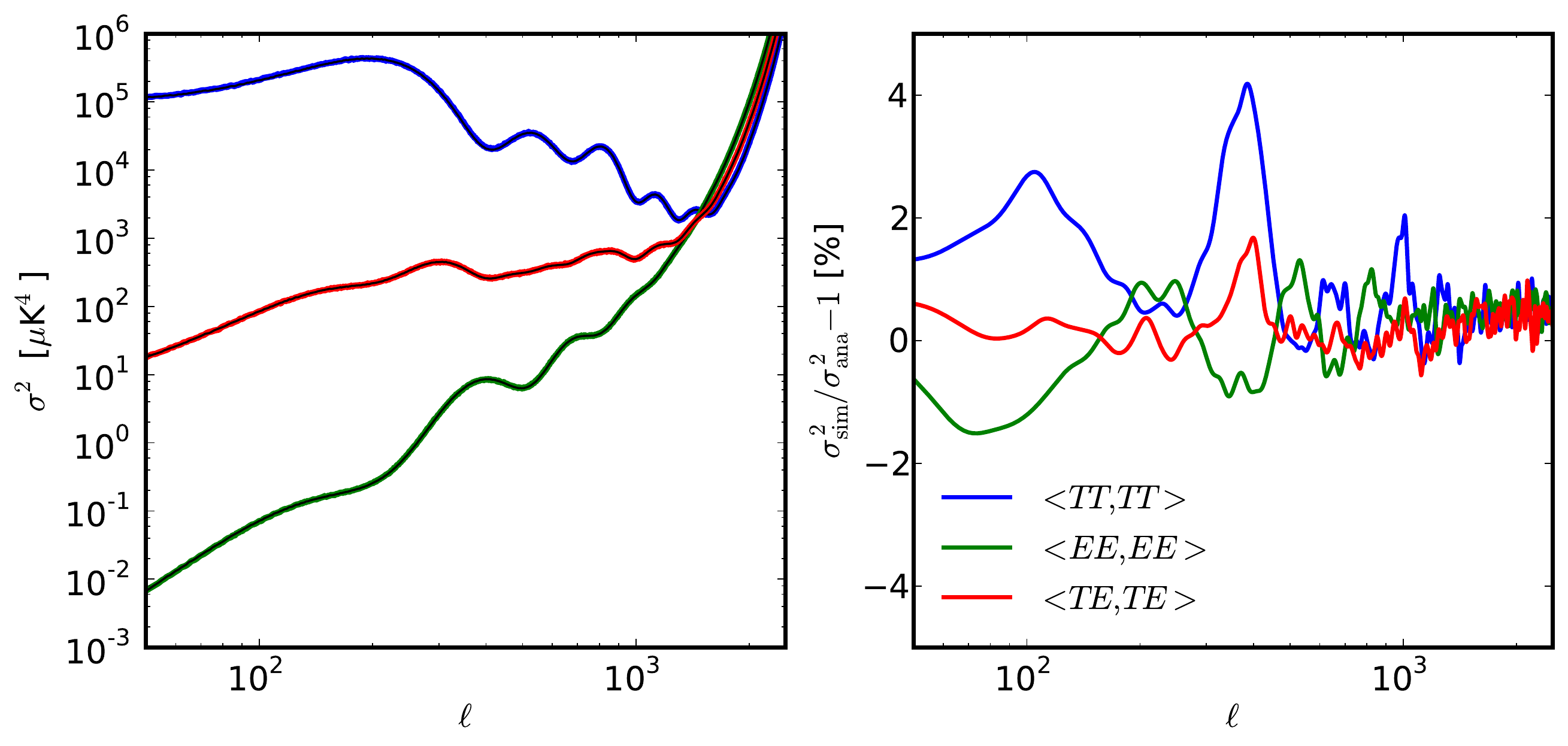}
        \includegraphics[draft=\draft,width=\columnwidth]{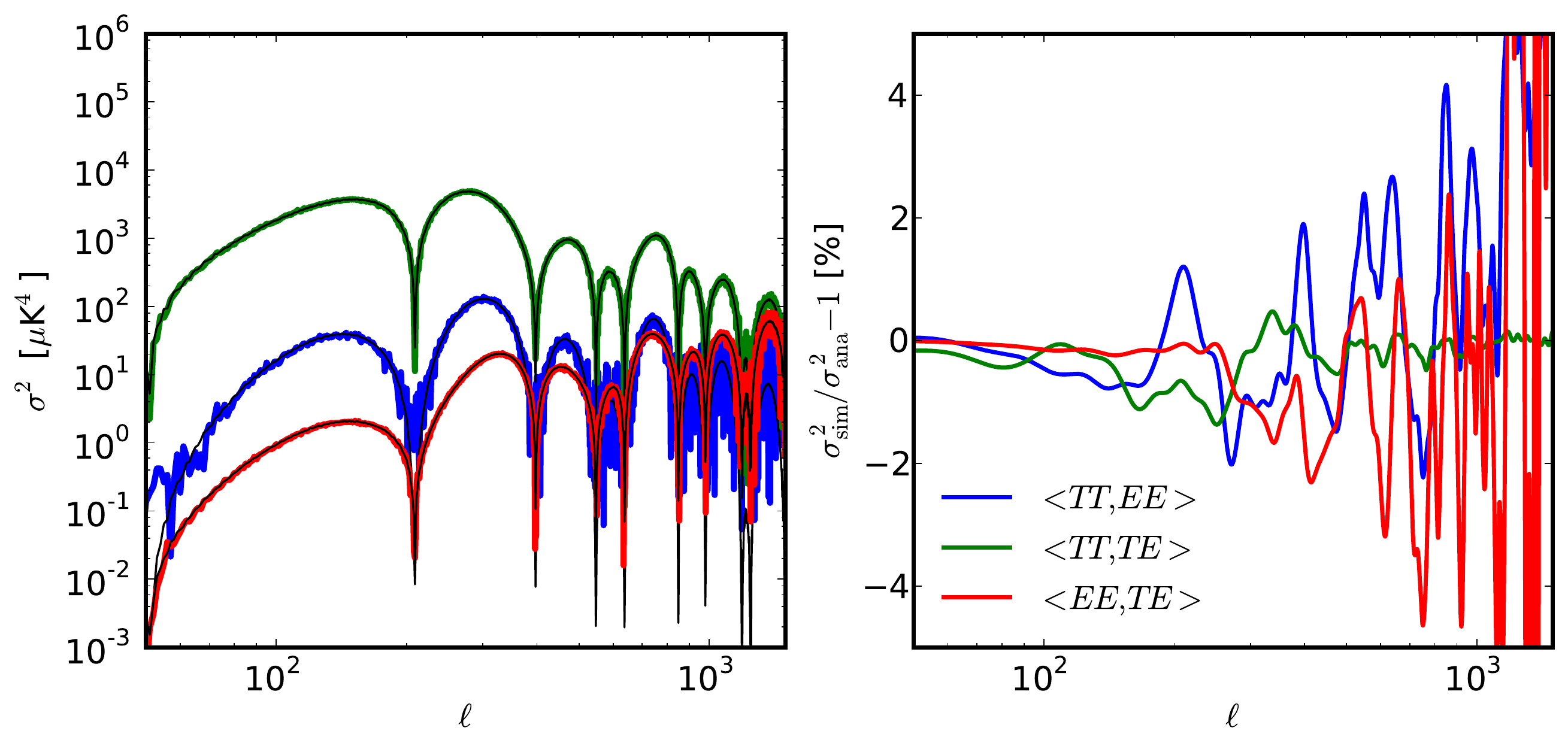}
        \caption{Diagonals of the $C_\ell$ covariance matrix $\tens{\Sigma}$ for the block 143A$\times$143B computed using the semi-analytical estimation ({\it coloured lines}) compared with the Monte Carlo ({\it black line}). {\it Top:} spectra auto-correlation. {\it Bottom:} spectra cross-correlation.}
        \label{fig:MatrixPrecision}
\end{figure}

When comparing the diagonal of the covariance matrix from the analytical estimation with the corresponding simulated variance, a precision better than a few per cent is found (Fig.~\ref{fig:MatrixPrecision}).
The residuals show some oscillations, essentially in temperature, which are introduced by the compact objects mask. Indeed, the large number of small holes with short apodization length induces structures in the harmonic window function which break the hypothesis used in the semi-analytical estimation of the $C_\ell$ covariance matrix.
However, the refined procedure used to construct our specific point source mask allows to keep the level of the impact to less than a few per  cent.

Since we are using a Gaussian approximation of the likelihood, the uncertainty of the covariance matrix will not bias the estimation of the  cosmological parameters. The per cent precision obtained here will then only propagate into a per cent error on the variance of the recovered cosmological model.

\subsection{Model}
\label{sec:lik:model}
We now present the model ($\vec{\hat C}_\ell$) used in the likelihood (Eq.~\ref{eq:likelihood}). The foreground emissions are mitigated by applying the masks (defined in Sect.~\ref{sec:data:maps}) and using an appropriate choice of multipole range. However, our likelihood function explicitly takes into account residuals of foreground emissions in power spectra together with CMB model and instrumental systematic effects. The model finally reads:
\begin{equation}
        \vec{\hat C}_\ell^{ij} =  A_{\rm pl}^2 c_i c_j \left(1+\beta^{ij}\mu_\ell^{ij}\right)^2 \left( \vec{C}_\ell^{\rm CMB} + \sum_{\rm fg} {A}^{ij}_{\rm fg} \vec{C}_\ell^{ij,\rm fg} \right)
        \label{eq:model}
\end{equation}
where $A_{\rm pl}$ is an absolute calibration factor, $c$ represents the inter-calibration of each map (normalized to the 143A map), $\beta$ is the amplitude of the beam uncertainty $\mu_\ell$, and $\tens{A}_{\rm fg}$ are the amplitudes of the foreground components $\vec{C}_\ell^{\rm fg}$.

The model for CMB, $\vec{C}^{\rm CMB}_\ell$, is computed solving numerically the background+perturbation equations for a specific cosmological model. In this paper, we consider a \lcdm\ model with six free parameters describing the current density of baryons ($\Omega_b$) and cold dark matter ($\Omega_{cdm}$); the angular size of sound horizon at recombination ($\theta$); the reionization optical depth ($\tau$); and the index and the amplitude of the primordial scalar spectrum ($n_{\rm s}$ and $A_{\rm s}$).

We include in the sum of the foregrounds for the temperature likelihood contributions from Galactic dust, cosmic infrared background (CIB), thermal (tSZ) and kinetic (kSZ) Sunyaev-Zel'dovich components, Poisson point sources (PS), and the correlation between infrared galaxies and the tSZ effect (tSZxCIB).
Only Galactic dust is considered in polarization. Synchrotron emission is known to be significantly polarized, but it is subdominant in the \planck-HFI channels and we can neglect its contribution in power spectra above $\ell=50$. The contribution from polarized point sources is also negligible in the $\ell$ range considered for polarized spectra~\citep{tucci:2012}.

In \hillipop, we use physically motivated templates of foreground emission power spectra, based on \planck\ measurements. We assume a $C_{\ell}$ template for each foreground with a fixed frequency spectrum and rescale it using a free parameter $\tens{A}^{\rm fg}$ normalized to one. 

The model is a function of the cosmological ($\mathbf{\Omega}$) and nuisance ($p$) parameters: $\vec{\hat C}_{\ell}^{\rm model}(\mathbf{\Omega}, p)$. 
The latter include instrumental parameters accounting for instrumental uncertainties and scaling parameters for each astrophysical foreground model as described in the following sections.
In the end, we have a total of 6 instrumental parameters (only calibration is considered, see Sect.~\ref{sssec:instru_syste}), 9 astrophysical parameters (7 for $TT$, 1 for $TE$, 1 for $EE$), and $6+$ cosmological parameters ($\Lambda$CDM and extensions), i.e. a total of $21+$ free parameters in the full likelihood function (see Appendix~\ref{hlp_params}).
We note that the \planck\ public likelihood depends on more nuisance parameters: 15 for $TT$ (compared to 13 for \hlp T), 9 for $TE$ (compared to 7 for \hlp X), and 9 for $EE$ (compared to 7 for \hlp E).

\subsubsection{Instrumental systematics}
\label{sssec:instru_syste}
The instrumental parameters of the \hillipop\ likelihood are the inter-calibration coefficients ($c$, which are measured relative to the 143A map), and the amplitudes ($\beta$) of the beam error modes ($\mu_\ell$). 
In practice, we have linearized Eq.~\ref{eq:model} for the coefficients $c$ and fit for small deviations around zero ($c \rightarrow 1+c$), while fixing $c_{\rm 143A}=0$ for normalization.
The uncertainty in the absolute calibration is propagated through a global rescaling factor $A_{\rm pl}$.

The effective beam window functions $B_{\ell}$ account for the scanning strategy and the weighted sum of individual detectors performed to obtained the combined maps \citep{planck2014-a08}. It is constructed from Monte Carlo simulations of CMB convolved with the measured beam on each time-ordered data sample. 
The uncertainties in the determination of the HFI effective beams come directly from simulations and is described in terms of the Monte Carlo eigenmodes $\mu_\ell$~\citep{planck2013-p08}. 
In the \planck\ 2013 analysis, it was found that, in practice, only the first beam eigenmode for the 100$\times$100 spectrum was relevant \citep{planck2013-p11}. For the 2015 analysis, \cite{planck2014-a13} found no evidence of beam error in their multipole range thanks to higher accuracy in the beam estimation, which reduced the amplitude of the beam uncertainty. As a consequence, in our analysis, we fixed their contribution to zero ($\beta=0$).

\subsubsection{Galactic dust}
\label{sec:dust_model}
\begin{figure}[!ht]
        \includegraphics[draft=\draft,width=\columnwidth]{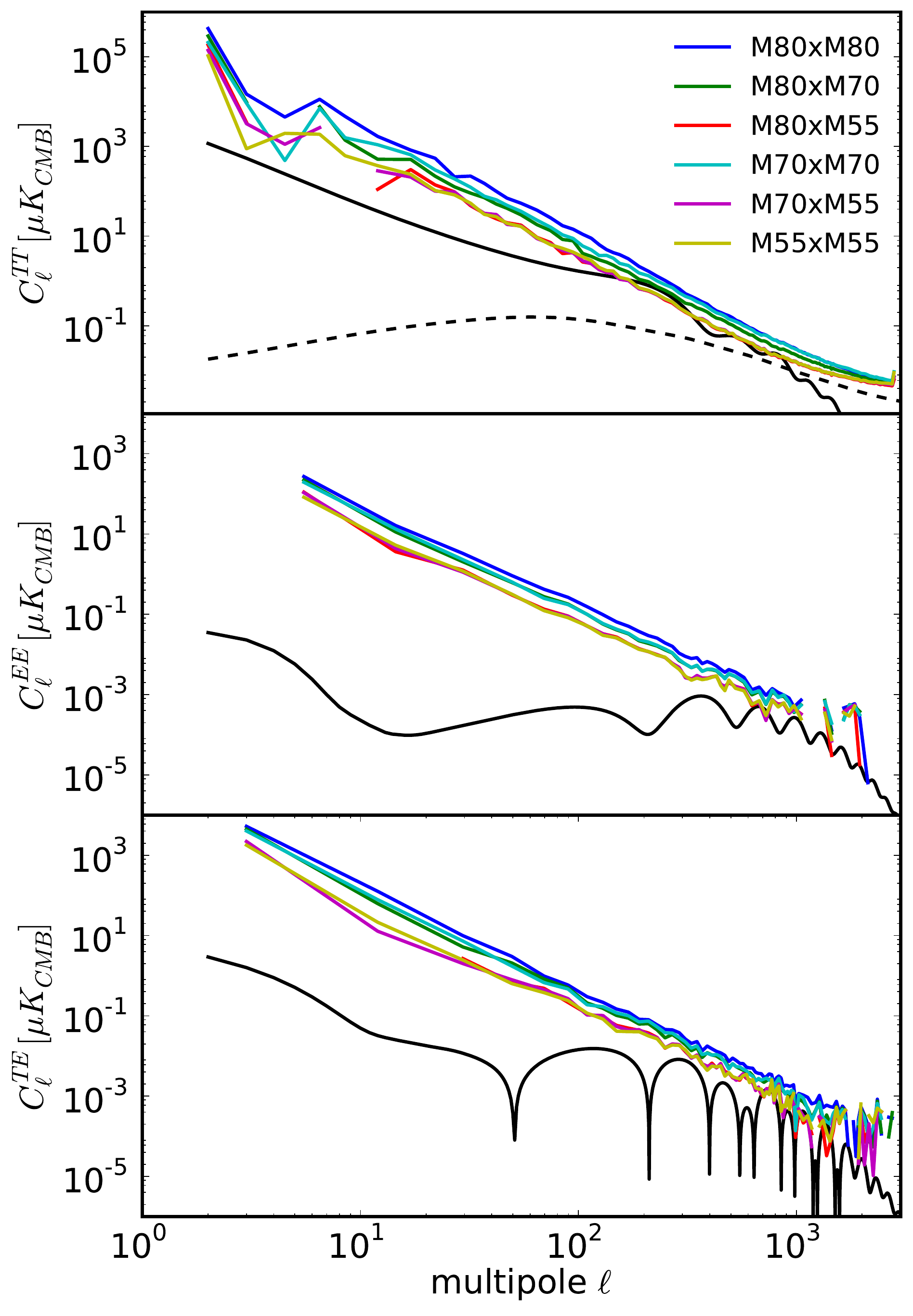}
        \caption{Dust power spectra at 353\GHz\ for $TT$ ({\it top}), $TE$ ({\it middle}), and $EE$ ({\it bottom}). The power spectra are computed from cross-correlation between half-mission maps for different sets of masks as defined in Sect.~\ref{sec:data:maps} and further corrected for CMB power spectrum ({\it solid black line}) and CIB power spectrum ({\it dashed black line}).}
        \label{fig:dust353}
\end{figure}

The $TT$, $EE$, and $TE$ Galactic dust $C_\ell$ templates are obtained from the cross-power spectra between half-mission maps at 353\GHz\ \citep[as in][]{planck2014-XXX}. This is repeated for each mask combination associated with the map data set. The estimated power spectra are then accordingly rescaled to each of the six cross-frequencies considered in this analysis. 
We compute the 353\GHz\ cross-spectra $\vec{\hat C}_\ell^{M_iM_j}$ for each pair of masks $(M_i,M_j)$ associated with the cross-spectra $i \times j$ (Fig.~\ref{fig:dust353}). We then subtract the \planck\ best-fit CMB power spectrum. For $TT$, we also subtract the CIB power spectrum \citep{planck2013-pip56}.
In addition to Galactic dust, unresolved point sources contribute to the $TT$ power spectra at 353\GHz. To construct the dust templates $\vec{C}_\ell^{M_iM_j,\rm dust}$ for our analysis, we thus fit a power-law model with a free constant $A\ell^\alpha+B$ in the range $\ell=[50,2500]$ for $TT$, while a simple power-law is used to fit the $EE$, $TE$ power spectra in the range $\ell=[50,1500]$.

Thanks to the use of the point source mask (described in Sect.~\ref{sec:data:maps}), our Galactic dust residual power spectrum is much simpler than in the case of the \planck\ official likelihood.
Indeed, the masks used in the \planck\ analysis remove some Galactic structures and bright cirrus, which induces an artificial knee in the residual dust power spectra around $\ell \sim 200$ \citep[Sect. 3.3.1 in][]{planck2014-a13}. In contrast, our Galactic dust power spectra are directly comparable to those derived in~\citet{planck2014-XXX}. Moreover, here we do not assume that the dust power spectra have the same spatial dependence across masks.

For each polarization mode ($TT$, $EE$, $TE$), we then extrapolate the dust templates at 353\GHz\ for each cross-mask to the cross-frequency considered
\begin{equation}
        \vec{C}_{\ell}^{ij,{\rm dust}} = A_{\rm dust} \, a^{\rm dust}_{\nu_i} a^{\rm dust}_{\nu_j} \vec{C}_\ell^{M_iM_j,{\rm dust}}
        \label{eq:dust_model}
,\end{equation}
where the $a^{dust}_{\nu} = f^{dust}(\nu) / f^{dust}(353\GHz)$ extrapolated factors are estimated for intensity or polarization maps. We use a greybody emission law with a mean dust temperature of $19.6$~K and spectral indices $\beta^T=1.59$ and $\beta_P=1.51$ as measured in \citet{planck2014-XXII}. The resulting $a^{dust}_\nu$ factors are $(0.0199,0.0387,0.1311)$ for total intensity and $(0.0179,0.0384,0.1263)$ for polarization at 100, 143, and 217~GHz, respectively.

In \hillipop, this results in three free parameters ($A_{\rm dust}^{TT}$, $A_{\rm dust}^{EE}$, $A_{\rm dust}^{TE}$) describing the amplitude of the dust residuals in each mode. This model based on \planck\ internal measurements is simpler than the one used in the \planck\ official likelihood, which allows the amplitude of each cross-frequency to vary (ending with a total of 16 free parameters) and puts constraints on the dust SED through the use of strong priors.

\subsubsection{Cosmic infrared background}
\label{sec:CIB_model}
The thermal radiation of dust heated by UV emission from young stars produces an extragalactic infrared background whose emission law is very close to the Galactic dust emission.
The Planck Collaboration has studied the CIB in detail in \citet{planck2013-pip56} and provides templates based on a model that associates  star forming galaxies with dark matter halos and their sub-halos, using a parametrized relation between the dust-processed infrared luminosity and (sub-)halo mass.
This model provides an accurate description of the Planck and IRAS CIB spectra from 3000\GHz\ down to 217\GHz. We extrapolate this model here, assuming it remains appropriate when describing the 143\GHz\ and 100\GHz\ data.

The halo model formalism, which is also used for the tSZ and the tSZ$\times$CIB models (see Sects.~\ref{sec:SZ_model} and \ref{sec:tSZxCIB_model}), has the general expression \citep{planck2014-a29}
\begin{equation}
        C_{\ell} = C^{{\rm AB, 1h}}_\ell + C^{{\rm AB, 2h}}_\ell,
\end{equation}
where A and B stand for tSZ effect or CIB emission, $C^{{\rm AB, 1h}}_\ell$ is the one-halo contribution, and $C^{{\rm AB, 2h}}_\ell$ is the two-halo term.
The one-halo term $C^{{\rm AB, 1h}}_\ell$ is computed as
\begin{equation}
        C_{\ell}^{\rm AB,{\rm 1h}} = 4 \pi \int {\rm d}z \frac{{\rm d}V}{{\rm d}z {\rm d}\Omega}\int{\rm d}M \frac{{\rm d^2N}}{{\rm d}M {\rm d}V} W^{\rm 1h}_{\rm A} W^{\rm 1h}_{\rm B} ,
\end{equation}
where $\frac{{\rm d^2N}}{{\rm d}M {\rm d}V}$ is the dark matter halo mass function from \citet{tinker:2008}, $\frac{{\rm d}V}{{\rm d}z {\rm d}\Omega}$ the comoving volume element, and $W^{\rm 1h}_{\rm A,B}$ is the window function that accounts for selection effects and total halo signal.
Instead, the contribution of the two-halo term, $C^{{\rm AB, 2h}}_\ell$, accounts  for correlation in the spatial distribution of halos over the sky.

For the CIB, the two-halo term (i.e. the term that considers galaxies belonging to two different halos) is dominant at low and intermediate multipoles and is very well constrained by \planck. The one-halo term is flat in $C_\ell$ and not well measured as it is degenerated with the shot noise. Hence, in \citet{planck2013-pip56} strong priors on the shot noises have been used to get the one-halo term. In \hillipop, we did not include any shot noise term in the CIB template to avoid degeneracies with the amplitude of  infrared sources (see Sect.~\ref{sec:ps_model}).

The power spectra template for each cross-frequency in $\mathrm{Jy}^2\mathrm{sr}^{-1}$ (with the IRAS convention $\nu I(\nu)=$cst) are then converted in $\mu {\rm K}_{\rm CMB}^2$ using a slightly revised version of Table~6 in \citet{planck2013-p03d}: $a^{\rm conv}_{100} = 1/244.06$, $a^{\rm conv}_{143} = 1/371.66$, and $a^{\rm conv}_{217} = 1/483.48$~K$_{\rm CMB}$/MJy.sr$^{-1}$ at 100, 143, and 217~GHz, respectively. Those coefficients account for the integration of the CIB emission law in the \planck\ bandwidth.

\begin{figure}[!ht]
        \centering
        \includegraphics[draft=\draft,width=\columnwidth]{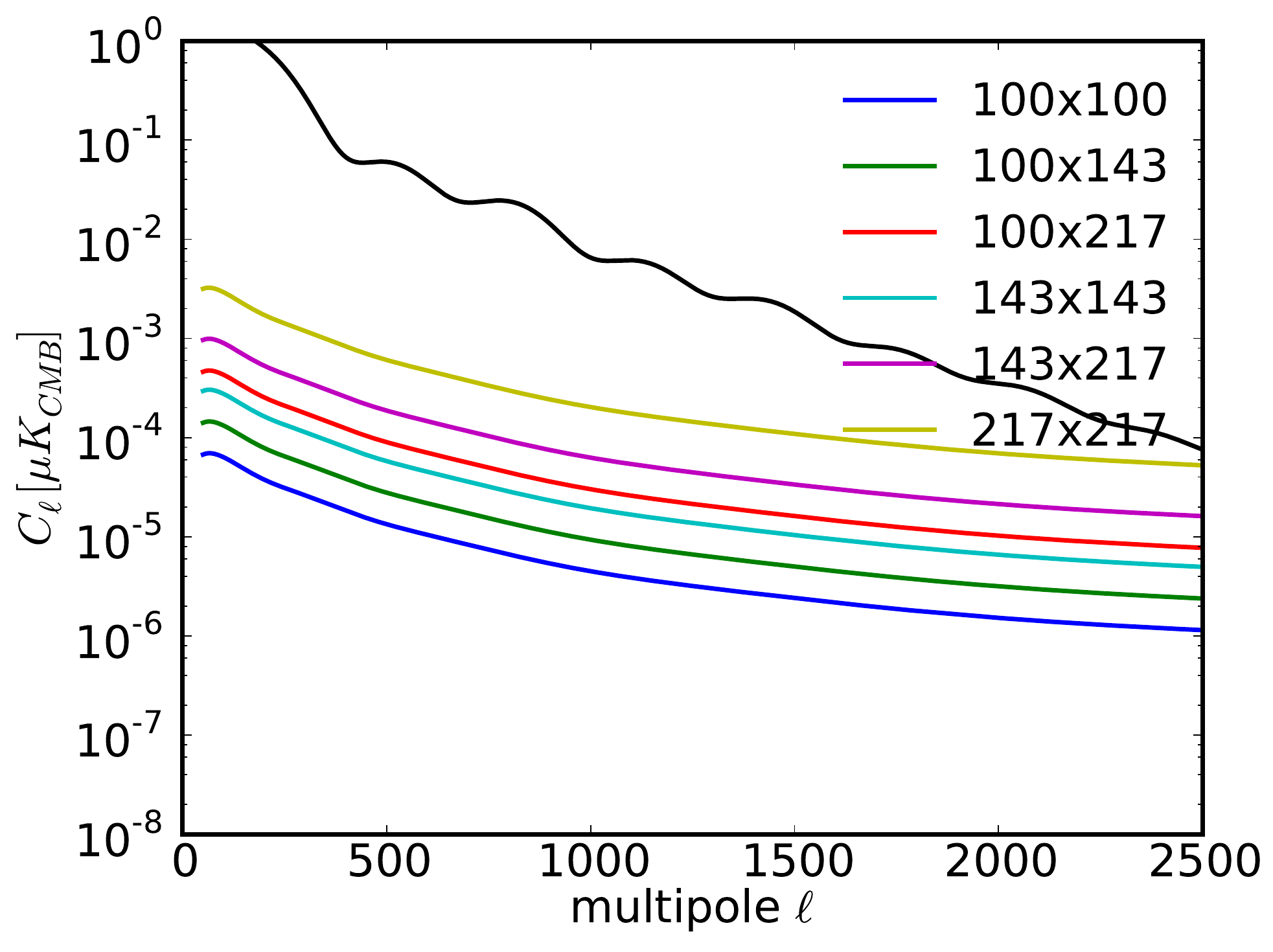}
        \caption{ CIB power spectra templates. The SED and the angular dependence is given by \citet{planck2013-pip56}. The CMB TT power spectrum is plotted in black.}
        \label{fig:CIB}
\end{figure}

The CIB templates used in \hillipop\ (Fig.~\ref{fig:CIB}) are then rescaled with a free single parameter $A_{\rm CIB}$:
\begin{equation}
        \vec{C}_{\ell}^{ij,{\rm CIB}}=A_{\rm CIB} \, a^{\rm conv}_{\nu_i} a^{\rm conv}_{\nu_j} C_{\ell}^{\nu_i\nu_j,{\rm temp}} \, .
        \label{eq:CIB_model}
\end{equation}
The same parametrization was finally adopted in the \planck\ official analysis for the 2015 release.

\subsubsection{Sunyaev-Zel'dovich effect}
\label{sec:SZ_model}
The thermal Sunyev-Zel'dovich emission (tSZ) is also parameterized by a single amplitude and a fixed template measured in \citet{planck2013-p05b} at 143\GHz,
\begin{equation}
        \vec{C}^{ij,\rm tSZ}_{\ell} = A_{\rm tSZ} \, a^{\rm tSZ}_{\nu_i} a^{\rm tSZ}_{\nu_j} C_{\ell}^{\rm tSZ} \,,
        \label{eq:tSZ_model}
\end{equation}
where $a^{\rm tSZ}_{\nu} = f^{\rm tSZ}(\nu)/f^{\rm tSZ}(143)$ is the thermal Sunyaev-Zel'dovich spectrum normalized at 143~GHz. 
We recall that, ignoring the bandpass corrections, the tSZ spectrum is given by
\begin{equation}
        f^{\rm tSZ}(\nu) = \left(x\coth\left(\frac{x}{2}\right)-4 \right) \quad \mbox{with}\ x=\frac{h\nu}{k_\mathrm{B} T_\mathrm{cmb}}.
        \label{eq:SZ_spectrum}
\end{equation}
After integrating over the instrumental bandpass, we obtain $f^{\rm tSZ} = -4.031, -2.785$, and  $0.187$ at 100, 143, and 217~GHz, respectively \citep[see Table 1 in][]{planck2014-a28}.
The \planck\ official likelihood uses the same parametrization but with an empirically motivated template power spectrum \citep{efstathiou:2012}.

\begin{figure}[!ht]
        \includegraphics[draft=\draft,width=\columnwidth]{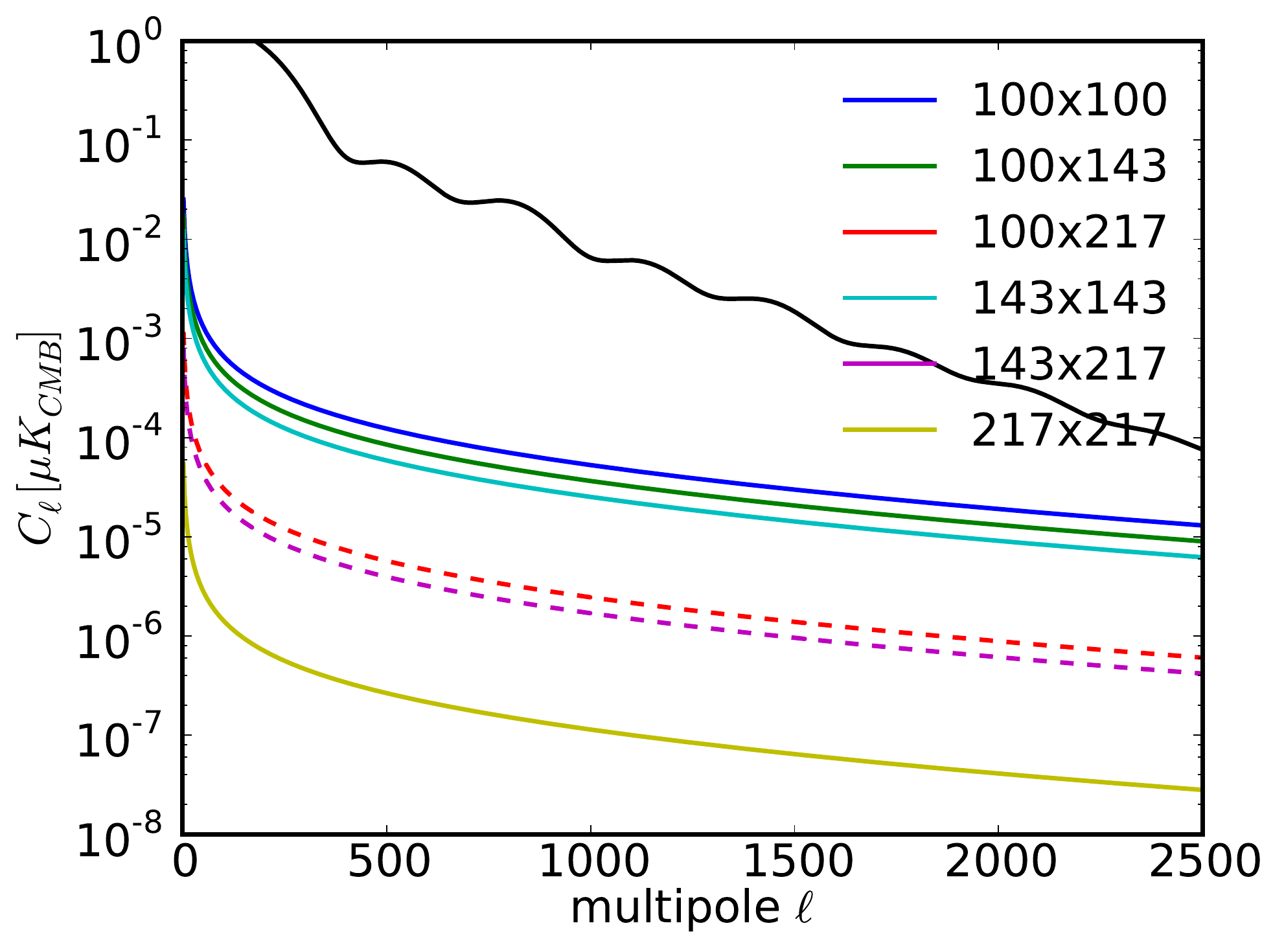}

        \includegraphics[draft=\draft,width=\columnwidth]{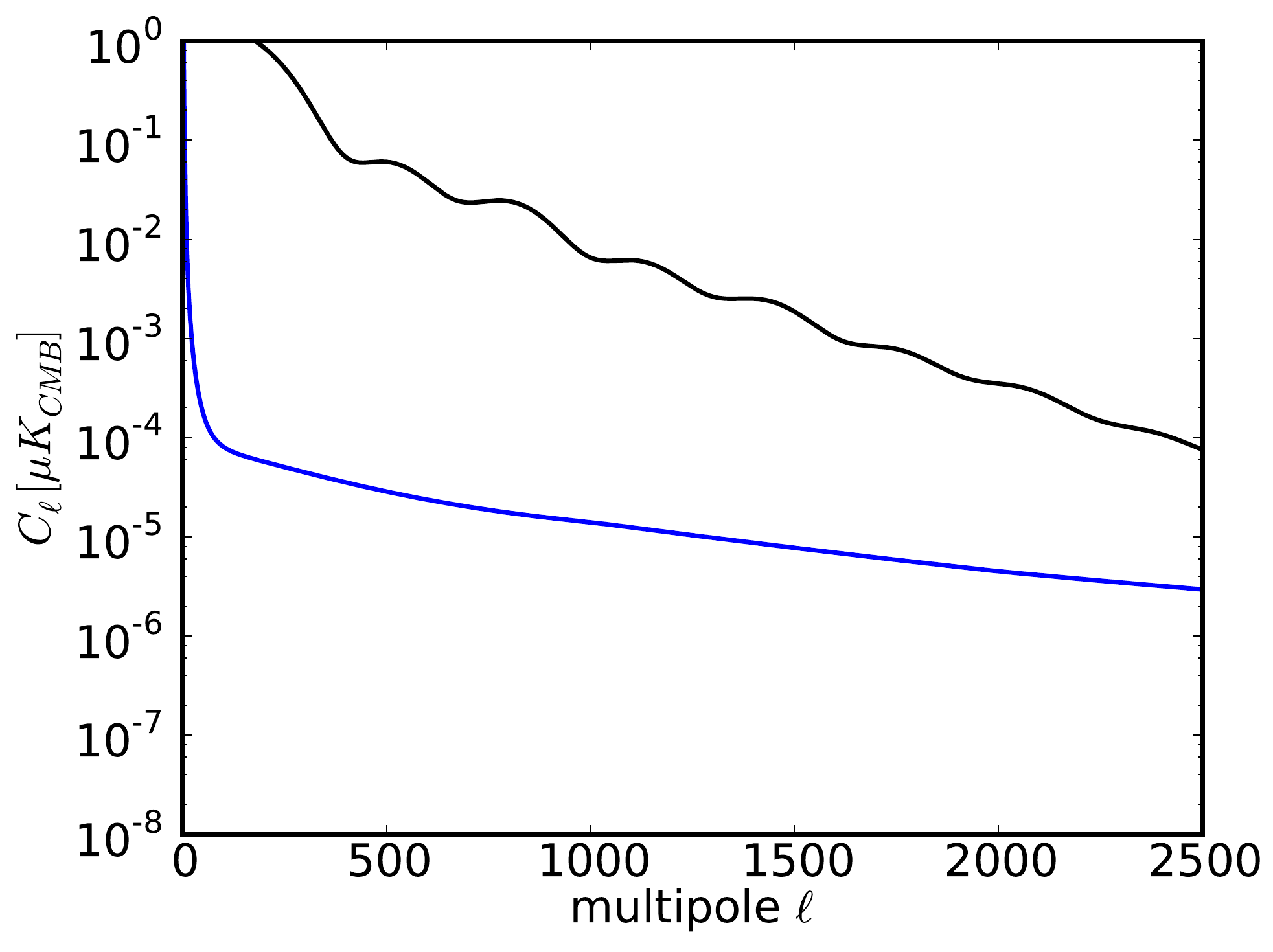}
        \caption{Top:  tSZ power spectra templates at each cross-frequency. Dashed lines are negative. SED are fixed and we fit the overall amplitude $A_{\rm tSZ}$. Bottom: Frequency independent kSZ template. The black line is the CMB power spectrum.}
        \label{fig:tSZ}
\end{figure}

The kinetic Sunyev-Zel'dovich (kSZ) is produced by the peculiar velocities of the clusters containing hot electron gas. We use power spectra extracted from reionization simulations. We supposed that the kSZ follows the same SED as the CMB and only fit a global free amplitude, $A_{\rm kSZ}$.
We chose a combination of templates coming from homogeneous and patchy reionization.
\begin{equation}
        \vec{C}_{\ell}^{ij,\rm kSZ} = A_{\rm kSZ} \, \left( C_{\ell}^{\rm hKSZ} + C_{\ell}^{\rm pKSZ} \right) \, .
        \label{eq:kSZ_model}
\end{equation}
For the homogeneous kSZ, we use a template power spectrum given by \citet{Shaw12} calibrated with a ``cooling and star formation'' simulation. For the patchy reionization kSZ we use the fiducial model of \citet{Battaglia13}.
Both templates are shown in Fig.~\ref{fig:tSZ}. The \planck\ official likelihood considers a template from homogeneous reionization only, but the impact on the cosmology is completely negligible.

\subsubsection{tSZxCIB correlation}
\label{sec:tSZxCIB_model}
The halo model can naturally account for the correlation between two different source populations, each tracing the underlying dark matter, but having different dependence on host halo properties \citep{addison:2012}.
An angular power spectrum can thus be extracted for the correlation between unresolved clusters contributing to the tSZ effect and the dusty sources that make up the CIB. While the latter has a peak in redshift distribution between $z \simeq 1$ and $z \simeq 2$, and is produced by galaxies in dark matter halos of $10^{11}$-$10^{13}$ ${\rm M_\odot}$, tSZ is mainly produced by local ($z < 1$) and massive dark matter halos (above $10^{14}$ ${\rm M_\odot}$). This implies that the CIB and tSZ distributions present a very small overlap for the angular scales probed by \planck, and their correlation is thus hard to detect \citep{planck2014-a29}.

\begin{figure}[!ht]
        \centering
        \includegraphics[draft=\draft,width=\columnwidth]{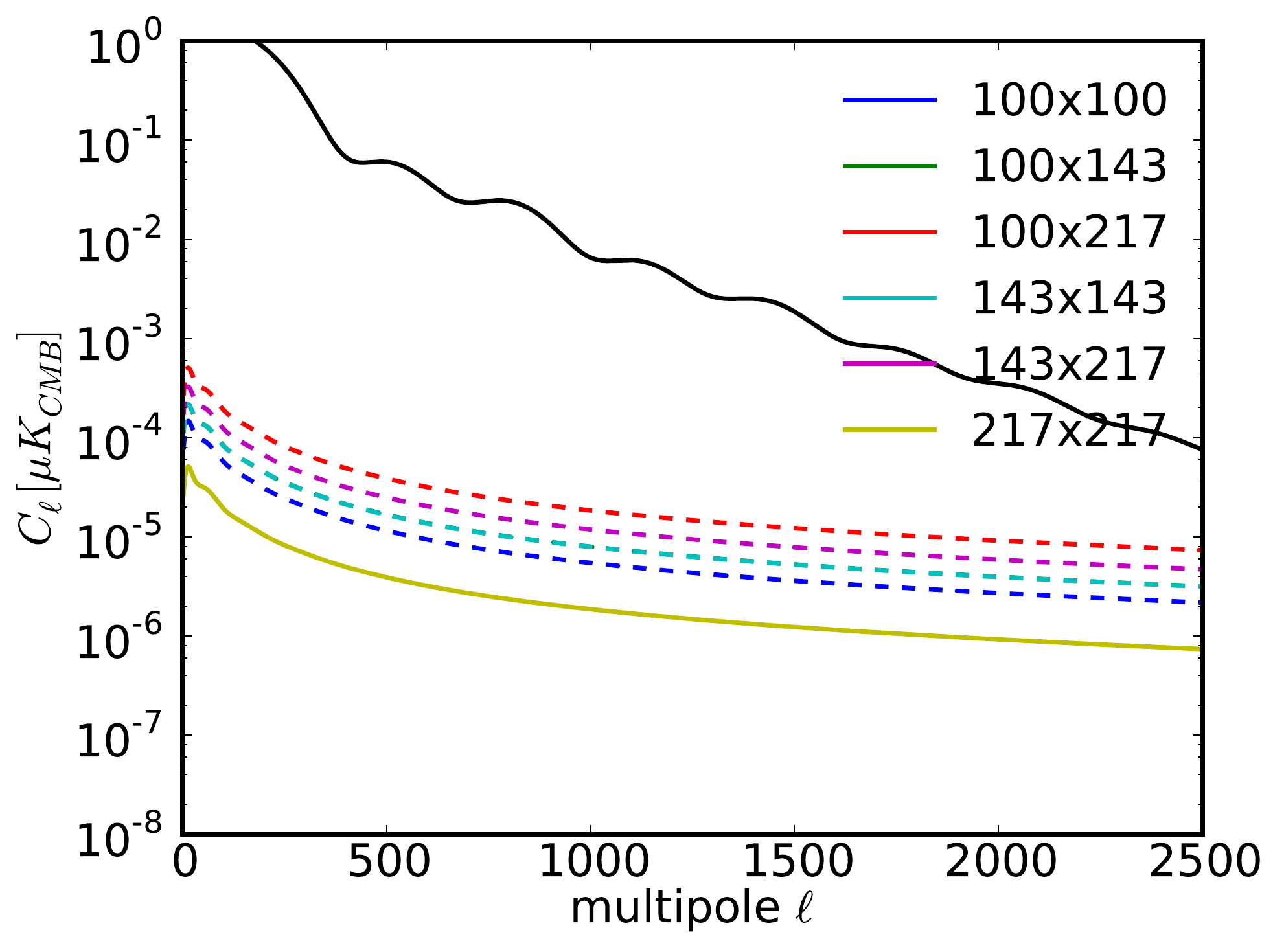}
        \caption{ tSZxCIB power spectra templates. The SED and the angular dependence are fixed. Dashed lines are negative.}
        \label{fig:tSZxCIB}
\end{figure}

We use the templates shown in Fig.~\ref{fig:tSZxCIB}, computed using a tSZ power spectrum template based on \citet{efstathiou:2012} and a CIB template as described in Sect.~\ref{sec:CIB_model}.
The power spectra templates in~$\mathrm{Jy}^2\mathrm{sr}^{-1}$ (with the convention $\nu I(\nu)$=cst) are then converted to~$\mu {\rm K}_{\rm CMB}^2$ using the same coefficients as for the CIB (Sect.~\ref{sec:CIB_model}).

As for the other foregrounds, we then allow for a global free amplitude, $A_{\rm tSZxCIB}$, and write
\begin{equation}
        \vec{C}_{\ell}^{ij,\rm tSZxCIB} = A_{\rm tSZxCIB} \, a^{\rm conv}_{\nu_i} a^{\rm conv}_{\nu_j} C_{\ell}^{\nu_i\nu_j,\rm temp} \, .
        \label{eq:tSZxCIB_model}
\end{equation}

\subsubsection{Unresolved PS}
\label{sec:ps_model}
At \planck\ frequencies, the unresolved point sources signal incorporates the contribution from extragalactic radio and infrared dusty galaxies \citep{tucci:2005}. We use a specific mask for each frequency to mitigate the impact of strong sources (see Sect.~\ref{sec:data:maps}).
\citet{planck2014-a13} gives the expected amplitudes for the Poisson shot noise from theoretical models that predict number counts $dN/dS$ for each frequency.
Their analyses take into account the details of the construction for the point source masks, such as the fact that the flux cut varies across the sky or the ``incompleteness'' of the catalogue from which the masks are built at each frequency.
We computed the expectation at each cross-frequency for the point source amplitudes ($a_{\nu_i,\nu_j}^{radio}$ for the radio sources and $a_{\nu_i,\nu_j}^{IR}$ for the infrared sources) based on the flux-cut considered for our own point sources masks (see Sect.~\ref{sec:data:maps}) using the model from \citet{tucci:2011} for the radio sources and from \citet{bethermin:2012} for dusty galaxies (see Table~\ref{tab:shot_noise}). We note that we found different prediction numbers for radio galaxies than those reported in Table 17 of \citet{planck2014-a13}.

We consider a flat Poisson-like power spectrum for each component and rescale by two free amplitudes $A^{\rm radio}_{\rm PS}$ and $A^{\rm IR}_{\rm PS}$:
\begin{equation}
        \vec{C}_{\ell}^{ij,\rm PS} = A_{\rm PS}^{\rm radio} \, a_{\nu_i,\nu_j}^{radio} + A_{\rm PS}^{\rm IR} \, a_{\nu_i,\nu_j}^{IR} \, .
        \label{eq:ps_model}
\end{equation}
In polarization, we neglect the point source contribution from both components~\citep{tucci:2004}.

It is important to notice that building a reliable multi-frequency model for the unresolved sources is difficult. Indeed, it depends on the flux-cut used to construct each mask, but also on the procedure used to identify spurious detections of high-latitude Galactic cirrus as point sources in the catalogue.  The uncertainty on the flux-cut estimation is particularly important in the case of radio sources as the flux-cuts considered for CMB analysis (typically around 200\,mJy) are close to the peak of the number count.
That is the main reason why the \planck\ public likelihood analysis  considers one amplitude for point sources per cross-spectrum.

\begin{figure*}[!ht]
        \center
        \includegraphics[width=0.9\textwidth]{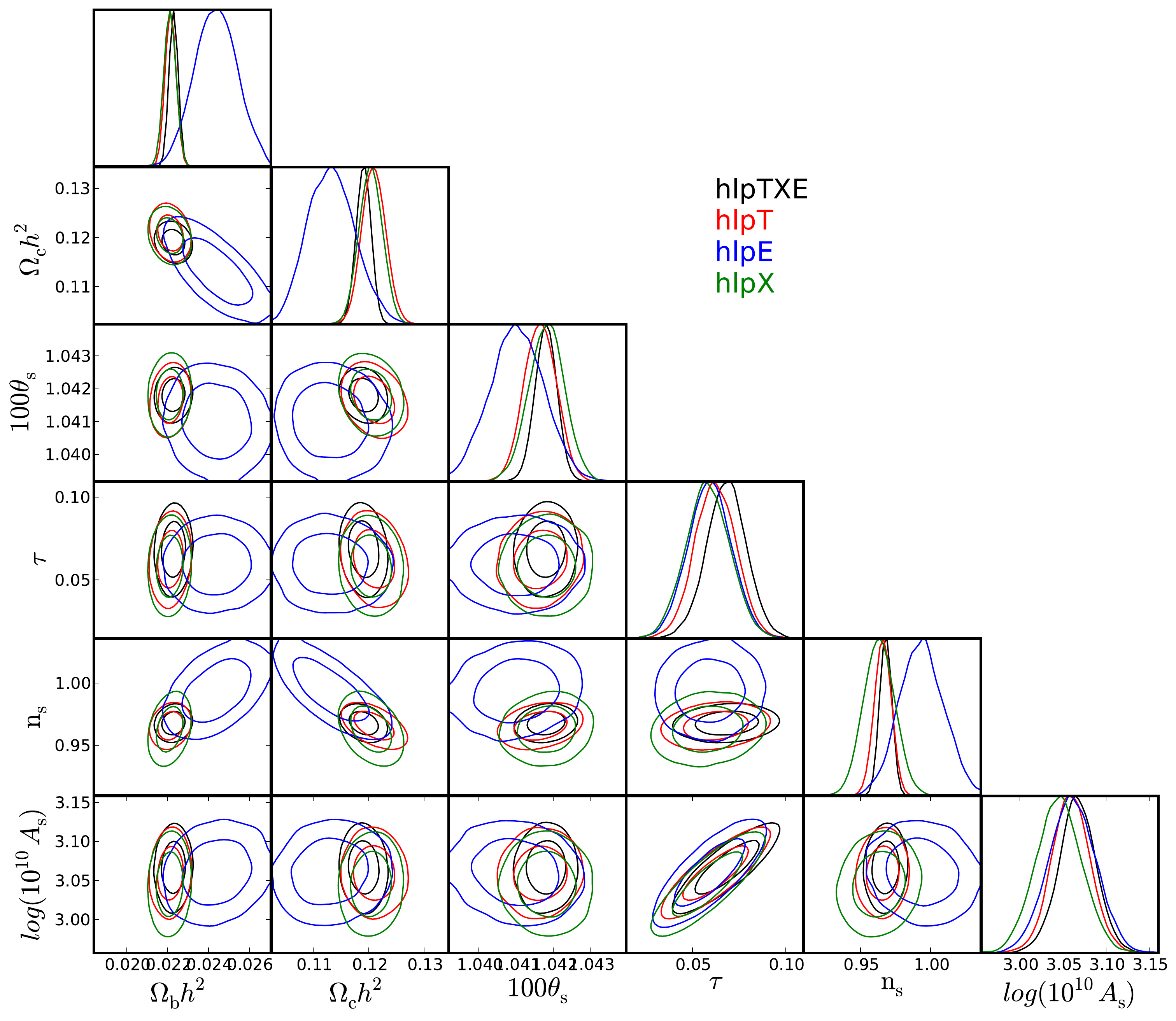}
        \caption{Posterior distribution for the six cosmological \lcdm\ parameters for \hillipop\ and a prior on $\tau$ ($0.058 \pm 0.012$).}
        \label{fig:cosmo}
\end{figure*}

\subsection{Additional priors}
\label{sec:lik:priors}
The various parameters considered in the model described in this section are not all well constrained by the CMB data themselves.
We complement our model with additional priors coming from external knowledge.

For the instrumental nuisances, Gaussian priors are applied on the calibration coefficients based on the uncertainty estimated in \citet{planck2014-a09}: $c_0 = c_1 = c_3 = 0\pm0.002$, $c_4 = c_5 = 0.002\pm0.004$ (Table 5), and $A_{\rm pl}=1\pm0.0025$.

Given its angular resolution, \planck\ is not equally able to constrain the different astrophysical emissions.
We choose to apply Gaussian priors on the dominant ones, including galactic dust, CIB, thermal-SZ, and point sources. The width of the priors is driven by the uncertainty of the foreground modelling. We recall that this model tries to capture residuals from highly non-Gaussian and non-isotropic emission using the template in $C_\ell$ with fixed spectral energy densities (SED). As a consequence, it is difficult to derive an accurate estimation of the expected amplitudes.
We used a Gaussian centred on one  with a 20\% width ($1.0 \pm 0.2$) as priors for the rescaling amplitudes of the five foregrounds ($A_{\rm dust}$, $A_{\rm CIB}$, $A_{\rm tSZ}$, $A_{\rm PS}^{\rm radio}$, and $A_{\rm PS}^{\rm IR}$).

The \planck\ collaboration suggests the addition of a 2D prior on both amplitudes of tSZ and kSZ in order to mimic the constraints from the high-resolution experiments ACT and SPT~\citep[see][]{planck2014-a13}. As demonstrated in \citet{couchot:2015}, this is not strictly equivalent, in particular for results on $A_{\rm L}$. We choose to leave the correlation free.

\section{\hillipop\ Results}
\label{sec:results}
This section is dedicated to the results derived with the \hillipop\ likelihood functions (\hlp TXE, \hlp T, \hlp E, and \hlp X). 
We discuss the cosmological parameters as well as the astrophysical foregrounds and instrumental nuisance. 
We pay particular attention to the difference between the results obtained with $TT$ spectra (\hlp T) and those obtained with $TE$ spectra (\hlp X).

We choose not to use any low-$\ell$ information and prefer to apply a simple prior on the optical reionization depth ($\tau = 0.058 \pm 0.012$) as given by the \emph{lollipop} likelihood in \citet{planck2014-a25}. We have checked that, for the \lcdm\ model, the parameters are undistinguishable when using the corresponding \planck\ low-$\ell$ likelihood. 
We use the Gaussian priors on the inter-calibration coefficients and on astrophysical rescaling factors (dust, CIB, tSZ, and point sources) as discussed in Sect.~\ref{sec:lik:priors}.

The results described here were  obtained using the adaptative-MCMC algorithm implemented in the CAMEL toolbox\footnote{available at \href{camel.in2p3.fr}{camel.in2p3.fr}}. We use the \texttt{CLASS}\footnote{\href{http://class-code.net}{http://class-code.net}} software to compute spectra models for a given cosmology.

\begin{table}[!ht]
        \begin{center}
        \begin{tabular}{lccc}
        \hline
        \hline
        Likelihood       & $\chi^2$     &  $n_{\rm dof}$        &  $\chi^2$/$n_{\rm dof}$  \\
        \hline
\hlp TXE & 27888.3 & 25597 & 1.090 \\
\hlp T & 9995.9 & 9543 & 1.047 \\
\hlp X & 9319.9 & 8799 & 1.059 \\
\hlp E & 7304.5 & 7249 & 1.008 \\
        \hline
        \end{tabular}
        \caption{$\chi^2$ values compared to the number of degree of freedom ($n_{\rm dof} = n_\ell - n_{\rm p}$) for each of the \hillipop\ likelihoods.}
        \label{tab:chi2}
        \end{center}
\end{table}
The $\chi^2$ values of the best fit for each \hillipop\ likelihood are given in Table~\ref{tab:chi2}. Using our simple foreground model, we are able to fit the \planck\ data with reasonable $\chi^2$ values and reduced-$\chi^2$ comparable to the Planck public likelihood (the absolute values are not directly comparable since the Planck public likelihood uses binned cross-power spectra and different foreground modelling). We note that \hlp T and \hlp X show comparable $\chi^2$ with a similar number of degrees of freedom.

\subsection{\lcdm\ cosmological results}
\label{sec:results:cosmo}

\begin{table*}
\begin{center}
\begin{tabular}{lcccc}
\hline
\hline
Parameters                              &              \hlp T                    &               \hlp X                  &            \hlp E                      &             \hlp TXE                  \\  
\hline
$\Omega_\mathrm{b}h^2$  & $0.02212 \pm 0.00021$ & $0.02210 \pm 0.00024$ & $0.02440 \pm 0.00106$ & $0.02227 \pm 0.00014$ \\ 
$\Omega_\mathrm{c}h^2$  & $0.1209 \pm 0.0021$   & $0.1204 \pm 0.0020$   & $0.1130 \pm 0.0043$   & $0.1191 \pm 0.0012$   \\ 
$100\theta_\mathrm{s}$  & $1.04164 \pm 0.00043$ & $1.04184 \pm 0.00047$ & $1.04101 \pm 0.00074$ & $1.04179 \pm 0.00028$ \\ 
$\tau$  & $0.062 \pm 0.011$     & $0.059 \pm 0.012$     & $0.059 \pm 0.012$     & $0.067 \pm 0.011$     \\ 
$\mathrm{n}_\mathrm{s}$ & $0.9649 \pm 0.0058$   & $0.9631 \pm 0.0108$   & $0.9939 \pm 0.0158$   & $0.9672 \pm 0.0037$   \\ 
$log(10^{10}A_\mathrm{s})$      & $3.058 \pm 0.022$     & $3.046 \pm 0.027$     & $3.061 \pm 0.027$     & $3.065 \pm 0.022$     \\ 
\hline
\end{tabular}
\caption{Central value and 68\% confidence limit for the base $\Lambda$CDM model with \hillipop\ likelihoods with a prior on $\tau$ ($0.058 \pm 0.012$).}
\label{tab:lcdm}
\end{center}
\end{table*}

Figure~\ref{fig:cosmo} shows the posterior distributions of the six \lcdm\ parameters reconstructed from each likelihood and their combinations, which are summarized in Table~\ref{tab:lcdm}.
We find very consistent results for cosmology between all the likelihoods.
For \hlp E, we find a $\sim2\sigma$ tension on both $n_s$ and $\Omega_b$, which is not related to the foregrounds or to the multipole range or the sky fraction.

Almost all parameters are compatible with the \planck\ results \citep{planck2014-a15} within 0.5$\sigma$ when considering the temperature data only or the full likelihood. 
Error bars from the \planck\ public likelihood and \hillipop\ as presented in this paper are nearly identical.
As discussed in detail in~\citet{couchot:2015}, the difference in $\tau$ and $A_{\rm s}$ can be understood as a preference of the \hillipop\ likelihood for a lower $A_{\rm L}$ (Sect.~\ref{sec:lcdm+}). In both cases, the shifted value for $A_{\rm L}$ comes from a tension between the high-$\ell$ and the $\tau$ constraint (either from lowTEB or from the prior), the likelihood for \hillipop\ alone showing almost no constraint on $\tau$ when $A_{\rm L}$ is free. 

The results are compatible with those presented in~\citet{couchot:2015}, where we used low-$\ell$ data from \planck-LFI (instead of a tighter prior on $\tau$ from the last results of \planck-HFI). We also now impose a model for the point source frequency spectrum (radio sources and infrared sources) which increased the sensitivity in $n_s$ by $\sim$15\%.

The \hlp X likelihood is almost as sensitive as \hlp T to \lcdm\ parameters, although  the signal-to-noise ratio is lower in the $TE$ spectra. As we  discuss in Sect.~\ref{sec:systematics}, this comes from the uncertainties on the foreground parameters which increase the width of the \hlp T posteriors. 
This is also the case for the Hubble parameter $H_0$ for which we find
\begin{eqs}
        H_0 &=& 67.09 \pm 0.86  \quad \text{(\hlp T)}\\
        H_0 &=& 67.16 \pm 0.89  \quad \text{(\hlp X)},
\end{eqs}
compatible with the low value reported by the \planck\ collaboration \citep{planck2014-a15}.
The only parameter which is significantly less constrained by the $TE$ data is $n_{\rm s}$. Indeed, for a cosmic-variance limited experiment, $TT$ and $TE$ show comparable sensitivity for $n_{\rm s}$; instead,  the \planck\ instrumental noise on $TE$ spectra increases the posterior width by a factor of  almost 2~\citep{galli:2014}.
As expected, the results based on the \hlp E likelihood are even less accurate.

\subsection{Instrumental nuisances}
\label{sec:results:instru}
In the likelihood function, the calibration uncertainties are modelled using an absolute rescaling $A_{\rm pl}$ and inter-calibration factors $c$. The parameter $A_{\rm pl}$ allows the  propagation of  an overall calibration error at the cross-spectra level (which principally translates into a larger error on the amplitude of the primordial power spectrum $A_s$). We apply the same calibration factors for temperature and polarization.

The constraints on inter-calibration coefficients from the \planck\ CMB data are much weaker than the external priors. Without priors, we found that the coefficients are recovered without any bias in all cases with posterior width of typically 1.5\%, 2\%, 7\%, and 5\% for \hlp TXE, \hlp T, \hlp E, and \hlp X, respectively.

\begin{figure}[!ht]
        \center
        \includegraphics[draft=\draft,width=\columnwidth]{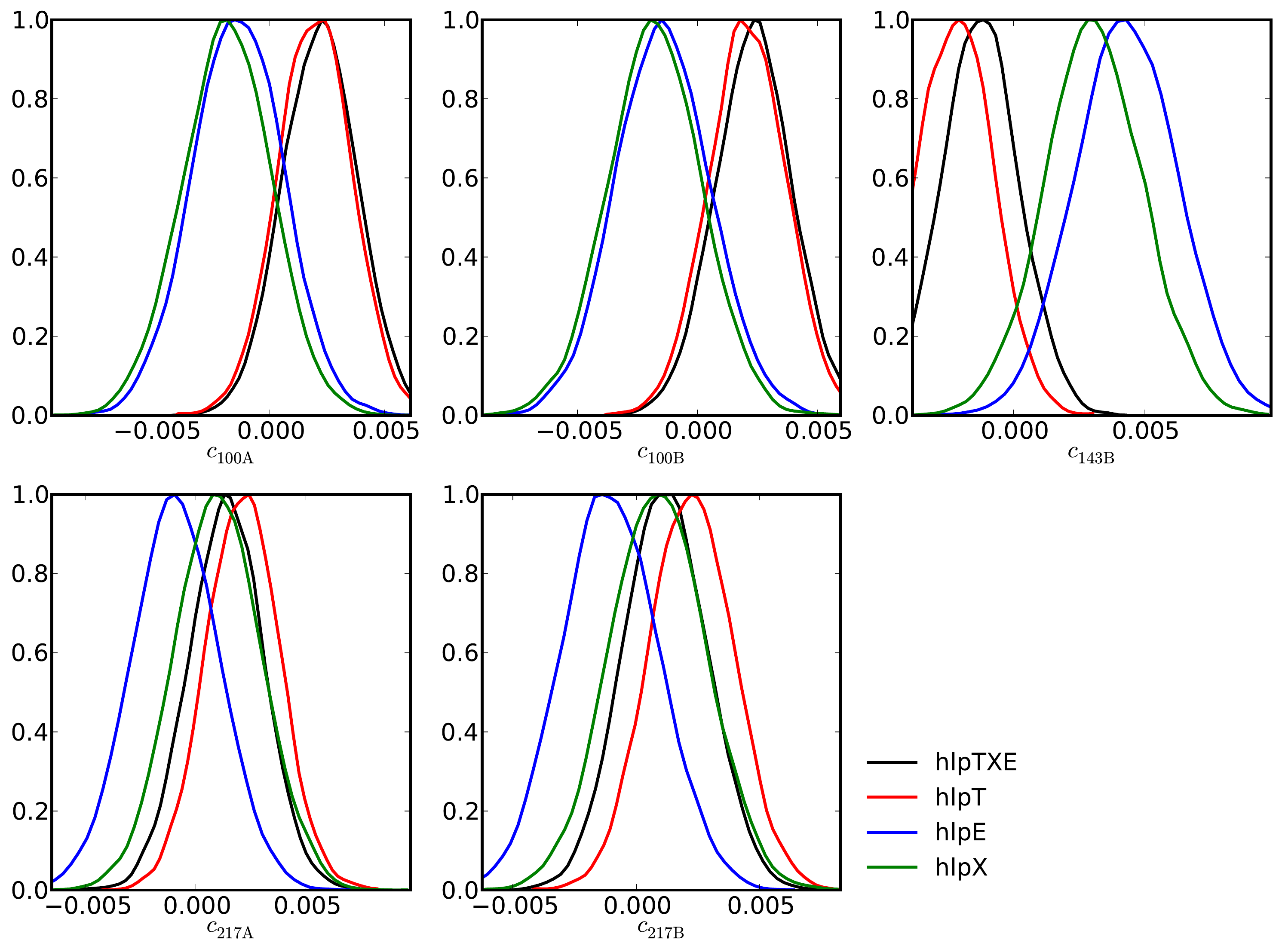}
        \caption{Posterior distribution of the five inter-calibration parameters for each of the \hillipop\ likelihoods (\hlp TXE, \hlp T, \hlp E, and \hlp X).}
        \label{fig:calib_calpriors}
\end{figure}

Figure~\ref{fig:calib_calpriors} shows the posterior distributions for the inter-calibration factors (including external priors described in Sect~\ref{sec:lik:priors}). 
We found a slight tension (less than $2\sigma$) between the calibration factors recovered from temperature and for polarization. 
The relatively bad $\chi_{\rm min}^2$ value of the full likelihood configuration (Table~\ref{tab:chi2}) is certainly partially due to this disagreement between calibrations. 
We tried to take into account the difference between temperature and polarization calibration. To this end we added, in the polarization case, additional new parameters $\epsilon$ (corresponding to the polarization efficiency) through the redefinition $c \rightarrow c(1+\epsilon)$ for the polarization maps. We checked the results with the \hlp X and \hlp TXE likelihoods. The calibrations in temperature are kept fixed and the $\epsilon_i$s are left free in the analysis.  We did not see any improvement of the $\chi_{\rm min}^2$ for the full likelihood. 
The level of the calibration shifts is of the order of one per mil. We have checked that it has a negligible impact on both the cosmology and the astrophysical parameters.

\subsection{Astrophysical results}
\label{sec:results:astro}
We recall that the foregrounds in the \hillipop\ likelihoods are modelled using fixed spectral energy densities (SED) and that, for each emission, the only free parameter is an overall rescaling amplitude (which should be one if the correct SED is used). 
\begin{figure}[!ht]
        \center
        \includegraphics[draft=\draft,width=\columnwidth]{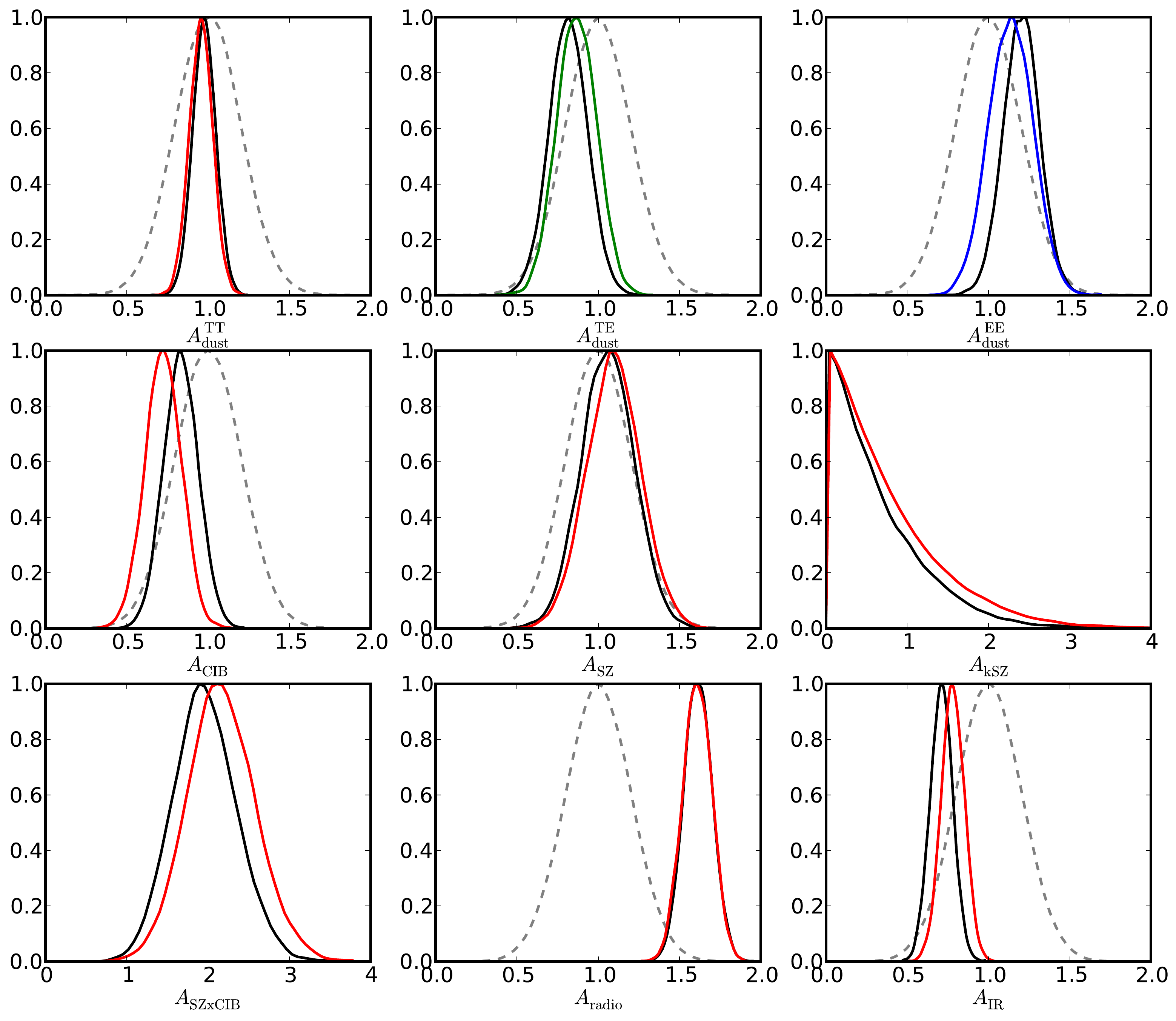}
        \caption{Astrophysical foreground amplitude posterior distributions for the \hillipop\ likelihoods: \hlp TXE ({\it black}), \hlp T ({\it red}), \hlp E ({\it blue}) and \hlp X ({\it green}). Priors are also plotted (grey dashed line).}
        \label{fig:astro}
\end{figure}
The compatibility with one for all foreground amplitudes is thus a good test for the consistency of the internal \planck\ templates. Figure~\ref{fig:astro} shows the posterior distributions for the astrophysical foreground amplitudes. 
We discuss the results in detail in the following sections. We check the stability of the cosmological results with respect to foreground parameters in Sect.~\ref{sec:systematics}.

\subsubsection*{Dust}
The emission of  galactic dust is the dominant residual foreground in the power spectra considered in this analysis. 

The recovered amplitudes for each case (and, in parentheses for the full likelihood) are
\begin{eqs}
        A^{TT}_{\rm dust} = 0.97 \pm 0.09       \quad (0.99 \pm 0.08)\\
        A^{TE}_{\rm dust} = 0.86 \pm 0.12       \quad (0.80 \pm 0.11)\\
        A^{EE}_{\rm dust} = 1.14 \pm 0.13       \quad (1.20 \pm 0.11)
\end{eqs}
The dust amplitude in temperature is recovered perfectly.
The amplitude for the $EE$ polarization mode is found to be slightly high at 1.5$\sigma$, while the $TE$ polarization mode is low at about 1.5$\sigma$.
When using the full \hillipop\ likelihood, the tension on the dust polarization modes $EE$ and $TE$ reaches 2$\sigma$ which is directly related to the small tension on calibration discussed in Sect.~\ref{sec:results:instru}.

\subsubsection*{CIB}
The second emission to which \planck\ $TT$ CMB power spectra are sensitive is the CIB. 
The $A_{\rm CIB}$ recovered for \hlp T and \hlp TXE are, respectively,
\begin{align}
        A_{\rm CIB} = 0.84 \pm 0.15     \quad (1.01 \pm 0.13) \, ,
\end{align}
which is perfectly compatible with the astrophysical measurement from \planck\ for \hlp TEX and at $1\sigma$ for \hlp T.

\subsubsection*{SZ}
\planck\ data are only mildly sensitive to SZ components. 
In particular, we have no constraint at all on the amplitude of the kSZ effect ($A_{\rm kSZ}$) and the correlation coefficient between SZ and CIB ($A_{\rm tSZxCIB}$).
When using astrophysical foreground information, the external prior on $A_{\rm tSZ}$ drives the final posterior:
\begin{align}
        A_{\rm SZ} = 1.00 \pm 0.20      \quad (0.94 \pm 0.19) \, .
\end{align}

\subsubsection*{Point Sources}
\label{sec:res:ps}

\begin{table*}[!ht]
        \center
        \begin{tabular}{lccc|cc}
        \hline
        \hline
                                        & Radio                 & IR            & Total   & \hlp T                        & \hlp TXE                      \\
        \hline
100$\times$100 & $7.8 \pm 1.6$ & ~~$0.2 \pm 0.0$  & ~~$7.9 \pm 1.6$ & $15.5 \pm 1.4$  & $15.8 \pm 0.9$ \\ 
100$\times$143 & $5.4 \pm 1.1$ & ~~$0.5 \pm 0.1$  & ~~$5.8 \pm 1.1$ & $10.4 \pm 1.5$  & $10.5 \pm 1.0$ \\ 
100$\times$217 & $4.3 \pm 0.9$ & ~~$1.9 \pm 0.4$  & ~~$6.2 \pm 1.0$ & $10.1 \pm 1.7$  & $10.0 \pm 1.4$ \\ 
143$\times$143 & $4.8 \pm 1.0$ & ~~$1.2 \pm 0.2$  & ~~$6.1 \pm 1.0$ & ~~$6.3 \pm 1.7$  & ~~$5.9 \pm 1.2$ \\ 
143$\times$217 & $3.6 \pm 0.8$ & ~~$5.1 \pm 1.0$  & ~~$8.7 \pm 1.3$ & ~~$6.2 \pm 1.8$  & ~~$5.3 \pm 1.5$ \\ 
217$\times$217 & $3.2 \pm 0.8$ & $21.0 \pm 3.8$  & $24.2 \pm 3.8$ & $16.7 \pm 2.2$  & $15.0 \pm 2.1$ \\ 
        \hline
        \end{tabular}
        \caption{Poisson amplitudes for radio galaxies \citep[model from][]{tucci:2011} and dusty galaxies \citep[model from][]{bethermin:2012} compared to \hillipop\ results. Units: Jy$^2$.sr$^{-1}$ ($\nu I_\nu = cte$).}
        \label{tab:shot_noise}
\end{table*}

\begin{figure}[!ht]
        \includegraphics[draft=\draft,width=\columnwidth]{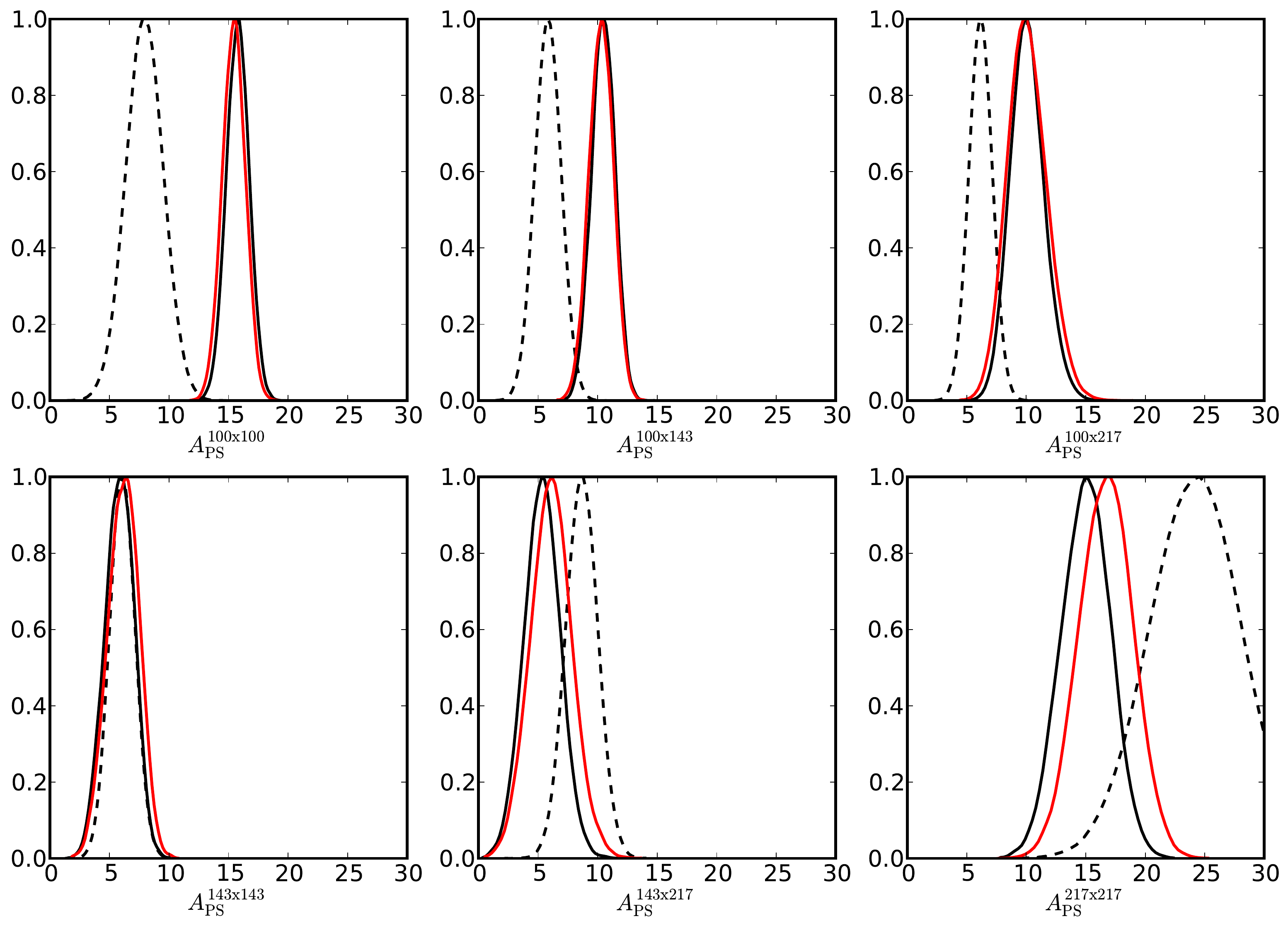}
        \caption{Posterior distributions for the six point sources amplitudes for \hlp TXE ({\it black line}) and \hlp T ({\it red line}) compared to model prediction ({\it dashed line}). Units: Jy$^2$.sr$^{-1}$ ($\nu I_\nu = cte$).}
        \label{fig:ps_distrib}
\end{figure}

We find more power in \planck\ power spectra for the radio sources than expected and a bit less for IR sources:
\begin{eqs}
        A_{\rm PS}^{\rm radio}  &=& 1.61 \pm 0.09       \quad (1.62 \pm 0.09) \\
        A_{\rm PS}^{\rm IR}             &=& 0.78 \pm 0.07       \quad (0.71 \pm 0.07) \, ,
\end{eqs}
with no impact on cosmology (see Sect.~\ref{sec:systematics}). We have identified that the tension comes essentially from the 100\GHz\ map which dominates the constraints for the radio source amplitude. Table~\ref{tab:shot_noise} shows the results when we fit one amplitude for each cross-spectra and when compared to the model expectation. The distribution of the posteriors for the point sources amplitudes are plotted in Fig.~\ref{fig:ps_distrib}. We find relatively good agreement between the predictions from source counts and the \hillipop\ results, with the exception of the 100$\times$100 where the measurement differ by up to 4$\sigma$ with the prediction. 
This is coherent with the results from the \planck\ collaboration \citep[discussed in Sect.~4.3 of][]{planck2014-a13}. It could be a sign for residual systematics in the data but we recall that an accurate point source modelling is very hard to obtain for a large sky coverage with inhomogeneous noise as such of \planck. This is particularly important for the estimation of the radio sources amplitudes which are sensitive to both catalogue completeness and flux cut estimation.

\section{$A_{\rm L}$ as a robustness test}
\label{sec:lcdm+}

As discussed in \citet{couchot:2015}, the measurement of the lensing effect in the angular power spectra of the CMB anisotropies provides a good internal consistency check for high-$\ell$ likelihoods. The \planck\ public likelihood shows an $A_{\rm L}$ different from one by up to 2.6$\sigma$. 

\begin{figure}[!ht]
        \center
        \includegraphics[width=0.8\columnwidth]{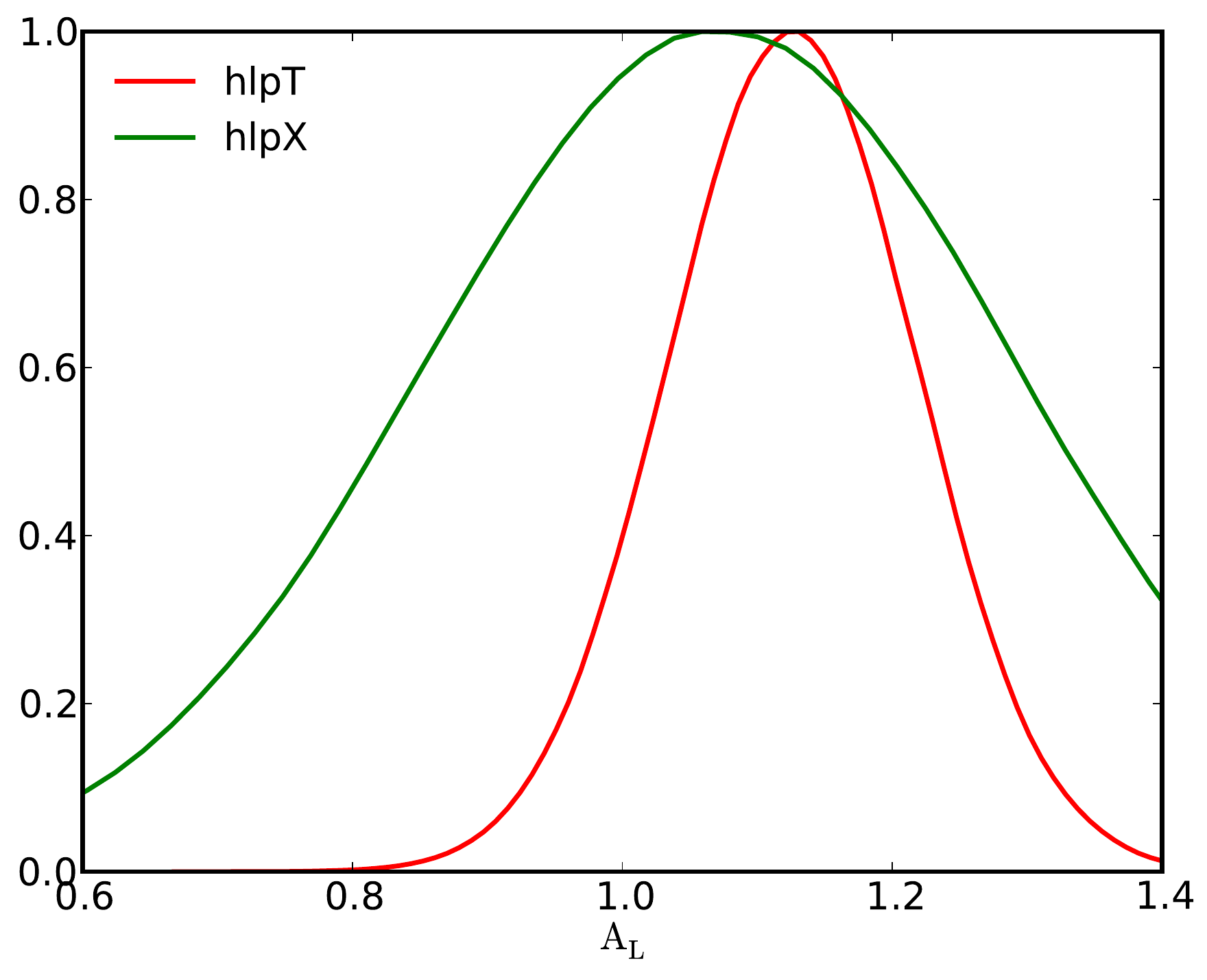}
        \caption{Posterior distribution for the $A_{\rm L}$ parameter for the temperature likelihood \hlp T ({\it red line}) and the temperature-polarization likelihood \hlp X ({\it green line}).}
        \label{fig:prof_alens}
\end{figure}
With \hillipop\ and the $\tau$-prior, the best fits for $A_{\rm L}$ (Fig.~\ref{fig:prof_alens}) are
\begin{eqs}
        A_{\rm L} &=& 1.12 \pm 0.09 \quad \text{(\hlp T + $\tau$ prior)} \\
        A_{\rm L} &=& 1.07 \pm 0.21 \quad \text{(\hlp X + $\tau$ prior)} \, ,
\end{eqs}
compatible with the standard expectation.
While the relative variation of the theoretical power spectra with $A_{\rm L}$ is more important for $TE$ than for $TT$, we find a weaker constraint for $TE$. This illustrates the fact that the noise level in the $TE$ power spectrum from \planck\ is unable to capture the information from the lensing of the CMB $TE$ at high multipoles.

\begin{figure}[!ht]
        \center
        \includegraphics[width=0.8\columnwidth]{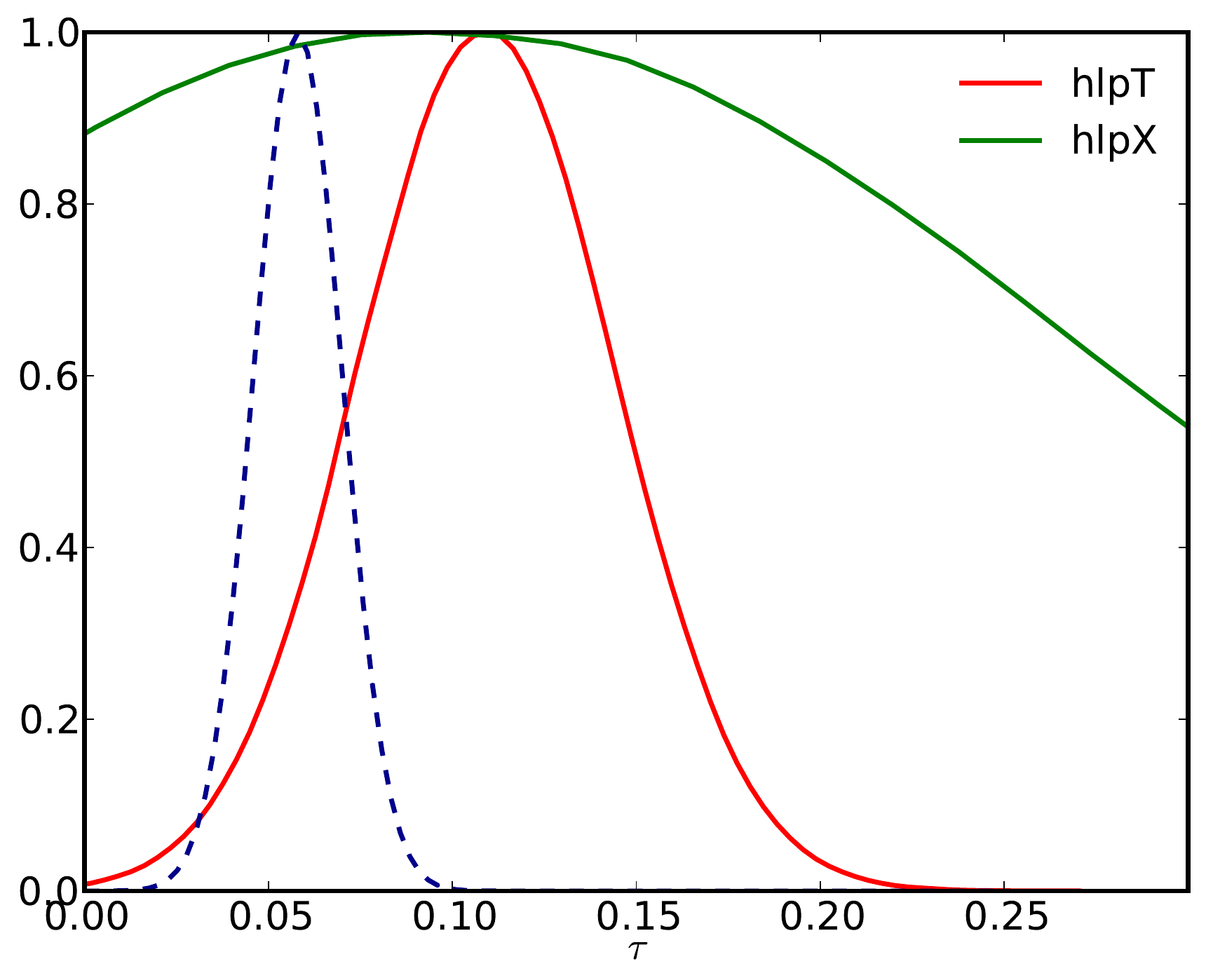}
        \caption{Posterior distribution for the reionization optical depth $\tau$ for \hlp T ({\it red line}) and \hlp X ({\it green line}) compared to the prior from \citet{planck2014-a25} used throughout this analysis ({\it dashed dark blue line}).}
        \label{fig:prof_tau}
\end{figure}

In \citet{couchot:2015}, we have shown that the \planck\ tension on $A_{\rm L}$ is directly related to the constraint on $\tau$. Indeed, the $\tau$ constraints from the \hillipop\ likelihoods (Fig.~\ref{fig:prof_tau}) are less in tension with the \planck\ low-$\ell$ likelihoods. 
The \hillipop\ only likelihoods give
\begin{eqs}
        \tau &=& 0.122 \pm 0.036        \quad \text{(\hlp T)}\\
        \tau &=& 0.103 \pm 0.081        \quad \text{(\hlp X)}
,\end{eqs}
which is, for \hlp T, at 1.7$\sigma$ from the HFI low-$\ell$ analysis $\tau = 0.058 \pm 0.012$ \citep{planck2014-a25}. The difference with the $\tau$ estimation derived in \citet{couchot:2015}  comes directly from the additional constraints in the point source sector. For \hlp X, the $\tau$ distribution is compatible with the \planck\ low-$\ell$ constraint, but the constraint is weaker.

We note that when adding the information from the measurement of the power spectrum of the lensing potential \citep[using the \planck\ lensing likelihood described in ][]{planck2014-a17} the constraints on $\tau$ from \hlp T and \hlp X become comparable
\begin{eqs}
        \tau &=& 0.077 \pm 0.028        \quad \text{(\hlp T + lensing)}\\
        \tau &=& 0.056 \pm 0.027        \quad \text{(\hlp X + lensing)},
\end{eqs}
and compatible with low-$\ell$-only results from both \planck-HFI \citep[$\tau=0.058 \pm 0.012$,][]{planck2014-a25} and \planck-LFI \citep[$\tau = 0.067 \pm 0.023$,][]{planck2014-a13}.

\section{Foreground robustness: TT vs. TE}
\label{sec:systematics}

In this section, we investigate the impact of foregrounds on the recovery of the \lcdm\ cosmological parameters.
We focus on the results from \hlp T and \hlp X.

\begin{figure}[!hb]
        \includegraphics[draft=\draft,width=1.025\columnwidth]{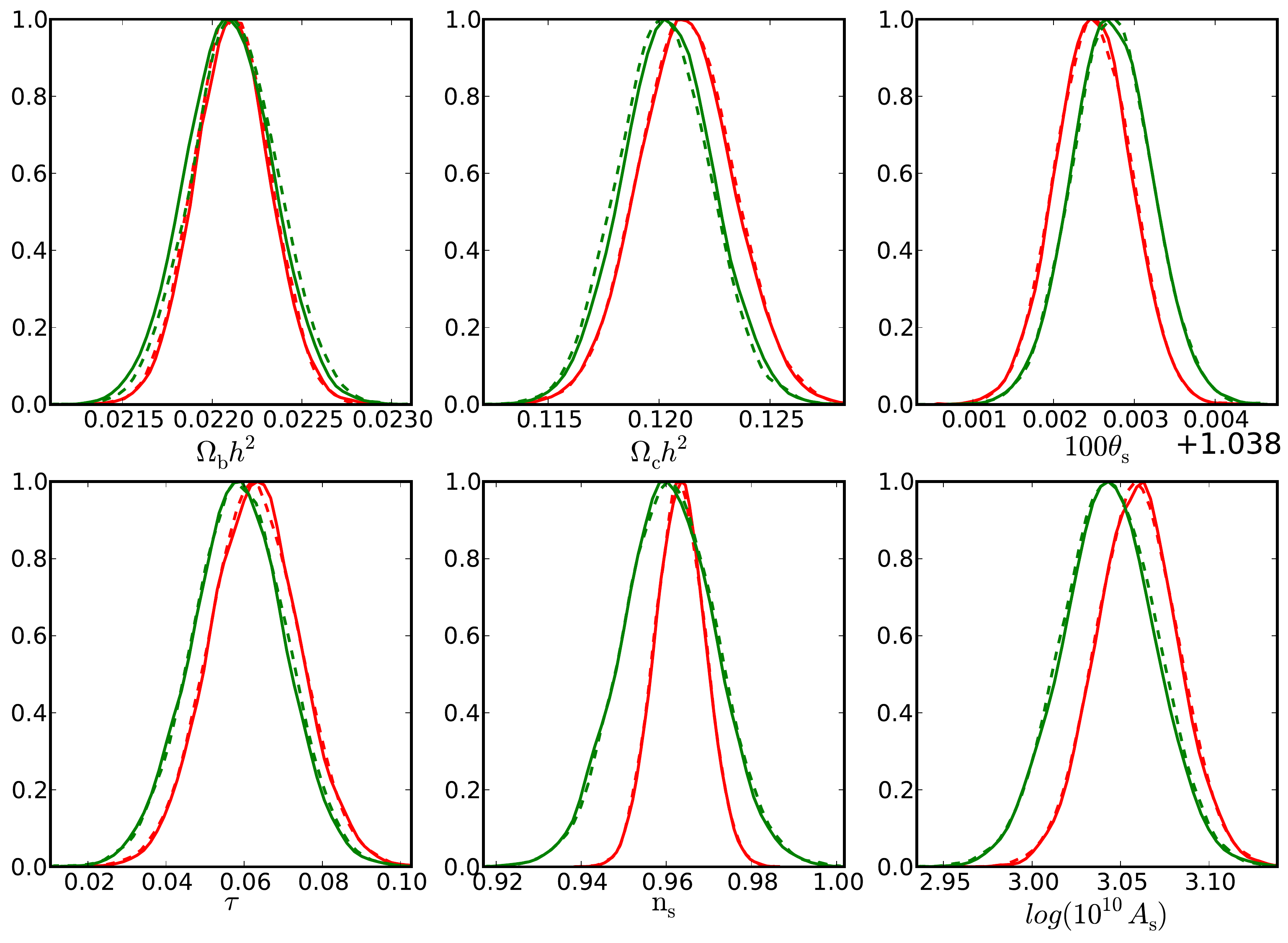}
        \caption{Posterior distributions for the six \lcdm\ parameters with ({\it solid lines}) and without ({\it dashed lines}) astrophysical foregrounds priors in the case of \hlp T ({\it red}) and \hlp X ({\it green}).}
        \label{fig:cosmo_nofg}
\end{figure}

\begin{figure}[!ht]
        \center
        \includegraphics[draft=\draft,width=\columnwidth]{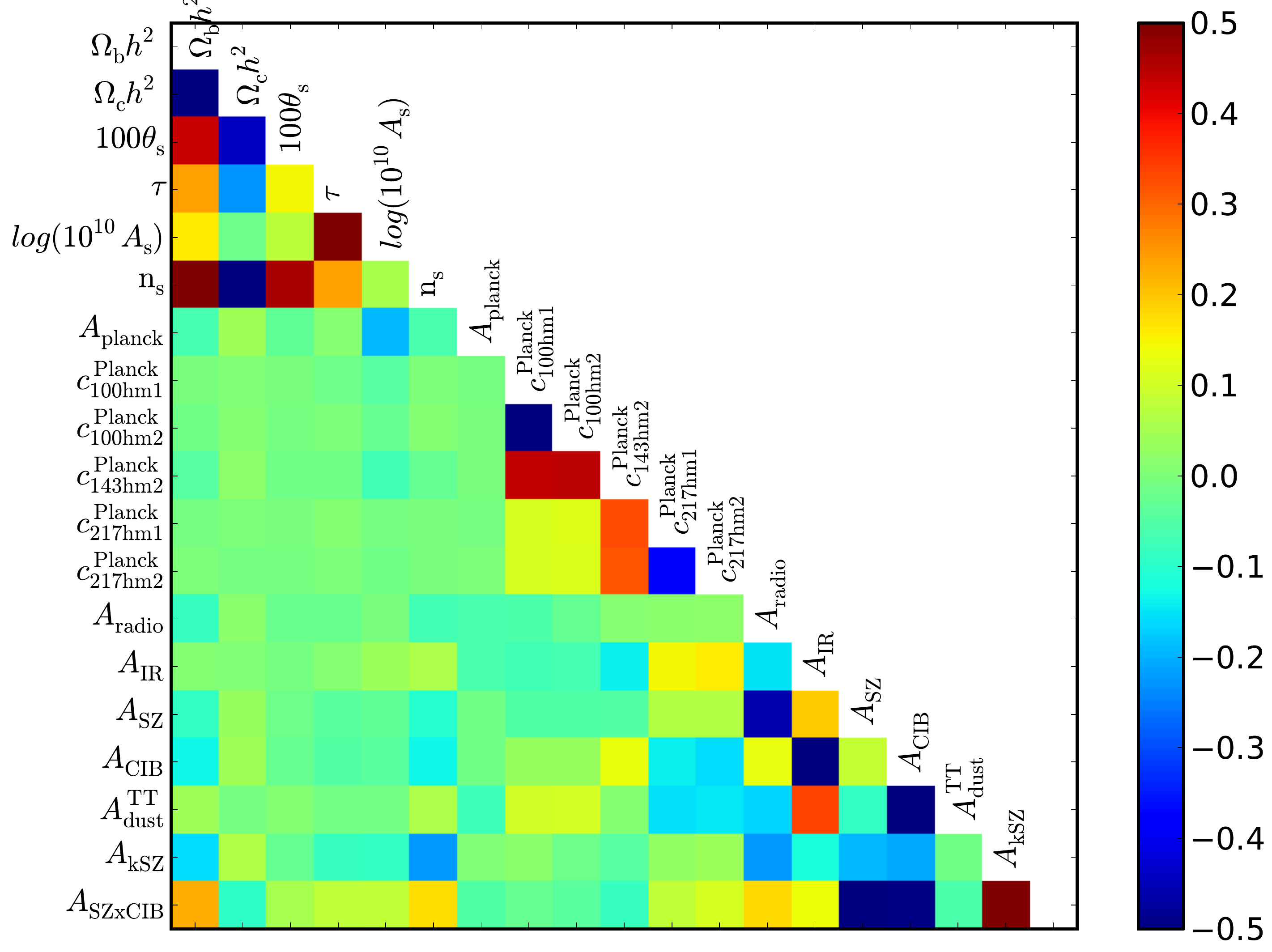}
        \includegraphics[draft=\draft,width=\columnwidth]{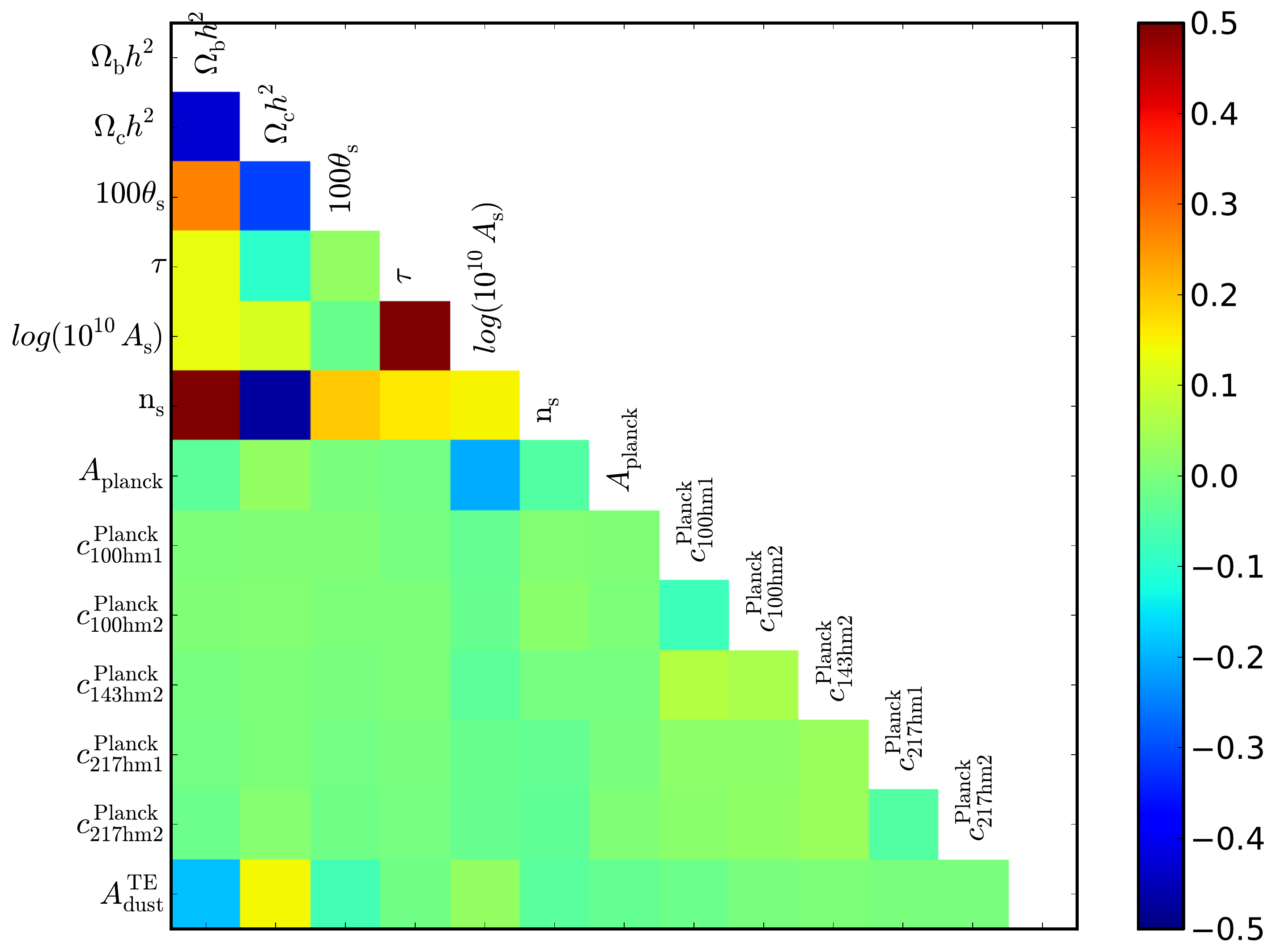}
        \caption{Correlation matrix of the likelihood parameters including \lcdm\ and nuisance parameters for \hlp T ({\it top}) and \hlp X ({\it bottom}). The colour scale is saturated at 50\%.}
        \label{fig:param_corr}
\end{figure}

First, we show in Fig.~\ref{fig:cosmo_nofg}, the posterior for the parameters with and without external foreground priors. These results demonstrate no impact of the priors on the final results, and suggest a low level of correlation between foreground parameters and cosmological parameters in the likelihood. Indeed, the statistics reconstructed from the MCMC samples (Fig.~\ref{fig:param_corr}) exhibit less than 15\% correlation between the two sets of parameters. 
In the case of temperature, we see strong correlations between the instrumental parameters on the one hand, and between the astrophysical parameters on the other hand. This is not the case for \hlp X, which, apart from the cosmological sector, exhibits less than 10\% correlation.

In a second step, we have estimated the contribution of the foreground parameters to the error budget of the cosmological parameters (Table~\ref{tab:errors}). This analysis assesses the degree to which  our uncertainties on the nuisance parameters impact the cosmological error budget. A parameter estimation is performed to assess the full error for each parameter. Then another parameter estimation is performed with the foreground parameters fixed to their best-fit values. The confidence intervals recovered in this last case give the \emph{statistical} uncertainties which are essentially driven by noise and cosmic variance (and which correspond to the errors on parameters if we knew the nuisance parameters perfectly). 
Finally the \emph{foreground} error is deduced by quadratically subtracting the statistical uncertainty from the total error following what was done in \citet{planck2014-a13}.\\
In the temperature case, we see a strong impact of the nuisances on the error of $\Omega_b h^2$ and $n_{s}$. 
The posterior width of the reionization optical depth $\tau$ is strongly dominated by the prior so it is marginally affected by foreground uncertainties.
Finally, even if the statistical uncertainty is larger in the case of $TE$, foreground uncertainties are negligible in the total error budget, which makes them competitive with $TT$ (except for $n_{\rm s}$).

\begin{table}[!ht] 
        \begin{center}
        \begin{tabular}{p{1.2cm}p{1.2cm}p{1.2cm}p{1.2cm}p{0.8cm}p{0.4cm}}
        \hline
        \hline
        Parameter               & Estimate      &       \multicolumn{3}{c}{Error}   \\
                                &                       & Full  & statistical       & foreground & \\
        \hline
        \multicolumn{5}{c}{\hlp T parameters} \\
        \hline
$\Omega_\mathrm{b}h^2$  & $0.02212$&     $0.00020$      & $0.00018$     & $0.00009$& ($27\%$)   \\ 
$\Omega_\mathrm{c}h^2$  & $0.1210$&      $0.0021$       & $0.0021$      & $0.0003$& ($ 3\%$)    \\ 
$100\theta_\mathrm{s}$  & $1.04164$&     $0.00043$      & $0.00044$     & $0.00000$& ($ 0\%$)   \\ 
$\tau$  & $0.062$&       $0.011$        & $0.011$       & $0.002$& ($ 5\%$)     \\ 
$\mathrm{n}_\mathrm{s}$ & $0.9649$&      $0.0058$       & $0.0052$      & $0.0025$& ($24\%$)    \\ 
$log(10^{10}A_\mathrm{s})$      & $3.058$&       $0.022$        & $0.022$       & $0.003$& ($ 2\%$)     \\ 
        \hline
        \multicolumn{5}{c}{\hlp X parameters} \\
        \hline
$\Omega_\mathrm{b}h^2$  & $0.02209$&     $0.00024$      & $0.00024$     & $0.00004$& ($ 3\%$)   \\ 
$\Omega_\mathrm{c}h^2$  & $0.1204$&      $0.0020$       & $0.0020$      & $0.0005$& ($ 6\%$)    \\ 
$100\theta_\mathrm{s}$  & $1.04184$&     $0.00047$      & $0.00047$     & $0.00003$& ($ 0\%$)   \\ 
$\tau$  & $0.058$&       $0.012$        & $0.012$       & $0.000$& ($ 0\%$)     \\ 
$\mathrm{n}_\mathrm{s}$ & $0.9630$&      $0.0111$       & $0.0107$      & $0.0026$& ($ 6\%$)    \\ 
$log(10^{10}A_\mathrm{s})$      & $3.046$&       $0.026$        & $0.027$       & $0.000$& ($ 0\%$)     \\ 
        \hline
        \end{tabular}
        \caption{Errors on cosmological parameters within the \lcdm\ model for \hlp T and \hlp X. The full error is split between statistical and foreground errors. Errors are given at 68\,\% confidence level.}
        \label{tab:errors}
        \end{center}
\end{table}

\begin{figure}[!ht]
        \includegraphics[width=\columnwidth]{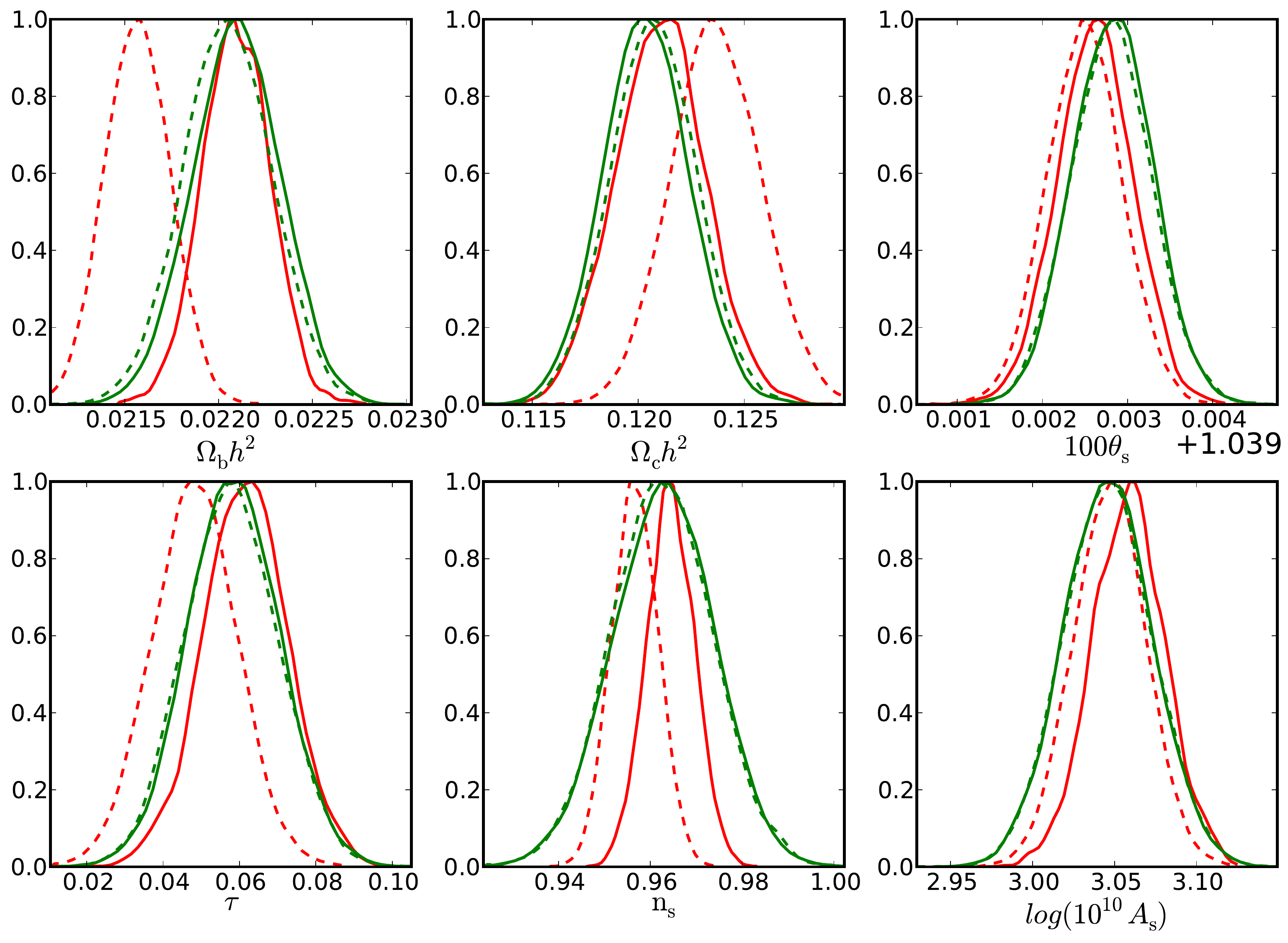}
        \caption{Posterior distribution for the cosmological parameters when foregrounds are fixed either to their best-fit value ({\it solid lines}) or to the expected astrophysical value ({\it dashed lines}) for \hlp T ({\it red}) and \hlp X ({\it green}).}
        \label{fig:fgbias}
\end{figure}

More important than increasing the error budget, nuisance uncertainties can also bias the cosmological parameters.
Figure~\ref{fig:fgbias} shows the results on the \lcdm\ parameters for \hlp T and \hlp X when nuisances are fixed either to their best fit or to the value expected by the astrophysical constraints (i.e. scaling parameters fixed to 1). This corresponds to the extreme case for the potential bias, where we supposed an exact knowledge of the characteristics of the complex spatial distribution of foregrounds and their spectra. The attempt here is to give an idea of the impact of foreground uncertainties on cosmological parameters.
Once again, we see a stronger impact on \hlp T than on \hlp X. In temperature, almost all parameters are shifted when changing the nuisance values, the strongest effect being for $\Omega_b h^2$, $\Omega_c h^2$, and $n_s$. 
On the contrary, we cannot see any impact of the $A_{\rm dust}^{TE}$ parameter shift even if its best-fit value is at $0.86$ compared to $1$.

\section{Discussion}
With the currently available CMB measurements, the sensitivity to \lcdm\ cosmological parameters is dominated by the \planck\ data in the $\ell$-range typically below $\ell=2000$ both in $TT$ and $TE$.
For $TE$, adding higher multipoles coming from the measurements of the South Pole Telescope~\citep{crites:2015} or the Atacama Cosmology Telescope~\citep{naess:2014}, we find almost identical results on \lcdm\ cosmological parameters without any reduction of parameter uncertainties. 
This is different from the temperature data for which high-resolution experiments help to reduce the uncertainty on foreground parameters, which indirectly reduces the posterior width for cosmological parameters through their correlation~\citep{couchot:2015}. 
On the low-$\ell$ side, measurements of $TE$ at $\ell$ < 20 give information about the reionization optical depth $\tau$ (although not equivalent to the low-$\ell$ from $EE$) and a longer lever arm for $n_s$.

We checked the results of the temperature-polarization cross-correlation likelihood on some basic extensions to the \lcdm\ model: essentially $A_{\rm L}$, $N_{\rm eff}$, and $\sum$$m_\nu$. Given \planck\ sensitivity, we do not find any competitive constraints compared to the temperature likelihood.
For example, we find an effective number of relativistic species $N_{\rm eff} = 2.45 \pm 0.45$ for \hlp X compared to $N_{\rm eff} = 2.95 \pm 0.32$ for \hlp T. Adding data from high-resolution experiments, we find $N_{\rm eff} = 2.84 \pm 0.43$, which does not help reduce the error  to the level of temperature data.

We combine CMB $TE$ data with complementary information from the late time evolution of the Universe geometry, coming from the Baryon Acoustic Oscillations scale evolution~\citep{alam:2016} and the SNIa magnitude-redshift measurements~\citep{betoule:2014}. We find very compatible results with a significantly better accuracy only on $\Omega_{\rm c} h^2$.

\section{Conclusion}
Building a coherent likelihood for CMB data given \planck\ sensitivity is difficult owing to the complexity of the foreground emissions modelling. In this paper, we have presented a full temperature and polarization likelihood based on cross-spectra (including $TT$, $TE$, and $EE$) over a wide range of multipoles (from $\ell=50$ to 2500). 
We have described in detail the foreground parametrization which relies on the \planck\ measurements for astrophysical modelling.

We found results on the \lcdm\ cosmological parameters consistent between the different likelihoods (\hlp T, \hlp X, \hlp E).
The cosmological constraints from this work are directly comparable to the \planck\ 2015 cosmological analysis \citep{planck2014-a15} despite the differences in the foreground modelling adopted in \hillipop.
Both instrumental and astrophysical nuisance parameters are compatible with expectations, with the exception of the point source amplitudes in temperature for which we found a small tension with the astrophysical expectations. This tension may be the sign of potential systematic residuals in \planck\ data and/or uncertainty in the foreground model in temperature (especially on the dust SED or the various $\ell$-shape of the foreground templates).

We investigated the robustness of the results with respect to the foreground and nuisance parameters. In particular, we demonstrated the impact of foreground uncertainties on the temperature power spectrum likelihood. We compared these data  to the results from the likelihood based on temperature-polarization cross-correlation which involves fewer foreground components, but is statistically less sensitive. 
We found that foreground uncertainties have a stronger impact on $TT$ than on $TE$ with comparable final errors (except for $n_s$). Moreover, the \hlp X likelihood function does include fewer nuisance parameters (only 7 compared to 13 for \hlp T) and shows less correlation in the nuisance/foregrounds sectors which,  in practice, allows much faster sampling.

This work illustrates the fact that $TE$ spectra provide an estimation of the cosmological parameters that are as accurate as $TT$ while being more robust with respect to foreground contaminations. The results from \planck\ in polarization are still limited by instrumental noise in $TE$, but as suggested in \citet{galli:2014}, future experiments only limited by cosmic variance over a wider range of multipoles will be able to constrain cosmology with $TE$ even better than with $TT$.

\begin{acknowledgement}
The authors thank G. Lagache for the work on the point source mask and the estimation of the infrared sources amplitude, M. Tucci for the estimation of the radio point source amplitude, and P. Serra for the work on the CIBxtSZ power spectra model.
\end{acknowledgement}

\bibliographystyle{aat} 
\bibliography{Bib/hillipop,Bib/Planck_bib}

\def\eprinttmppp@#1arXiv:@{#1}
\providecommand{\arxivlink[1]}{\href{http://arxiv.org/abs/#1}{arXiv:#1}}
\def\eprinttmp@#1arXiv:#2 [#3]#4@{\ifthenelse{\equal{#3}{x}}{\ifthenelse{
\equal{#1}{}}{\arxivlink{\eprinttmppp@#2@}}{\arxivlink{#1}}}{\arxivlink{#2}
  [#3]}}
\providecommand{\eprintlink}[1]{\eprinttmp@#1arXiv: [x]@}
\providecommand{\eprint}[1]{\eprintlink{#1}}
\providecommand{\adsurl}[1]{\href{#1}{ADS}}
\begin{thebibliography}{42}
\expandafter\ifx\csname natexlab\endcsname\relax\def\natexlab#1{#1}\fi

\bibitem[{{Addison} {et~al.}(2012){Addison}, {Dunkley}, \&
  {Spergel}}]{addison:2012}
{Addison}, G.~E., {Dunkley}, J., \& {Spergel}, D.~N., {Modelling the
  correlation between the thermal Sunyaev Zel'dovich effect and the cosmic
  infrared background}. 2012, \mnras, 427, 1741, \eprint{1204.5927}

\bibitem[{{Alam} {et~al.}(2016){Alam}, {Ata}, {Bailey}, {Beutler}, {Bizyaev},
  {Blazek}, {Bolton}, {Brownstein}, {Burden}, {Chuang}, {Comparat}, {Cuesta},
  {Dawson}, {Eisenstein}, {Escoffier}, {Gil-Mar{\'{\i}}n}, {Grieb}, {Hand},
  {Ho}, {Kinemuchi}, {Kirkby}, {Kitaura}, {Malanushenko}, {Malanushenko},
  {Maraston}, {McBride}, {Nichol}, {Olmstead}, {Oravetz}, {Padmanabhan},
  {Palanque-Delabrouille}, {Pan}, {Pellejero-Ibanez}, {Percival}, {Petitjean},
  {Prada}, {Price-Whelan}, {Reid}, {Rodr{\'{\i}}guez-Torres}, {Roe}, {Ross},
  {Ross}, {Rossi}, {Rubi{\~n}o-Mart{\'{\i}}n}, {S{\'a}nchez}, {Saito},
  {Salazar-Albornoz}, {Samushia}, {Satpathy}, {Sc{\'o}ccola}, {Schlegel},
  {Schneider}, {Seo}, {Simmons}, {Slosar}, {Strauss}, {Swanson}, {Thomas},
  {Tinker}, {Tojeiro}, {Vargas Maga{\~n}a}, {Vazquez}, {Verde}, {Wake}, {Wang},
  {Weinberg}, {White}, {Wood-Vasey}, {Y{\`e}che}, {Zehavi}, {Zhai}, \&
  {Zhao}}]{alam:2016}
{Alam}, S., {Ata}, M., {Bailey}, S., {et~al.}, {The clustering of galaxies in
  the completed SDSS-III Baryon Oscillation Spectroscopic Survey: cosmological
  analysis of the DR12 galaxy sample}. 2016, ArXiv e-prints,
  \eprint{1607.03155}

\bibitem[{{Battaglia} {et~al.}(2013){Battaglia}, {Natarajan}, {Trac}, {Cen}, \&
  {Loeb}}]{Battaglia13}
{Battaglia}, N., {Natarajan}, A., {Trac}, H., {Cen}, R., \& {Loeb}, A.,
  {Reionization on Large Scales. III. Predictions for Low-l Cosmic Microwave
  Background Polarization and High-l Kinetic Sunyaev-Zel'dovich Observables}.
  2013, \apj, 776, 83, \eprint{1211.2832}

\bibitem[{{B{\'e}thermin} {et~al.}(2012){B{\'e}thermin}, {Daddi}, {Magdis},
  {Sargent}, {Hezaveh}, {Elbaz}, {Le Borgne}, {Mullaney}, {Pannella}, {Buat},
  {Charmandaris}, {Lagache}, \& {Scott}}]{bethermin:2012}
{B{\'e}thermin}, M., {Daddi}, E., {Magdis}, G., {et~al.}, {A Unified Empirical
  Model for Infrared Galaxy Counts Based on the Observed Physical Evolution of
  Distant Galaxies}. 2012, \apjl, 757, L23, \eprint{1208.6512}

\bibitem[{{Betoule} {et~al.}(2014){Betoule}, {Kessler}, {Guy}, {Mosher},
  {Hardin}, {Biswas}, {Astier}, {El-Hage}, {Konig}, {Kuhlmann}, {Marriner},
  {Pain}, {Regnault}, {Balland}, {Bassett}, {Brown}, {Campbell}, {Carlberg},
  {Cellier-Holzem}, {Cinabro}, {Conley}, {D'Andrea}, {DePoy}, {Doi}, {Ellis},
  {Fabbro}, {Filippenko}, {Foley}, {Frieman}, {Fouchez}, {Galbany}, {Goobar},
  {Gupta}, {Hill}, {Hlozek}, {Hogan}, {Hook}, {Howell}, {Jha}, {Le Guillou},
  {Leloudas}, {Lidman}, {Marshall}, {M{\"o}ller}, {Mour{\~a}o}, {Neveu},
  {Nichol}, {Olmstead}, {Palanque-Delabrouille}, {Perlmutter}, {Prieto},
  {Pritchet}, {Richmond}, {Riess}, {Ruhlmann-Kleider}, {Sako}, {Schahmaneche},
  {Schneider}, {Smith}, {Sollerman}, {Sullivan}, {Walton}, \&
  {Wheeler}}]{betoule:2014}
{Betoule}, M., {Kessler}, R., {Guy}, J., {et~al.}, {Improved cosmological
  constraints from a joint analysis of the SDSS-II and SNLS supernova samples}.
  2014, \aap, 568, A22, \eprint{1401.4064}

\bibitem[{{Couchot} {et~al.}(2017){Couchot}, {Henrot-Versill{\'e}},
  {Perdereau}, {Plaszczynski}, {Rouill{\'e} d'Orfeuil}, {Spinelli}, \&
  {Tristram}}]{couchot:2015}
{Couchot}, F., {Henrot-Versill{\'e}}, S., {Perdereau}, O., {et~al.}, {Relieving
  tensions related to the lensing of the cosmic microwave background
  temperature power spectra}. 2017, \aap, 597, A126, \eprint{1510.07600}

\bibitem[{{Crites} {et~al.}(2015){Crites}, {Henning}, {Ade}, {Aird},
  {Austermann}, {Beall}, {Bender}, {Benson}, {Bleem}, {Carlstrom}, {Chang},
  {Chiang}, {Cho}, {Citron}, {Crawford}, {de Haan}, {Dobbs}, {Everett},
  {Gallicchio}, {Gao}, {George}, {Gilbert}, {Halverson}, {Hanson},
  {Harrington}, {Hilton}, {Holder}, {Holzapfel}, {Hoover}, {Hou}, {Hrubes},
  {Huang}, {Hubmayr}, {Irwin}, {Keisler}, {Knox}, {Lee}, {Leitch}, {Li},
  {Liang}, {Luong-Van}, {McMahon}, {Mehl}, {Meyer}, {Mocanu}, {Montroy},
  {Natoli}, {Nibarger}, {Novosad}, {Padin}, {Pryke}, {Reichardt}, {Ruhl},
  {Saliwanchik}, {Sayre}, {Schaffer}, {Smecher}, {Stark}, {Story}, {Tucker},
  {Vanderlinde}, {Vieira}, {Wang}, {Whitehorn}, {Yefremenko}, \&
  {Zahn}}]{crites:2015}
{Crites}, A.~T., {Henning}, J.~W., {Ade}, P.~A.~R., {et~al.}, {Measurements of
  E-Mode Polarization and Temperature-E-Mode Correlation in the Cosmic
  Microwave Background from 100 Square Degrees of SPTpol Data}. 2015, \apj,
  805, 36, \eprint{1411.1042}

\bibitem[{{Efstathiou} \& {Migliaccio}(2012)}]{efstathiou:2012}
{Efstathiou}, G. \& {Migliaccio}, M., {A simple empirically motivated template
  for the thermal Sunyaev-Zel'dovich effect}. 2012, \mnras, 423, 2492,
  \eprint{1106.3208}

\bibitem[{Efstathiou(2006)}]{efstathiou:2006}
Efstathiou, G.~P., {Hybrid estimation of cosmic microwave background
  polarization power spectra}. 2006, MNRAS, 370, 343

\bibitem[{{Galli} {et~al.}(2014){Galli}, {Benabed}, {Bouchet}, {Cardoso},
  {Elsner}, {Hivon}, {Mangilli}, {Prunet}, \& {Wandelt}}]{galli:2014}
{Galli}, S., {Benabed}, K., {Bouchet}, F., {et~al.}, {CMB polarization can
  constrain cosmology better than CMB temperature}. 2014, \prd, 90, 063504,
  \eprint{1403.5271}

\bibitem[{{Hivon} {et~al.}(2002){Hivon}, {G{\'o}rski}, {Netterfield}, {Crill},
  {Prunet}, \& {Hansen}}]{hivon2002}
{Hivon}, E., {G{\'o}rski}, K.~M., {Netterfield}, C.~B., {et~al.}, {MASTER of
  the Cosmic Microwave Background Anisotropy Power Spectrum: A Fast Method for
  Statistical Analysis of Large and Complex Cosmic Microwave Background Data
  Sets}. 2002, \apj, 567, 2, \eprint{astro-ph/0105302}

\bibitem[{{Kogut} {et~al.}(2003){Kogut}, {Spergel}, {Barnes}, {Bennett},
  {Halpern}, {Hinshaw}, {Jarosik}, {Limon}, {Meyer}, {Page}, {Tucker},
  {Wollack}, \& {Wright}}]{kogut2003}
{Kogut}, A., {Spergel}, D.~N., {Barnes}, C., {et~al.}, {First-Year Wilkinson
  Microwave Anisotropy Probe (WMAP) Observations: Temperature-Polarization
  Correlation}. 2003, \apjs, 148, 161, \eprint{astro-ph/0302213}

\bibitem[{{Mangilli} {et~al.}(2015){Mangilli}, {Plaszczynski}, \&
  {Tristram}}]{mangilli:2015}
{Mangilli}, A., {Plaszczynski}, S., \& {Tristram}, M., {Large-scale cosmic
  microwave background temperature and polarization cross-spectra likelihoods}.
  2015, \mnras, 453, 3174, \eprint{1503.01347}

\bibitem[{{Naess} {et~al.}(2014){Naess}, {Hasselfield}, {McMahon}, {Niemack},
  {Addison}, {Ade}, {Allison}, {Amiri}, {Battaglia}, {Beall}, {de Bernardis},
  {Bond}, {Britton}, {Calabrese}, {Cho}, {Coughlin}, {Crichton}, {Das},
  {Datta}, {Devlin}, {Dicker}, {Dunkley}, {D{\"u}nner}, {Fowler}, {Fox},
  {Gallardo}, {Grace}, {Gralla}, {Hajian}, {Halpern}, {Henderson}, {Hill},
  {Hilton}, {Hilton}, {Hincks}, {Hlozek}, {Ho}, {Hubmayr}, {Huffenberger},
  {Hughes}, {Infante}, {Irwin}, {Jackson}, {Muya Kasanda}, {Klein}, {Koopman},
  {Kosowsky}, {Li}, {Louis}, {Lungu}, {Madhavacheril}, {Marriage}, {Maurin},
  {Menanteau}, {Moodley}, {Munson}, {Newburgh}, {Nibarger}, {Nolta}, {Page},
  {Pappas}, {Partridge}, {Rojas}, {Schmitt}, {Sehgal}, {Sherwin}, {Sievers},
  {Simon}, {Spergel}, {Staggs}, {Switzer}, {Thornton}, {Trac}, {Tucker},
  {Uehara}, {Van Engelen}, {Ward}, \& {Wollack}}]{naess:2014}
{Naess}, S., {Hasselfield}, M., {McMahon}, J., {et~al.}, {The Atacama Cosmology
  Telescope: CMB polarization at 200 < l < 9000}. 2014, \jcap, 10, 007,
  \eprint{1405.5524}

\bibitem[{{Peebles}(1973)}]{peebles:1973}
{Peebles}, P.~J.~E., {Statistical Analysis of Catalogs of Extragalactic
  Objects. I. Theory}. 1973, \apj, 185, 413

\bibitem[{{\sorthelp{Planck Collaboration 2014I}}{Planck Collaboration
  IX}(2014)}]{planck2013-p03d}
{\sorthelp{Planck Collaboration 2014I}}{Planck Collaboration IX},
  {\textit{Planck} 2013 results. IX. HFI spectral response}. 2014, \aap, 571,
  A9, \eprint{1303.5070}

\bibitem[{{\sorthelp{Planck Collaboration 2014M}}{Planck Collaboration
  XIII}(2014)}]{planck2013-p03a}
{\sorthelp{Planck Collaboration 2014M}}{Planck Collaboration XIII},
  {\textit{Planck} 2013 results. XIII. Galactic CO emission}. 2014, \aap, 571,
  A13, \eprint{1303.5073}

\bibitem[{{\sorthelp{Planck Collaboration 2014O}}{Planck Collaboration
  XV}(2014)}]{planck2013-p08}
{\sorthelp{Planck Collaboration 2014O}}{Planck Collaboration XV},
  {\textit{Planck} 2013 results. XV. CMB power spectra and likelihood}. 2014,
  \aap, 571, A15, \eprint{1303.5075}

\bibitem[{{\sorthelp{Planck Collaboration 2014P}}{Planck Collaboration
  XVI}(2014)}]{planck2013-p11}
{\sorthelp{Planck Collaboration 2014P}}{Planck Collaboration XVI},
  {\textit{Planck} 2013 results. XVI. Cosmological parameters}. 2014, \aap,
  571, A16, \eprint{1303.5076}

\bibitem[{{\sorthelp{Planck Collaboration 2014U}}{Planck Collaboration
  XXI}(2014)}]{planck2013-p05b}
{\sorthelp{Planck Collaboration 2014U}}{Planck Collaboration XXI},
  {\textit{Planck} 2013 results. XXI. Power spectrum and high-order statistics
  of the \textit{Planck} all-sky Compton parameter map}. 2014, \aap, 571, A21,
  \eprint{1303.5081}

\bibitem[{{\sorthelp{Planck Collaboration 2014ZE}}{Planck Collaboration
  XXX}(2014)}]{planck2013-pip56}
{\sorthelp{Planck Collaboration 2014ZE}}{Planck Collaboration XXX},
  {\textit{Planck} 2013 results. XXX. Cosmic infrared background measurements
  and implications for star formation}. 2014, \aap, 571, A30,
  \eprint{1309.0382}

\bibitem[{{\sorthelp{Planck Collaboration 2015G}}{Planck Collaboration
  VII}(2016)}]{planck2014-a08}
{\sorthelp{Planck Collaboration 2015G}}{Planck Collaboration VII},
  {\textit{Planck} 2015 results. VII. High Frequency Instrument data
  processing: Time-ordered information and beam processing}. 2016, \aap, 594,
  A7, \eprint{1502.01586}

\bibitem[{{\sorthelp{Planck Collaboration 2015H}}{Planck Collaboration
  VIII}(2016)}]{planck2014-a09}
{\sorthelp{Planck Collaboration 2015H}}{Planck Collaboration VIII},
  {\textit{Planck} 2015 results. VIII. High Frequency Instrument data
  processing: Calibration and maps}. 2016, \aap, 594, A8, \eprint{1502.01587}

\bibitem[{{\sorthelp{Planck Collaboration 2015K}}{Planck Collaboration
  XI}(2016)}]{planck2014-a13}
{\sorthelp{Planck Collaboration 2015K}}{Planck Collaboration XI},
  {\textit{Planck} 2015 results. XI. CMB power spectra, likelihoods, and
  robustness of parameters}. 2016, \aap, 594, A11, \eprint{1507.02704}

\bibitem[{{\sorthelp{Planck Collaboration 2015L}}{Planck Collaboration
  XII}(2016)}]{planck2014-a14}
{\sorthelp{Planck Collaboration 2015L}}{Planck Collaboration XII},
  {\textit{Planck} 2015 results. XII. Full Focal Plane simulations}. 2016,
  \aap, 594, A12, \eprint{1509.06348}

\bibitem[{{\sorthelp{Planck Collaboration 2015M}}{Planck Collaboration
  XIII}(2016)}]{planck2014-a15}
{\sorthelp{Planck Collaboration 2015M}}{Planck Collaboration XIII},
  {\textit{Planck} 2015 results. XIII. Cosmological parameters}. 2016, \aap,
  594, A13, \eprint{1502.01589}

\bibitem[{{\sorthelp{Planck Collaboration 2015O}}{Planck Collaboration
  XV}(2016)}]{planck2014-a17}
{\sorthelp{Planck Collaboration 2015O}}{Planck Collaboration XV},
  {\textit{Planck} 2015 results. XV. Gravitational lensing}. 2016, \aap, 594,
  A15, \eprint{1502.01591}

\bibitem[{{\sorthelp{Planck Collaboration 2015V}}{Planck Collaboration
  XXII}(2016)}]{planck2014-a28}
{\sorthelp{Planck Collaboration 2015V}}{Planck Collaboration XXII},
  {\textit{Planck} 2015 results. XXII. A map of the thermal Sunyaev-Zeldovich
  effect}. 2016, \aap, 594, A22, \eprint{1502.01596}

\bibitem[{{\sorthelp{Planck Collaboration 2015W}}{Planck Collaboration
  XXIII}(2016)}]{planck2014-a29}
{\sorthelp{Planck Collaboration 2015W}}{Planck Collaboration XXIII},
  {\textit{Planck} 2015 results. XXIII. The thermal Sunyaev-Zeldovich
  effect--cosmic infrared background correlation}. 2016, \aap, 594, A23,
  \eprint{1509.06555}

\bibitem[{{\sorthelp{Planck Collaboration 2015ZA}}{Planck Collaboration
  XXVI}(2016)}]{planck2014-a35}
{\sorthelp{Planck Collaboration 2015ZA}}{Planck Collaboration XXVI},
  {\textit{Planck} 2015 results. XXVI. The Second Planck Catalogue of Compact
  Sources}. 2016, \aap, 594, A26, \eprint{1507.02058}

\bibitem[{{\sorthelp{Planck Collaboration IntS}}{Planck Collaboration Int.
  XIX}(2015)}]{planck2014-XIX}
{\sorthelp{Planck Collaboration IntS}}{Planck Collaboration Int. XIX},
  {\textit{Planck} intermediate results. XIX. An overview of the polarized
  thermal emission from Galactic dust}. 2015, \aap, 576, A104,
  \eprint{1405.0871}

\bibitem[{{\sorthelp{Planck Collaboration IntV}}{Planck Collaboration Int.
  XXII}(2015)}]{planck2014-XXII}
{\sorthelp{Planck Collaboration IntV}}{Planck Collaboration Int. XXII},
  {\textit{Planck} intermediate results. XXII. Frequency dependence of thermal
  emission from Galactic dust in intensity and polarization}. 2015, \aap, 576,
  A107, \eprint{1405.0874}

\bibitem[{{\sorthelp{Planck Collaboration IntZE}}{Planck Collaboration Int.
  XXX}(2016)}]{planck2014-XXX}
{\sorthelp{Planck Collaboration IntZE}}{Planck Collaboration Int. XXX},
  {\textit{Planck} intermediate results. XXX. The angular power spectrum of
  polarized dust emission at intermediate and high Galactic latitudes}. 2016,
  \aap, 586, A133, \eprint{1409.5738}

\bibitem[{{\sorthelp{Planck Collaboration IntZV}}{Planck Collaboration Int.
  XLVII}(2016)}]{planck2014-a25}
{\sorthelp{Planck Collaboration IntZV}}{Planck Collaboration Int. XLVII},
  {\textit{Planck} intermediate results. XLVII. Constraints on reionization
  history}. 2016, \aap, 596, A108, \eprint{1605.03507}

\bibitem[{{Shaw} {et~al.}(2012){Shaw}, {Rudd}, \& {Nagai}}]{Shaw12}
{Shaw}, L.~D., {Rudd}, D.~H., \& {Nagai}, D., {Deconstructing the Kinetic SZ
  Power Spectrum}. 2012, \apj, 756, 15, \eprint{1109.0553}

\bibitem[{{Tinker} {et~al.}(2008){Tinker}, {Kravtsov}, {Klypin}, {Abazajian},
  {Warren}, {Yepes}, {Gottl{\"o}ber}, \& {Holz}}]{tinker:2008}
{Tinker}, J., {Kravtsov}, A.~V., {Klypin}, A., {et~al.}, {Toward a Halo Mass
  Function for Precision Cosmology: The Limits of Universality}. 2008, \apj,
  688, 709, \eprint{0803.2706}

\bibitem[{{Tristram} {et~al.}(2005){Tristram}, {Mac{\'{\i}}as-P{\'e}rez},
  {Renault}, \& {Santos}}]{tristram:2005}
{Tristram}, M., {Mac{\'{\i}}as-P{\'e}rez}, J.~F., {Renault}, C., \& {Santos},
  D., {XSPECT, estimation of the angular power spectrum by computing
  cross-power spectra with analytical error bars}. 2005, \mnras, 358, 833,
  \eprint{astro-ph/0405575}

\bibitem[{{Tucci} {et~al.}(2004){Tucci}, {Mart{\'{\i}}nez-Gonz{\'a}lez},
  {Toffolatti}, {Gonz{\'a}lez-Nuevo}, \& {De Zotti}}]{tucci:2004}
{Tucci}, M., {Mart{\'{\i}}nez-Gonz{\'a}lez}, E., {Toffolatti}, L.,
  {Gonz{\'a}lez-Nuevo}, J., \& {De Zotti}, G., {Predictions on the
  high-frequency polarization properties of extragalactic radio sources and
  implications for polarization measurements of the cosmic microwave
  background}. 2004, \mnras, 349, 1267, \eprint{astro-ph/0307073}

\bibitem[{Tucci {et~al.}(2005)Tucci, Martinez-Gonzalez, Vielva, \&
  Delabrouille}]{tucci:2005}
Tucci, M., Martinez-Gonzalez, E., Vielva, P., \& Delabrouille, J., {Limits on
  the detectability of the CMB B-mode polarization imposed by foregrounds}.
  2005, \mnras, 360, 935, \eprint{astro-ph/0411567}

\bibitem[{{Tucci} \& {Toffolatti}(2012)}]{tucci:2012}
{Tucci}, M. \& {Toffolatti}, L., {The Impact of Polarized Extragalactic Radio
  Sources on the Detection of CMB Anisotropies in Polarization}. 2012, Advances
  in Astronomy, 2012, 52, \eprint{1204.0427}

\bibitem[{Tucci {et~al.}(2011)Tucci, Toffolatti, de~Zotti, \&
  Mart{\'\i}nez-Gonz{\'a}lez}]{tucci:2011}
Tucci, M., Toffolatti, L., de~Zotti, G., \& Mart{\'\i}nez-Gonz{\'a}lez, E.,
  {High-frequency predictions for number counts and spectral properties of
  extragalactic radio sources. New evidence of a break at mm wavelengths in
  spectra of bright blazar sources}. 2011, A{\&}A, 533, 57

\bibitem[{Varshalovich {et~al.}(1988)Varshalovich, Moskalev, \&
  Khersonskii}]{varshalovich:1988}
Varshalovich, D.~A., Moskalev, A.~N., \& Khersonskii, V.~K. 1988, Quantum
  Theory of Angular Momentum (Singapore: World Scientific)

\end{thebibliography}

\clearpage
\onecolumn
\small
\begin{appendix}

\section{\xpol\ computations}
\label{ann:xpol_covariance}

\newcommand{\R}{\Re\textrm{e}}
\newcommand{\I}{\Im\textrm{m}}

\def\E#1{{ \operatorname{E}[#1] }}
\newcommand{\Y}[2]{{\ {}_{#1}Y^{}_{#2}}}
\newcommand{\Ystar}[2]{{\ {}_{#1}Y^{*}_{#2}}}
\newcommand{\K}[3]{{\ {}_{#1}K^{#2}_{#3}}}

\subsection{Harmonic decomposition}
Following notations from \citet{hivon2002}, the decomposition into spherical harmonics reads 
\begin{eqnarray}
        T(\hat{\bf n}) & = & \sum_{\lm}a^T_{\lm} Y_{\lm}(\hat{\bf n}) \\
        Q(\hat{\bf n}) \pm iU(\hat{\bf n}) & = & \sum_{\lm} \ _{\mp2}a_{\lm} \Y{\mp2}{\lm}(\hat{\bf n})
\end{eqnarray}

\begin{equation}
\label{alm_spin}
        \left\{
        \begin{array}{rcl}
        \ _{2}a_{\lm} &=& a^E_{\lm}+ia^B_{\lm} \\
        \ _{-2}a_{\lm} &=& a^E_{\lm}-ia^B_{\lm} 
        \end{array} \right.
\end{equation}
with 
\begin{equation}
\label{elm}
        a^E_{\lm} \equiv \frac{1}{2} \int \Big\{
                \big( Q(\hat{\bf n})-iU(\hat{\bf n}) \big) \Ystar{2}{\lm}(\hat{\bf n}) + 
                \big( Q(\hat{\bf n})+iU(\hat{\bf n}) \big) \Ystar{-2}{\lm}(\hat{\bf n})
        \Big\}d\hat{\bf n} 
\end{equation}
\begin{equation}
\label{blm}
        a^B_{\lm} \equiv -\frac{i}{2} \int \Big\{
                \big( Q(\hat{\bf n})-iU(\hat{\bf n}) \big) \Ystar{2}{\lm}(\hat{\bf n}) - 
                \big( Q(\hat{\bf n})+iU(\hat{\bf n}) \big)\Ystar{-2}{\lm}(\hat{\bf n})
        \Big\}d\hat{\bf n} 
.\end{equation}

When including a sky window function (or mask) $w$, then
\begin{eqnarray}
\label{that}
        T_{\lm}
        & = & 
        \sum_{\lmun} a^{T}_{\lmun} \sum_{\lmdeux} w^T_{\lmdeux} 
                \int \Ystar{0}{\lm}(\hat{\bf n}) \Y{0}{\lmun}(\hat{\bf n}) \Y{0}{\lmdeux}(\hat{\bf n}) d\hat{\bf n} \\
\label{ehat} 
        E_{\lm}
        & = &
        \frac{1}{2} \sum_{\lmun} \bigg\{
                (a^E_{\lmun}+ia^B_{\lmun}) \sum_{\lmdeux} w^E_{\lmdeux}
                \int \Ystar{2}{\lm}(\hat{\bf n}) \Y{2}{\lmun}(\hat{\bf n}) \Y{0}{\lmdeux} (\hat{\bf n}) d\hat{\bf n}
                + (a^E_{\lmun}-ia^B_{\lmun}) \sum_{\lmdeux} w^E_{\lmdeux} 
                \int \Ystar{-2}{\lm}(\hat{\bf n}) \Y{-2}{\lmun}(\hat{\bf n}) \Y{0}{\lmdeux}(\hat{\bf n})
        \bigg\}
        \\
\label{bhat}
        B_{\lm}
        & = &
        -\frac{i}{2} \sum_{\lmun} \bigg\{
                (a^E_{\lmun}+ia^B_{\lmun}) \sum_{\lmdeux} w^B_{\lmdeux}
                \int \Ystar{2}{\lm}(\hat{\bf n}) \Y{2}{\lmun}(\hat{\bf n}) \Y{0}{\lmdeux}(\hat{\bf n}) d\hat{\bf n}
                - (a^E_{\lmun}-ia^B_{\lmun}) \sum_{\lmdeux} w^B_{\lmdeux}
                \int \Ystar{-2}{\lm}(\hat{\bf n}) \Y{-2}{\lmun}(\hat{\bf n}) \Y{0}{\lmdeux}(\hat{\bf n}),
        \bigg\}
\end{eqnarray}
where the window function is decomposed into spin-0 spherical harmonics ($w = \sum_{\lm} a^W_{\lm} \Y{0}{\lm}$).

We define $\K{s}{X}{\lmun\lmdeux}$ with $s \in \{-2,0,2\}$ (X=T,E,B),
\begin{equation}
\label{defk}
        \K{s}{X}{\lmun\lmdeux}
        \equiv 
        \int \Ystar{s}{\lmun}(\hat{\bf n}) \Y{s}{\lmdeux}(\hat{\bf n}) w^X(\hat{\bf n}) d\hat{\bf n}
        =
        \sum_{\lmtrois} w^X_{\lmtrois} \int \Ystar{s}{\lmun}(\hat{\bf n}) \Y{s}{\lmdeux}(\hat{\bf n}) \Y{0}{\lmtrois}(\hat{\bf n}) d\hat{\bf n};
\end{equation}
we can then rewrite $E_{\lm}$ and $B_{\lm}$ as
\begin{eqnarray}
\label{thatk}
T_{\lm} & = & \sum_{\lmun}a^T_{\lmun} \K{0}{T}{\lm\lmun}\\
\label{ehatk}
E_{\lm} & = & \phantom{-}  \frac{1}{2} \sum_{\lmun} \Bigg\{ a^E_{\lmun} \Big[ \K{2}{E}{\lm\lmun} + \K{-2}{E}{\lm\lmun} \Big] + i a^B_{\lmun} \Big[ \K{2}{E}{\lm\lmun} - \K{-2}{E}{\lm\lmun} \Big] \Bigg\} \\
\label{bhatk}
B_{\lm} & = & - \frac{i}{2} \sum_{\lmun} \Bigg\{ a^E_{\lmun} \Big[ \K{2}{B}{\lm\lmun} - \K{-2}{B}{\lm\lmun} \Big] + i a^B_{\lmun} \Big[ \K{2}{B}{\lm\lmun} + \K{-2}{B}{\lm\lmun} \Big] \bigg\}
\end{eqnarray}

We define $W_{\lmun\lmdeux}$, 
\begin{eqnarray}
\label{defw}
        \left\{
        \begin{array}{rcl}
        \W{0}{X}{\lmun}{\lmdeux} & \equiv & \K{0}{X}{\lmun\lmdeux}\\
        \W{\pm}{X}{\lmun}{\lmdeux} & \equiv & \K{2}{X}{\lmun\lmdeux} \pm \K{-2}{X}{\lmun\lmdeux}
        \end{array} \right.
.\end{eqnarray}

Thus, finally
\begin{eqnarray}
\label{thatw}
        \left(
        \begin{array}{c}
                T_{\lm}\\
                E_{\lm}\\
                B_{\lm}
        \end{array}
        \right)
        & = & 
        \sum_{\lmun} \left(
        \begin{array}{ccc}
                \W{0}{T}{\lm}{\lmun} & 0 & 0\\
                0 & \frac{1}{2} \W{+}{E}{\lm}{\lmun} & \frac{i}{2} \W{-}{E}{\lm}{\lmun}\\
                0 & -\frac{i}{2} \W{-}{B}{\lm}{\lmun} & \frac{1}{2} \W{+}{B}{\lm}{\lmun}
        \end{array}
        \right)
        \left(
        \begin{array}{c}
                a^T_{\lmun}\\
                a^E_{\lmun}\\
                a^B_{\lmun}
        \end{array}
        \right)
.\end{eqnarray}

\subsection{Power spectra (first order $\mathbf{a_\lm}$ correlation)}

We suppose independent data sets ($i,j$) for which we have (I,Q,U) maps and compute the spherical transform to obtain $X_\lm$ (for $X \in \{T,E,B\}$) coefficients.
The cross-power spectra are thus defined as
\begin{equation}
\label{eq:cell}
        \tilde C_\ell^{X_aY_b} = \frac{1}{2\ell+1} \sum_{m=-\ell}^{\ell} \VEV{ X^a_{\lm}Y^{b*}_{\lm}}
.\end{equation}

We have to compute $\VEV{ X^a_{\lm}Y^{b*}_{\lmprime}}$ for each set of $(X,Y) \in {T,E,B}$ using (\ref{thatw}).
In this section, we will neglect the terms in $EB$ and $TB$ with respect to other mode correlation.

\subsubsection{Temperature}

$$
        \VEV{ T^a_{\lm}T^{b*}_{\lmprime}} = 
        \sum_{\lmun}\sum_{\lmdeux}
                \VEV{ a^{T_a}_{\lmun}a^{T_b*}_{\lmdeux}} \W{0}{T_a}{\lm}{\lmun} \W{0}{T_b*}{\lmprime}{\lmdeux} 
$$

with 
\begin{equation}
\label{defcell}
        \VEV{ a^{X_a}_{\lmun}a^{X_b*}_{\lmdeux}} = \delta_{\ell_1\ell_2}\delta_{m_1m_2}C^{X_aX_b}_{\ell_1}\qquad (x=T,E,B)
\end{equation}

Thus, 
\begin{equation}
\label{eq:tt}
        \framebox{
        $\displaystyle
        \VEV{ T^a_{\lm}T^{b*}_{\lmprime}} = \sum_{\lmun} C^{T_aT_b}_{\ell_1} \W{0}{T_a}{\lm}{\lmun} \W{0}{T_b*}{\lmprime}{\lmun}
        $}
\end{equation}

\subsubsection{Polarization}

$$
        \VEV{ E^{a*}_{\lm}E^{b}_{\lmprime}} = 
        \frac{1}{4} \sum_\lmun \sum_\lmdeux \Bigg\langle
                \bigg\{ a^{E_a*}_{\lmun} \W{+}{E_a*}{\lm}{\lmun} - ia^{B_a*}_{\lmun} \W{-}{E_a*}{\lm}{\lmun} \bigg\} \times
                \bigg\{ a^{E_b}_{\lmdeux} \W{+}{E_b}{\lmprime}{\lmdeux} + ia^{B_b}_{\lmdeux} \W{-}{E_b}{\lmprime}{\lmdeux} \bigg\}
        \Bigg\rangle
$$

using (\ref{defcell}):
\begin{eqnarray}
        \VEV{ E^{a*}_{\lm} E^{b}_{\lmprime}} 
        & = & 
        \frac{1}{4} \sum_\lmun \bigg\{
                C^{E_aE_b}_{\ell_1} \W{+}{E_a*}{\lm}{\lmun} \W{+}{E_b}{\lmprime}{\lmun} + 
                iC^{E_aB_b}_{\ell_1} \W{+}{E_a*}{\lm}{\lmun} \W{-}{E_b}{\lmprime}{\lmun} \nonumber \\
        & & - iC^{B_aE_b}_{\ell_1}W^{-,E_a*}_{\lm\lmun}W^{+,E_b}_{\lmprime\lmun}+C^{B_aB_b}_{\ell_1}W^{-,E_b*}_{\lm\lmun}W^{-,E_b}_{\lmprime\lmun}\bigg\} \nonumber 
\end{eqnarray}

We neglect $C_\ell^{EB}$, so that $\VEV{ E^{a*}_{\lm} E^{b}_{\lmprime}}$ is real and reads
\begin{equation}
\label{eq:ee}
\framebox{
$\displaystyle
        \VEV{ E^{a*}_{\lm} E^{b}_{\lmprime}} = \frac{1}{4} \sum_\lmun
                \left\{
                        C^{E_aE_b}_{\ell_1} \W{+}{E_a*}{\lm}{\lmun} \W{+}{E_b}{\lmprime}{\lmun} +
                        C^{B_aB_b}_{\ell_1} \W{-}{E_a*}{\lm}{\lmun} \W{-}{E_b}{\lmprime}{\lmun}
                \right\}
$}
\end{equation}

Analogously,
$$
        \VEV{ B^{a*}_{\lm} B^{b}_{\lmprime} } = 
        \frac{1}{4} \sum_\lmun \sum_\lmdeux \VEV{
                \left\{ a^{B_a*}_{\lmun} \W{+}{B_a*}{\lm}{\lmun} + i a^{E_a*}_{\lmun} \W{-}{B_a*}{\lm}{\lmun} \right\}
                \times
                \left\{ a^{B_b}_{\lmdeux} \W{+}{B_b}{\lmprime}{\lmdeux} - i a^{E_b}_{\lmdeux} \W{-}{B_b}{\lmprime}{\lmdeux} \right\}
        }
$$

\begin{equation}
\label{eq:bb}
\framebox{
$\displaystyle
        \VEV{ B^{a*}_{\lm} B^{b}_{\lmprime}} = \frac{1}{4} \sum_\lmun
                \left\{
                        C^{B_a B_b}_{\ell_1} \W{+}{B_a*}{\lm}{\lmun} \W{+}{B_b}{\lmprime}{\lmun} +
                        C^{E_a E_b}_{\ell_1} \W{-}{B_a*}{\lm}{\lmun} \W{-}{B_b}{\lmprime}{\lmun}
                \right\}
$}
\end{equation}

\subsubsection{Cross modes}

$$
        \VEV{ T^{a*}_{\lm}E^{b}_{\lmprime} } = 
        \frac{1}{2} \sum_\lmun \sum_\lmdeux \VEV{
                \left\{ a^{T_a*}_{\lmun} \W{0}{T_a*}{\lm}{\lmun} \right\}  \times
                \left\{ a^{E_b}_{\lmdeux} \W{+}{E_b}{\lmprime}{\lmdeux} + ia^{B_b}_{\lmdeux} \W{-}{E_b}{\lmprime}{\lmdeux} \right\}
        }
$$

We neglect $C_\ell^{TB}$ null, so that $\VEV{ T^{a*}_{\lm} E^{b}_{\lmprime}}$ is real and reads
\begin{equation}
\label{eq:te}
\framebox{
$\displaystyle
        \VEV{T^{a*}_{\lm} E^{b}_{\lmprime}} = \VEV{E^{b*}_{\lm} T^{a}_{\lmprime}} = \frac{1}{2} \sum_\lmun
                \left\{
                        C^{T_a E_b}_{\ell_1} \W{0}{T_a*}{\lm}{\lmun} \W{+}{E_b}{\lmprime}{\lmun}
                \right\}
$}
\end{equation}

and
$$
        \VEV{ T^{a*}_{\lm} B^{b}_{\lmprime} } = 
        \frac{1}{2} \sum_\lmun \sum_\lmdeux \VEV{
                \left\{ a^{T_a*}_{\lmun} \W{0}{T_a*}{\lm}{\lmun} \right\}  \times
                \left\{ a^{B_b}_{\lmdeux} \W{+}{B_b}{\lmprime}{\lmdeux} - ia^{E_b}_{\lmdeux} \W{-}{B_b}{\lmprime}{\lmdeux} \right\}
        }
$$
\begin{equation}
\label{eq:tb}
\framebox{
$\displaystyle
        \VEV{T^{a*}_{\lm} B^{b}_{\lmprime}} = \VEV{B^{b*}_{\lm} T^{a}_{\lmprime}} = \frac{1}{2} \sum_\lmun
                \left\{
                        C^{T_a B_b}_{\ell_1} \W{0}{T_a*}{\lm}{\lmun} \W{+}{B_b}{\lmprime}{\lmun} - i C^{T_a E_b}_{\ell_1} \W{0}{T_a*}{\lm}{\lmun} \W{-}{E_b}{\lmprime}{\lmun}
                \right\}
$}
\end{equation}
\\

$$
        \VEV{ E^{a*}_{\lm} B^{b}_{\lmprime} } = 
        \frac{1}{4} \sum_\lmun \sum_\lmdeux \VEV{
                \left\{ a^{E_a*}_{\lmun} \W{+}{E_a*}{\lm}{\lmun} - ia^{B_a*}_{\lmun} \W{-}{E_a*}{\lm}{\lmun} \right\}  \times
                \left\{ a^{B_b}_{\lmdeux} \W{+}{B_b}{\lmprime}{\lmdeux} - ia^{E_b}_{\lmdeux} \W{-}{B_b}{\lmprime}{\lmdeux} \right\}
        }
$$
\begin{equation}
\label{eq:eb}
\framebox{$\displaystyle
        \begin{array}{rcl}
        \VEV{E^{a*}_{\lm} B^{b}_{\lmprime}} = \VEV{B^{b*}_{\lm} E^{a}_{\lmprime}} = \frac{1}{4} \sum_\lmun
        &&
                \left\{
                        C^{B_a E_b}_{\ell_1} \W{-}{E_a*}{\lm}{\lmun} \W{-}{B_b}{\lmprime}{\lmun} -
                        C^{E_a B_b}_{\ell_1} \W{+}{E_a*}{\lm}{\lmun} \W{+}{B_b}{\lmprime}{\lmun}
                \right\}\\
        &- i& 
                \left\{
                        C^{E_a E_b}_{\ell_1} \W{+}{E_a*}{\lm}{\lmun} \W{-}{B_b}{\lmprime}{\lmun} +
                        C^{B_a B_b}_{\ell_1} \W{-}{E_a*}{\lm}{\lmun} \W{+}{B_b}{\lmprime}{\lmun}
                \right\}
        \end{array}
$}
\end{equation}

\subsubsection{Application to $\mathbf{C_\ell}$}

For Eq.~\ref{eq:cell}, we need to compute the product of 2 $W^{\oplus,X}_{\ell m \ell' m'}$ as in Eqs~(\ref{eq:tt}, \ref{eq:ee}, \ref{eq:bb}, \ref{eq:te}, \ref{eq:tb}, \ref{eq:eb}).

Here, we wrote integrals as 3j-Wigner symbols using
\begin{eqnarray}
        \int d\hat n \ _s Y_{lm}^{*}(\hat n)\ _{s'} Y_{\ell'm'}(\hat n) \ _{s''} Y_{\ell''m''} (\hat n)
        & = &
        (-1)^{s+m} \left[ \frac{(2\ell+1)(2\ell'+1)(2\ell''+1)}{4\pi} \right]^{1/2}
        \wigner{\ell}{\ell'}{\ell''}{-s}{s'}{s''} \wigner{\ell}{\ell'}{\ell''}{-m}{m'}{m''}
\end{eqnarray}
and make use of the orthogonality for the spinned-harmonics:
\begin{eqnarray}
        \left\{
        \begin{array}{rcl}
        \sum_{\ell m} (2\ell+1) \wigner{\ell_1}{\ell_2}{\ell}{m_1}{m_2}{m} \wigner{\ell_1}{\ell_2}{\ell}{m_1'}{m_2'}{m} &=& \delta_{m_1m_1'}\delta_{m_2m_2'}\\
        \sum_{m_1 m_2} (2\ell+1) \wigner{\ell_1}{\ell_2}{\ell}{m_1}{m_2}{m} \wigner{\ell_1}{\ell_2}{\ell'}{m_1}{m_2}{m'} &=& \delta_{\ell\ell'}\delta_{mm'}
        \end{array}
        \right.
.\end{eqnarray}

We then define
\begin{equation}
\label{eq:def_chi}
\everymath={\displaystyle}
        \left\{
        \begin{array}{rclcl}
                \tens{M}_{TT,TT}(\ell_1, \ell_2; a,b) &\equiv& L_{\ell_1 \ell_2} \sum_{m_1m_2} \W{0}{a}{\lmun}{\lmdeux} \W{0}{b*}{\lmun}{\lmdeux}&=&  \frac{1}{4\pi} \sum_{\ell_3 }  (2 \ell_3 + 1)C^{W_a,W_b}_{\ell_3} \wigner{\ell_1}{\ell_2}{\ell_3}{0}{0}{0}^2\\
                \tens{M}_{TE,TE}(\ell_1, \ell_2; a,b) &\equiv& \frac{L_{\ell_1 \ell_2}}{2} \sum_{m_1m_2} \W{0}{a}{\lmun}{\lmdeux} \W{+}{b*}{\lmun}{\lmdeux}&=&  \frac{1}{8 \pi} \sum_{\ell_3 } (2 \ell_3 + 1) C^{W_a,W_b}_{\ell_3} (1 + (-1)^L) \wigner{\ell_1}{\ell_2}{\ell_3}{0}{0}{0}\wigner{\ell_1}{\ell_2}{\ell_3}{-2}{2}{0}\\

                \tens{M}_{EE,EE}(\ell_1, \ell_2; a,b) &\equiv& \frac{L_{\ell_1 \ell_2}}{4} \sum_{m_1m_2} \W{+}{a}{\lmun}{\lmdeux} \W{+}{b*}{\lmun}{\lmdeux}&=& \frac{1}{16 \pi} \sum_{\ell_3 }  (2 \ell_3 + 1) C^{W_a,W_b}_{\ell_3} (1 + (-1)^L)^2 \wigner{\ell_1}{\ell_2}{\ell_3}{-2}{2}{0}^2\\
                \tens{M}_{EE,BB}(\ell_1, \ell_2; a,b) &\equiv& \frac{L_{\ell_1 \ell_2}}{4} \sum_{m_1m_2} \W{-}{a}{\lmun}{\lmdeux} \W{-}{b*}{\lmun}{\lmdeux}&=& \frac{1}{16 \pi} \sum_{\ell_3 }  (2 \ell_3 + 1) C^{W_a,W_b}_{\ell_3} (1 - (-1)^L)^2 \wigner{\ell_1}{\ell_2}{\ell_3}{-2}{2}{0}^2
        \end{array}
        \right.
\end{equation}
with $L_{\ell\ell'} \equiv \frac{1}{(2\ell+1)(2\ell'+1)}$, $L = \ell_1 + \ell_2 + \ell_3$,
which depend only on the scalar cross-power spectrum of the masks $C_\ell^{W_a,W_b} = \sum_m w^a_{\ell m} w^{b*}_{\ell m} / (2\ell+1)$.

Finally, for $\vec{C}^{ab}_\ell = \left(C_\ell^{T_aT_b},C_\ell^{E_aE_b},C_\ell^{B_aB_b},C_\ell^{T_aE_b}\right)$, the relation between pseudo-$C_\ell$ ($\vec{\tilde C_\ell}$) and $\vec{C}_\ell$ is given using the general coupling matrix
\begin{equation}
\framebox{$\displaystyle
        \vec{\tilde C}^{ab}_\ell = (2\ell'+1) \tens{M}^{a \times b}_{\ell\ell'} \, \vec{C}^{ab}_{\ell'}
$}
,\end{equation}
and the coupling matrix $\tens{M}$ that translate pseudo-spectra to power spectra reads \citep[see][]{kogut2003}
\begin{equation}
        \scriptsize
        \tens{M}^{a \times b}_{\ell\ell'} = 
        \left(\begin{array}{cccc}
                \tens{M}_{TT,TT} & 0 & 0 & 0 \\
                0 & \tens{M}_{EE,EE} & \tens{M}_{EE,BB} & 0 \\
                0 & \tens{M}_{EE,BB} & \tens{M}_{EE,EE} & 0 \\
                0 & 0 & 0 & \tens{M}_{TE,TE} \\
        \end{array}\right) \normalsize
        (\ell, \ell'; w_a,w_b)
.\end{equation}

\subsection{Covariance matrix (second-order $\mathbf{a_\lm}$ correlation)}

We want to write the correlation matrix $\tens{\Sigma}^{ab,cd}_{\ell \ell'}$ that gives the correlation between cross-spectra $(ab)$ and $(cd)$ and between multipoles $\ell$ and $\ell'$
\begin{equation}
        \tens{\Sigma}^{ab,cd}_{\ell\ell'}
        \equiv \VEV{ \Delta\vec{C}^{ab}_\ell \Delta\vec{C}^{cd*}_{\ell'}}
        = \left( \tens{M}^{ab}_{\ell \ell_1} \right)^{-1} \VEV{ \Delta\vec{\tilde C}^{ab}_{\ell_1} \Delta\vec{\tilde C}^{cd*}_{\ell_2}} \left( \tens{M}^{cd*}_{\ell' \ell_2} \right)^{-1}
\label{xi}
\end{equation}

with the \emph{pseudo}-covariance matrix $\tens{\tilde \Sigma}$:
\begin{equation}
\label{def}
        \tens{\tilde \Sigma}^{ab,cd}_{\ell\ell'} = \VEV{ \Delta\vec{\tilde C}^{ab}_\ell \Delta\vec{\tilde C}^{cd*}_{\ell'}} = \VEV{\vec{\tilde C}^{ab}_\ell\vec{\tilde C}^{cd*}_{\ell'}} - \vec{C}^{ab}_\ell \vec{C}^{cd*}_{\ell'}
\end{equation}

We write the 4-$a_\lm$ correlations~:
\begin{equation}
\label{clhat}
        \VEV{\vec{\tilde C}^{X_aX_b}_\ell\vec{\tilde C}^{X_cX_d*}_{\ell'}} = 
        \frac{1}{(2\ell+1)(2\ell'+1)} \sum_{mm'} \VEV{ X^a_{\lm} X^{b*}_{\lm} X^{c*}_{\lmprime} X^d_{\lmprime} }
        \qquad
        (X_{\lm}=T_{\lm},E_{\lm},B_{\lm})
\end{equation}

We use  Isserlis' formula (or Wick's theorem), which gives for Gaussian variables
\begin{equation}
\label{gauss}
\VEV{ x_ix_jx_kx_l}=\VEV{ x_ix_j}\VEV{ x_kx_l}+\VEV{ x_ix_k}\VEV{ x_jx_l}+\VEV{ x_ix_l}\VEV{ x_jx_k}
.\end{equation}

Thus (\ref{clhat}) reads
\begin{eqnarray}
        \VEV{ \tilde{C}^{X_aX_b}_\ell \tilde{C}^{X_cX_d*}_{\ell'}}
        & = & L_{\ell\ell'} \sum_{mm'} \bigg\{
                \VEV{ X^a_{\lm}X^{b*}_{\lm}} \VEV{ X^{c*}_{\lmprime}X^{d}_{\lmprime}} + 
                \VEV{ X^a_{\lm}X^{c*}_{\lmprime}} \VEV{ X^{b*}_{\lm}X^{d}_{\lmprime}} +
                \VEV{ X^a_{\lm}X^{d}_{\lmprime}} \VEV{ X^{b*}_{\lm}X^{c*}_{\lmprime}} \bigg\} \nonumber \\
        & = &
        C^{X_aX_b}_{\ell} C^{X_cX_d*}_{\ell'} +
        L_{\ell\ell'} \sum_{mm'} \bigg\{
                \VEV{ X^a_{\lm}X^{c*}_{\lmprime}} \VEV{ X^{b*}_{\lm}X^{d}_{\lmprime}} + 
                \VEV{ X^a_{\lm}X^{d}_{\lmprime}} \VEV{ X^{b*}_{\lm}X^{c*}_{\lmprime}} \bigg\} \\
\label{cross}
        \VEV{\Delta\tilde{C}^{X_aX_b}_\ell\Delta\tilde{C}^{X_cX_d*}_{\ell'}} 
        & = & 
        L_{\ell\ell'} \sum_{mm'} \bigg\{
                \VEV{ X^a_{\lm}X^{c*}_{\lmprime}} \VEV{ X^{b*}_{\lm}X^{d}_{\lmprime}} + 
                \VEV{ X^a_{\lm}X^{d}_{\lmprime}}\VEV{ X^{b*}_{\lm}X^{c*}_{\lmprime}}
                \bigg\}
\end{eqnarray}

We know that 
\begin{equation}
\label{alm}
X^*_\lm=(-1)^m X_{\ell-m}
.\end{equation}

We can thus replace $m'$ by $-m'$ in the sum of the right-hand side of (\ref{cross}) and use (\ref{alm})
$$
\VEV{ X^a_{\lm}X^{d}_{\lmprime}} \VEV{ X^{b*}_{\lm}X^{c*}_{\lmprime}} = (-1)^{-2m'} \VEV{ X^{a}_{\lm}X^{d*}_{\lmprime}}\VEV{ X^{b*}_{\lm}X^{c}_{\lmprime}}.
$$

And finally, elements of the pseudo-covariance matrix $\tilde{\tens{\Sigma}}$ reads
\begin{equation}
\label{xifinal}
        \framebox{
                $\displaystyle \VEV{\Delta\vec{\tilde C}^{X_aX_b}_\ell\Delta\vec{\tilde C}^{X_cX_d*}_{\ell'}} = 
                        L_{\ell\ell'} \sum_{mm'} \bigg\{
                                \VEV{ X^{a}_{\lm}X^{c*}_{\lmprime}} \VEV{ X^{b*}_{\lm}X^{d}_{\lmprime}} + 
                                \VEV{ X^{a}_{\lm}X^{d*}_{\lmprime}} \VEV{ X^{b*}_{\lm}X^{c}_{\lmprime}}
                        \bigg\}
                $}
.\end{equation}


\subsubsection{Basic properties}

We recall some basic properties:
\begin{equation}
\label{sphericalproperties}
{\ }_sY_{\lm}=(-1)^{s+m} {\ }_{-s}Y^*_{\ell-m}
,\end{equation}

which leads to
\begin{eqnarray}
        \bigg[\int \Ystar{s}{\lmun}(\hat{\bf n}) \Y{s}{\lmdeux}(\hat{\bf n}) \Y{0}{\lmtrois}(\hat{\bf n}) d\hat{\bf n}\bigg]^* 
        & = & 
        \int \Y{s}{\lmun}(\hat{\bf n}) \Ystar{s}{\lmdeux}(\hat{\bf n}) \Ystar{0}{\lmtrois}(\hat{\bf n}) d\hat{\bf n}
        \nonumber \\
        & = &
        (-1)^{(s+s+0)} \int \Ystar{-s}{\ell_1-m_1}(\hat{\bf n}) \Y{-s}{\ell_2-m_2}(\hat{\bf n}) \Y{0}{\ell_3-m_3}(\hat{\bf n})d\hat{\bf n}
        \nonumber \\
        & = &
        \int \Ystar{-s}{\ell_1-m_1}(\hat{\bf n}) \Y{-s}{\ell_2-m_2}(\hat{\bf n}) \Y{0}{\ell_3-m_3}(\hat{\bf n})d\hat{\bf n}.
        \nonumber
\end{eqnarray}

From this we deduce that
$$
        \left( \K{\pm2}{X}{\lmun\lmdeux} \right)^*= \K{\mp2}{X}{\ell_1 -m_1 \ell_2 -m_2}.
$$

Thus for $\W{\oplus}{X}{\lmun}{\lmdeux}$ we have $\left( \W{0}{X}{\lmun}{\lmdeux} \right)^{*} = \W{0}{X}{\ell_1-m_1}{\ell_2-m_2}$, $\left(\W{+}{X}{\lmun}{\lmdeux}\right)^* = \W{+}{X}{\ell_1-m_1}{\ell_2-m_2}$, and $\left(\W{-}{X}{\lmun}{\lmdeux}\right)^* = -\W{-}{X}{\ell_1-m_1}{\ell_2-m_2}$.
We  also recall that the spin-lowering and spin-raising derivative reads 
\begin{eqnarray}
\label{eq:spin_raise}
        {}_sY_{\ell m} &=& \sqrt{\frac{(\ell-s)!}{(\ell+s)!}}\ \eth^s Y_{\ell m},\ \ 0\leq s \leq \ell\\
        {}_sY_{\ell m} &=& \sqrt{\frac{(\ell+s)!}{(\ell-s)!}}\ (-1)^s \bar\eth^{-s} Y_{\ell m},\ \ -\ell\leq s \leq 0
\label{eq:spin_lower}
\end{eqnarray}

and
\begin{equation}
        (\eth^{s})^* = \bar\eth^{s}
\label{eq:deriv_ylm}
\end{equation}

Other important properties include the following:
\begin{eqnarray}
    \eth\left({}_sY_{\ell m}\right) &=& +\sqrt{(\ell-s)(\ell+s+1)}\ {}_{s+1}Y_{\ell m}\\
    \bar\eth\left({}_sY_{\ell m}\right) &=& -\sqrt{(\ell+s)(\ell-s+1)}\ {}_{s-1}Y_{\ell m}
\end{eqnarray}

Using spin-raising (resp. spin-lowering) operators (\ref{eq:spin_raise}) and (\ref{eq:spin_lower}) on $Y_{\ell m}({\hat n}_j)$ and then integrating twice by part, we notice that
\begin{eqnarray*}
        \int \Ystar{-2}{\lmprime}({\hat n}_j) \Y{-2}{\lmun}({\hat n}_j) w_j d{\hat n}_j 
        &=& 
        \int \sqrt{\frac{(\ell'-2)!}{(\ell'+2)!}} \ \left(  \bar\eth^2 \Y{}{\lmprime}({\hat n}_j) \right)^* \Y{-2}{\lmun}({\hat n}_j) \ w_j \ d{\hat n}_j \\
        &=& 
        \int \sqrt{\frac{(\ell'-2)!}{(\ell'+2)!}} \ \eth^2 \left(  \Ystar{}{\lmprime}({\hat n}_j) \right) \Y{-2}{\lmun}({\hat n}_j) \ w_j \ d{\hat n}_j \\
        &=& 
        \int \sqrt{\frac{(\ell'-2)!}{(\ell'+2)!}} \Ystar{}{\lmprime}({\hat n}_j) \ \eth^2 \left( \Y{-2}{\lmun}({\hat n}_j) \ w_j \right) \ d{\hat n}_j \\
        &\simeq& 
        \int \sqrt{\frac{(\ell'-2)!}{(\ell'+2)!}} \Ystar{}{\lmprime}({\hat n}_j) \ \eth^2 \left( \Y{-2}{\lmun}({\hat n}_j) \right) \ w_j \ d{\hat n}_j \\
        &=& 
        \int \sqrt{\frac{(\ell'-2)!}{(\ell'+2)!}} \sqrt{\frac{(\ell_1+2)!}{(\ell_1-2)!}} \Ystar{}{\lmprime}({\hat n}_j) \Y{}{\lmun}({\hat n}_j) \ w_j \ d{\hat n}_j\\
        &=& 
        \int \sqrt{\frac{(\ell'-2)!}{(\ell'+2)!}} \Ystar{}{\lmprime}({\hat n}_j) \bar\eth^2 \left( \Y{2}{\lmun}({\hat n}_j) \right) \ w_j \ d{\hat n}_j\\
        &\simeq& 
        \int \sqrt{\frac{(\ell'-2)!}{(\ell'+2)!}} \left( \bar\eth^2 \Ystar{}{\lmprime}({\hat n}_j) \ w_j \right)   \Y{+2}{\lmun}({\hat n}_j) \ d{\hat n}_j \\
        &=& 
        \int \sqrt{\frac{(\ell'-2)!}{(\ell'+2)!}} \left( \eth^2 \Y{}{\lmprime}({\hat n}_j) \right)^* \ w_j    \Y{+2}{\lmun}({\hat n}_j) \ d{\hat n}_j \\
        &\simeq& 
        \int \Ystar{+2}{\lmprime}({\hat n}_j) \Y{+2}{\lmun}({\hat n}_j)  \ w_j \ d{\hat n}_j,
\end{eqnarray*}
where we neglected gradients of the window function $w_j = w({\hat n}_j)$.

Finally, we also have the completeness relation for spherical harmonics \citep{varshalovich:1988}:
\begin{equation}
        \sum_{\lm} {}_sY_{\lm}({\hat n}_i) {}_sY_{\lm}^*({\hat n}_j) = \delta({\hat n}_i-{\hat n}_j)
\label{eq:ylm_orthogonality}
.\end{equation}

\subsubsection{Product of 2 $\W{\oplus}{X}{\lmun}{\lmdeux}$ \label{sec:twoW}}

We neglect gradients of the window function and apply the completeness relation for spherical harmonics (Eq.~\ref{eq:ylm_orthogonality})

\begin{eqnarray}
        \sum_\lmun W^{0,X*}_{\lm \lmun} W^{0,Y}_{\lmprime \lmun}
        & = &
        \sum_{\ell_1 m_1} \int_{ij} w^X_i w^Y_j d{\hat n}_i d{\hat n}_j Y_{\ell m}({\hat n}_i) Y_{\ell_1 m_1}^*({\hat n}_i) Y_{\ell' m'}^*({\hat n}_j) Y_{\ell_1 m_1}({\hat n}_j)  \nonumber \\
        & = &
        \int_{i} \left(w^X_i w^Y_i\right) d{\hat n}_i Y_{\ell m}({\hat n}_i) Y_{\ell' m'}^*({\hat n}_i) \nonumber \\
        & = &   
        W^{0}_{\lm\lmprime}(w^Xw^Y) \equiv W_{\lm\lmprime}^{0,XY}
\end{eqnarray}

\begin{eqnarray}
        \sum_\lmun W^{+,X*}_{\lm \lmun} W^{+,Y}_{\lmprime \lmun}
        &=& 
        \everymath={\displaystyle}
        \begin{array}{rcl} 
        \sum_{\ell_1 m_1} \int_{ij} w^X_i w^Y_j d{\hat n}_i d{\hat n}_j \bigg[ 
        \Y{+2}{\lm}({\hat n}_i) \Ystar{+2}{\lmun}({\hat n}_i) \Ystar{+2}{\lmprime}({\hat n}_j) \Y{+2}{\lmun}({\hat n}_j)
        &+& \Y{+2}{\lm}({\hat n}_i) \Ystar{+2}{\lmun}({\hat n}_i) \Ystar{-2}{\lmprime}({\hat n}_j) \Y{-2}{\lmun}({\hat n}_j) \\
        + \Y{-2}{\lm}({\hat n}_i) \Ystar{-2}{\lmun}({\hat n}_i) \Ystar{+2}{\lmprime}({\hat n}_j) \Y{+2}{\lmun}({\hat n}_j)
        &+& \Y{-2}{\lm}({\hat n}_i) \Ystar{-2}{\lmun}({\hat n}_i) \Ystar{-2}{\lmprime}({\hat n}_j) \Y{-2}{\lmun}({\hat n}_j)
        \bigg] \end{array}   \nonumber \\
        &\simeq&
        2\ \sum_{\ell_1 m_1} \int_{ij} w^X_i w^Y_j d{\hat n}_i d{\hat n}_j \bigg[ 
        \Y{+2}{\lm}({\hat n}_i) \Ystar{+2}{\lmun}({\hat n}_i) \Ystar{+2}{\lmprime}({\hat n}_j) \Y{+2}{\lmun}({\hat n}_j)
        +\Y{-2}{\lm}({\hat n}_i) \Ystar{-2}{\lmun}({\hat n}_i) \Ystar{-2}{\lmprime}({\hat n}_j) \Y{-2}{\lmun}({\hat n}_j)
        \bigg]   \nonumber\\        
        & \simeq &
        2 \int_{i} \left(w^X_i w^Y_i\right) d{\hat n}_i \bigg[
                \Y{2}{\lm}({\hat n}_i) \Ystar{2}{\lmprime}({\hat n}_i)  + \Y{-2}{\lm}({\hat n}_i) \Ystar{-2}{\lmprime}({\hat n}_i)
                \bigg]  \nonumber \\
        & \simeq &
        2 \ W^{+,XY}_{\lm\lmprime}
\end{eqnarray}

\begin{eqnarray}
        \sum_\lmun W^{-,X*}_{\lm \lmun} W^{-,Y}_{\lmprime \lmun}
        & = & 
        \everymath={\displaystyle}
        \begin{array}{rcl} 
        \sum_{\ell_1 m_1} \int_{ij} w^X_i w^Y_j d{\hat n}_i d{\hat n}_j \bigg[
        \Y{+2}{\lm}({\hat n}_i) \Ystar{+2}{\lmun}({\hat n}_i) \Ystar{+2}{\lmprime}({\hat n}_j) \Y{+2}{\lmun}({\hat n}_j)
        &-& \Y{+2}{\lm}({\hat n}_i) \Ystar{+2}{\lmun}({\hat n}_i) \Ystar{-2}{\lmprime}({\hat n}_j) \Y{-2}{\lmun}({\hat n}_j) \\
        - \Y{-2}{\lm}({\hat n}_i) \Ystar{-2}{\lmun}({\hat n}_i) \Ystar{+2}{\lmprime}({\hat n}_j) \Y{+2}{\lmun}({\hat n}_j)
        &+& \Y{-2}{\lm}({\hat n}_i) \Ystar{-2}{\lmun}({\hat n}_i) \Ystar{-2}{\lmprime}({\hat n}_j) \Y{-2}{\lmun}({\hat n}_j)
        \bigg] \end{array}   \nonumber \\
        & \simeq & 0
\end{eqnarray}

\begin{eqnarray}
        \sum_\lmun W^{0,X*}_{\lm \lmun} W^{+,Y}_{\lmprime \lmun}
        & = &
        \sum_{\ell_1 m_1} \int_{ij} w^X_i w^Y_j d{\hat n}_i d{\hat n}_j Y_{\ell m}^{} ({\hat n}_i) Y_{\ell_1 m_1}^*({\hat n}_i) 
                \left[ \Y{2}{\lmprime}^*({\hat n}_j) \Y{2}{\lmun}({\hat n}_j) + \Y{-2}{\lmprime}^*({\hat n}_j) \Y{-2}{\lmun}({\hat n}_j)  \right]     \nonumber \\
        & \simeq &
        2\ \sum_{\ell_1 m_1} \int_{ij} w^X_i w^Y_j d{\hat n}_i d{\hat n}_j Y_{\ell m}^{} ({\hat n}_i) Y_{\ell_1 m_1}^*({\hat n}_i) \Y{}{\lmprime}^*({\hat n}_j) \Y{}{\lmun}({\hat n}_j)
        \nonumber \\
        & \simeq &
        2\ \int_{i} (w^X_i w^Y_i) d{\hat n}_i Y_{\ell m}^{} ({\hat n}_i) \Y{}{\lmprime}^*({\hat n}_i)  \nonumber \\
        & \simeq &
        2\ W^{0,XY}_{\lm\lmprime}
\end{eqnarray}

\begin{eqnarray}
        \sum_\lmun W^{0,X*}_{\lm \lmun} W^{-,Y}_{\lmprime \lmun}
        & = &
        \sum_{\ell_1 m_1} \int_{ij} w^X_i w^Y_j d{\hat n}_i d{\hat n}_j Y_{\ell m}^{} ({\hat n}_i) Y_{\ell_1 m_1}^*({\hat n}_i) 
                \left[ \Y{2}{\lmprime}^*({\hat n}_j) \Y{2}{\lmun}({\hat n}_j) - \Y{-2}{\lmprime}^*({\hat n}_j) \Y{-2}{\lmun}({\hat n}_j)  \right]     \nonumber \\
        & \simeq &
        0
\end{eqnarray}

\begin{eqnarray}
        \sum_\lmun W^{+,X*}_{\lm \lmun} W^{-,Y}_{\lmprime \lmun}
        &=& 
        \everymath={\displaystyle}
        \begin{array}{rcl} 
        \sum_{\ell_1 m_1} \int_{ij} w^X_i w^Y_j d{\hat n}_i d{\hat n}_j \bigg[ 
        \Y{+2}{\lm}({\hat n}_i) \Ystar{+2}{\lmun}({\hat n}_i) \Ystar{+2}{\lmprime}({\hat n}_j) \Y{+2}{\lmun}({\hat n}_j)
        &-& \Y{+2}{\lm}({\hat n}_i) \Ystar{+2}{\lmun}({\hat n}_i) \Ystar{-2}{\lmprime}({\hat n}_j) \Y{-2}{\lmun}({\hat n}_j) \\
        + \Y{-2}{\lm}({\hat n}_i) \Ystar{-2}{\lmun}({\hat n}_i) \Ystar{+2}{\lmprime}({\hat n}_j) \Y{+2}{\lmun}({\hat n}_j)
        &-& \Y{-2}{\lm}({\hat n}_i) \Ystar{-2}{\lmun}({\hat n}_i) \Ystar{-2}{\lmprime}({\hat n}_j) \Y{-2}{\lmun}({\hat n}_j)
        \bigg] \end{array}   \nonumber \\
        &\simeq&
        2\ \sum_{\ell_1 m_1} \int_{ij} w^X_i w^Y_j d{\hat n}_i d{\hat n}_j \bigg[ 
        \Y{+2}{\lm}({\hat n}_i) \Ystar{+2}{\lmun}({\hat n}_i) \Ystar{+2}{\lmprime}({\hat n}_j) \Y{+2}{\lmun}({\hat n}_j) -
        \Y{-2}{\lm}({\hat n}_i) \Ystar{-2}{\lmun}({\hat n}_i) \Ystar{-2}{\lmprime}({\hat n}_j) \Y{-2}{\lmun}({\hat n}_j)
        \bigg]   \nonumber \\       
        & \simeq &
        2 \int_{i} \left(w^X_i w^Y_i\right) d{\hat n}_i \bigg[
                \Y{2}{\lm}({\hat n}_i) \Ystar{2}{\lmprime}({\hat n}_i)  - \Y{-2}{\lm}({\hat n}_i) \Ystar{-2}{\lmprime}({\hat n}_i)
                \bigg]  \nonumber \\
        & \simeq &
        2 \ W^{-,XY}_{\lm\lmprime}
\end{eqnarray}

\subsubsection{Variance of pseudo-$C_\ell$}

We now write elements of the pseudo-covariance matrix $\tens{\tilde \Sigma}$ (Eq.~\ref{xifinal}).
We consider approximation of high multipoles (greater than the width of the window function $W_\ell$) for which, if the Galactic cut is sufficiently narrow, we can the replace the product $C^X_{\ell_1}C^Y_{\ell_2}$ by $C^X_{\ell\ell'}C^Y_{\ell\ell'} = \sqrt{C^X_\ell C^X_{\ell'}C^Y_\ell C^Y_{\ell'}}$ allowing the matrix to be symmetric. Then we apply the relation to the product of $W^{0,\pm,X}_{\lm \lmprime}$ (Section~\ref{sec:twoW}) on $(\ell_1,m_1)$ and $(\ell_2,m_2)$ successively. Finally, we identify the kernels $\tens{M}(\ell,\ell',a,b)$ as defined in Eqs.~(\ref{eq:def_chi}).

\begin{eqnarray}
        \VEV{ \Delta\tilde{C}^{T_aT_b}_\ell\Delta\tilde{C}^{T_cT_d\star}_{\ell^\prime}} 
        &=& 
        L_{\ell\ell'} \sum_{mm'} \bigg\{
                \VEV{ T^{a}_{\lm}T^{c*}_{\lmprime}} \VEV{ T^{b*}_{\lm}T^{d}_{\lmprime}} + 
                \VEV{ T^{a}_{\lm}T^{d*}_{\lmprime}} \VEV{ T^{b*}_{\lm}T^{c}_{\lmprime}}
        \bigg\}\\
        &=& 
        L_{\ell\ell^\prime}\sum_{mm^\prime}\sum_\lmun\sum_\lmdeux \bigg\{
        C_{\ell_1}^{T_aT_c} C_{\ell_2}^{T_bT_d}W^{0,T_a}_{\lm\lmun}W^{0,T_c\star}_{\lmprime\lmun}W^{0,T_b\star}_{\lm\lmdeux}W^{0,T_d}_{\lmprime\lmdeux}
        +C_{\ell_1}^{T_aT_d}C_{\ell_2}^{T_bT_c}W^{0,T_a}_{\lm\lmun}W^{0,T_d\star}_{\lmprime\lmun}W^{0,T_b\star}_{\lmprime\lmdeux}W^{0,T_c}_{\lm\lmdeux}
        \bigg\}
         \nonumber \\
         &\simeq&
        L_{\ell\ell^\prime} C_{\ell\ell'}^{T_aT_c}C_{\ell\ell'}^{T_bT_d} \sum_{mm^\prime} W^{0,T_aT_c}_{\lm\lmprime} W^{0,T_bT_d\star}_{\lm\lmprime}
        + L_{\ell\ell^\prime} C_{\ell\ell'}^{T_aT_d}C_{\ell\ell'}^{T_bT_c} \sum_{mm^\prime} W^{0,T_aT_d}_{\lm\lmprime} W^{0,T_bT_c\star}_{\lm\lmprime}      
        \nonumber
\end{eqnarray}
\begin{equation}
\framebox{
$\displaystyle
        \VEV{ \Delta\tilde{C}^{T_aT_b}_\ell\Delta\tilde{C}^{T_cT_d\star}_{\ell^\prime}}
        \simeq
        C_{\ell\ell'}^{T_aT_c}C_{\ell\ell'}^{T_bT_d} \tens{M}_{TT,TT}(\ell, \ell'; w^T_aw^T_c, w^T_bw^T_d) + C_{\ell\ell'}^{T_aT_d}C_{\ell\ell'}^{T_bT_c} \tens{M}_{TT,TT}(\ell, \ell'; w^T_aw^T_d,w^T_bw^T_c)
$}
\end{equation}

\begin{eqnarray}
        \VEV{ \Delta\tilde{C}^{E_aE_b}_\ell\Delta\tilde{C}^{E_cE_d\star}_{\ell^\prime}} 
        &=& 
        L_{\ell\ell'} \sum_{mm'} \bigg\{
                \VEV{ E^{a}_{\lm}E^{c*}_{\lmprime}} \VEV{ E^{b*}_{\lm}E^{d}_{\lmprime}} + 
                \VEV{ E^{a}_{\lm}E^{d*}_{\lmprime}} \VEV{ E^{b*}_{\lm}E^{c}_{\lmprime}}
        \bigg\}\\
        &=& 
        \frac{1}{16} L_{\ell\ell^\prime}\sum_{mm^\prime}\sum_\lmun\sum_\lmdeux \bigg\{ 
        C_{\ell_1}^{E_aE_c} C_{\ell_2}^{E_bE_d}W^{+,E_a}_{\lm\lmun}W^{+,E_c\star}_{\lmprime\lmun}W^{+,E_b\star}_{\lm\lmdeux}W^{+,E_d}_{\lmprime\lmdeux}
        +C_{\ell_1}^{E_aE_d}C_{\ell_2}^{E_bE_c}W^{+,E_a}_{\lm\lmun}W^{+,E_d\star}_{\lmprime\lmun}W^{+,E_b\star}_{\lmprime\lmdeux}W^{+,E_c}_{\lm\lmdeux}
        \bigg\} \nonumber
        \nonumber \\
        &\simeq&
        \frac{1}{4} L_{\ell\ell^\prime} C_{\ell\ell'}^{E_aE_c}C_{\ell\ell'}^{E_bE_d} \sum_{mm^\prime} W^{+,E_aE_c}_{\lm\lmprime} W^{+,E_bE_d\star}_{\lm\lmprime}
        + \frac{1}{4} L_{\ell\ell^\prime} C_{\ell\ell'}^{E_aE_d}C_{\ell\ell'}^{E_bE_c} \sum_{mm^\prime} W^{+,E_aE_d}_{\lm\lmprime} W^{+,E_bE_c\star}_{\lm\lmprime}      
        \nonumber
\end{eqnarray}
\begin{equation}
\framebox{
$\displaystyle
        \VEV{ \Delta\tilde{C}^{E_aE_b}_\ell\Delta\tilde{C}^{E_cE_d\star}_{\ell^\prime}} 
        \simeq
        C_{\ell\ell'}^{E_aE_c}C_{\ell\ell'}^{E_bE_d} \tens{M}_{EE,EE}(\ell, \ell'; w^P_aw^P_c, w^P_bw^P_d) + C_{\ell\ell'}^{E_aE_d}C_{\ell\ell'}^{E_bE_c} \tens{M}_{EE,EE}(\ell, \ell'; w^P_aw^P_d,w^P_bw^P_c)
$}
\end{equation}

\begin{eqnarray}
        \VEV{ \Delta\tilde{C}^{B_aB_b}_\ell\Delta\tilde{C}^{B_cB_d\star}_{\ell^\prime}} 
        &=& 
        L_{\ell\ell'} \sum_{mm'} \bigg\{
                \VEV{ B^{a}_{\lm}B^{c*}_{\lmprime}} \VEV{ B^{b*}_{\lm}B^{d}_{\lmprime}} + 
                \VEV{ B^{a}_{\lm}B^{d*}_{\lmprime}} \VEV{ B^{b*}_{\lm}B^{c}_{\lmprime}}
        \bigg\}\\
        &=& 
        \frac{1}{16} L_{\ell\ell^\prime}\sum_{mm^\prime}\sum_\lmun\sum_\lmdeux \bigg\{ 
        C_{\ell_1}^{B_aB_c} C_{\ell_2}^{B_bB_d}W^{+,B_a}_{\lm\lmun}W^{+,B_c\star}_{\lmprime\lmun}W^{+,B_b\star}_{\lm\lmdeux}W^{+,B_d}_{\lmprime\lmdeux}
        +C_{\ell_1}^{B_aB_d}C_{\ell_2}^{B_bB_c}W^{+,B_a}_{\lm\lmun}W^{+,B_d\star}_{\lmprime\lmun}W^{+,B_b\star}_{\lmprime\lmdeux}W^{+,B_c}_{\lm\lmdeux}
        \bigg\} \nonumber
        \nonumber \\
        &\simeq&
        \frac{1}{4} L_{\ell\ell^\prime} C_{\ell\ell'}^{B_aB_c}C_{\ell\ell'}^{B_bB_d} \sum_{mm^\prime} W^{+,B_aB_c}_{\lm\lmprime} W^{+,B_bB_d\star}_{\lm\lmprime}
        + \frac{1}{4} L_{\ell\ell^\prime} C_{\ell\ell'}^{B_aB_d}C_{\ell\ell'}^{B_bB_c} \sum_{mm^\prime} W^{+,B_aB_d}_{\lm\lmprime} W^{+,B_bB_c\star}_{\lm\lmprime}      
        \nonumber
\end{eqnarray}
\begin{equation}
\framebox{
$\displaystyle
        \VEV{ \Delta\tilde{C}^{B_aB_b}_\ell\Delta\tilde{C}^{B_cB_d\star}_{\ell^\prime}} 
        \simeq
        C_{\ell\ell'}^{B_aB_c}C_{\ell\ell'}^{B_bB_d} \tens{M}_{EE,EE}(\ell, \ell'; w^P_aw^P_c, w^P_bw^P_d) + C_{\ell\ell'}^{B_aB_d}C_{\ell\ell'}^{B_bB_c} \tens{M}_{EE,EE}(\ell, \ell'; w^P_aw^P_d,w^P_bw^P_c)
$}
\end{equation}

\begin{eqnarray}
        \VEV{ \Delta\tilde{C}^{E_aE_b}_\ell\Delta\tilde{C}^{B_cB_d\star}_{\ell^\prime}} 
        &=& 
        L_{\ell\ell'} \sum_{mm'} \bigg\{
                \VEV{ E^{a}_{\lm}B^{c*}_{\lmprime}} \VEV{ E^{b*}_{\lm}B^{d}_{\lmprime}} + 
                \VEV{ E^{a}_{\lm}B^{d*}_{\lmprime}} \VEV{ E^{b*}_{\lm}B^{c}_{\lmprime}}
        \bigg\}\\
        &=& 
        \begin{array}{ccc}
                \frac{1}{16} L_{\ell\ell^\prime}\sum_{mm^\prime}\sum_\lmun\sum_\lmdeux \bigg\{ 
                &  & C_{\ell_1}^{E_aE_c} C_{\ell_2}^{E_bE_d}W^{+,E_a}_{\lm\lmun}W^{-,B_c\star}_{\lmprime\lmun}W^{+,E_b\star}_{\lm\lmdeux}W^{-,B_d}_{\lmprime\lmdeux}\ \\
                &+& C_{\ell_1}^{E_aE_c} C_{\ell_2}^{B_bB_d}W^{+,E_a}_{\lm\lmun}W^{-,B_c\star}_{\lmprime\lmun}W^{-,E_b\star}_{\lm\lmdeux}W^{+,B_d}_{\lmprime\lmdeux}\ \\
                &+& C_{\ell_1}^{B_aB_c} C_{\ell_2}^{E_bE_d}W^{-,E_a}_{\lm\lmun}W^{+,B_c\star}_{\lmprime\lmun}W^{+,E_b\star}_{\lm\lmdeux}W^{-,B_d}_{\lmprime\lmdeux}\ \\
                &+& C_{\ell_1}^{B_aB_c} C_{\ell_2}^{B_bB_d}W^{-,E_a}_{\lm\lmun}W^{+,B_c\star}_{\lmprime\lmun}W^{-,E_b\star}_{\lm\lmdeux}W^{+,B_d}_{\lmprime\lmdeux}\ \\
                &+& C_{\ell_1}^{E_aE_d} C_{\ell_2}^{E_bE_c}W^{+,E_a}_{\lm\lmun}W^{-,B_d\star}_{\lmprime\lmun}W^{+,E_b\star}_{\lm\lmdeux}W^{-,B_c}_{\lmprime\lmdeux}\ \\
                &+& C_{\ell_1}^{E_aE_d} C_{\ell_2}^{B_bB_c}W^{+,E_a}_{\lm\lmun}W^{-,B_d\star}_{\lmprime\lmun}W^{-,E_b\star}_{\lm\lmdeux}W^{+,B_c}_{\lmprime\lmdeux}\ \\
                &+& C_{\ell_1}^{B_aB_d} C_{\ell_2}^{E_bE_c}W^{-,E_a}_{\lm\lmun}W^{+,B_d\star}_{\lmprime\lmun}W^{+,E_b\star}_{\lm\lmdeux}W^{-,B_c}_{\lmprime\lmdeux}\ \\
                &+& C_{\ell_1}^{B_aB_d} C_{\ell_2}^{B_bB_c}W^{-,E_a}_{\lm\lmun}W^{+,B_d\star}_{\lmprime\lmun}W^{-,E_b\star}_{\lm\lmdeux}W^{+,B_c}_{\lmprime\lmdeux}
                \bigg\} \nonumber
        \end{array}
        \nonumber \\
        &\simeq&
        \begin{array}{cl}
                & \frac{1}{4} L_{\ell\ell^\prime} ({C_{\ell\ell'}^{E_aE_c}C_{\ell\ell'}^{E_bE_d}}+{C_{\ell\ell'}^{E_aE_c}C_{\ell\ell'}^{B_bB_d}}+{C_{\ell\ell'}^{B_aB_c}C_{\ell\ell'}^{E_bE_d}}+{C_{\ell\ell'}^{B_aB_c}C_{\ell\ell'}^{B_bB_d}}) \sum_{mm^\prime} W^{-,E_aB_c}_{\lm\lmprime} W^{-,E_bB_d\star}_{\lm\lmprime}\\
                +& \frac{1}{4} L_{\ell\ell^\prime} ({C_{\ell\ell'}^{E_aE_d}C_{\ell\ell'}^{E_bE_c}}+{C_{\ell\ell'}^{E_aE_d}C_{\ell\ell'}^{B_bB_c}}+{C_{\ell\ell'}^{B_aB_d}C_{\ell\ell'}^{E_bE_c}}+{C_{\ell\ell'}^{B_aB_d}C_{\ell\ell'}^{B_bB_c}}) \sum_{mm^\prime} W^{-,E_aB_d}_{\lm\lmprime} W^{-,E_bB_c\star}_{\lm\lmprime}
        \end{array}     \nonumber
\end{eqnarray}
\begin{equation}
\framebox{
$\displaystyle
        \begin{array}{rcl}
        \VEV{ \Delta\tilde{C}^{E_aE_b}_\ell\Delta\tilde{C}^{B_cB_d\star}_{\ell^\prime}} 
        &\simeq&
        (       {C_{\ell\ell'}^{E_aE_c}C_{\ell\ell'}^{E_bE_d}}+{C_{\ell\ell'}^{E_aE_c}C_{\ell\ell'}^{B_bB_d}}+
                {C_{\ell\ell'}^{B_aB_c}C_{\ell\ell'}^{E_bE_d}}+{C_{\ell\ell'}^{B_aB_c}C_{\ell\ell'}^{B_bB_d}}   ) 
                \tens{M}_{EE,BB}(\ell, \ell'; w^P_aw^P_c, w^P_bw^P_d) + \\
        && 
        (       {C_{\ell\ell'}^{E_aE_d}C_{\ell\ell'}^{E_bE_c}}+{C_{\ell\ell'}^{E_aE_d}C_{\ell\ell'}^{B_bB_c}}+
                {C_{\ell\ell'}^{B_aB_d}C_{\ell\ell'}^{E_bE_c}}+{C_{\ell\ell'}^{B_aB_d}C_{\ell\ell'}^{B_bB_c}}   ) 
                \tens{M}_{EE,BB}(\ell, \ell'; w^P_aw^P_d,w^P_bw^P_c)
        \end{array}
$}
\end{equation}

\begin{eqnarray}
        \VEV{ \Delta\tilde{C}^{T_aE_b}_\ell\Delta\tilde{C}^{T_cE_d\star}_{\ell^\prime}} 
        &=& 
        L_{\ell\ell'} \sum_{mm'} \bigg\{
                \VEV{ T^{a}_{\lm}T^{c*}_{\lmprime}} \VEV{ E^{b*}_{\lm}E^{d}_{\lmprime}} + 
                \VEV{ T^{a}_{\lm}E^{d*}_{\lmprime}} \VEV{ E^{b*}_{\lm}T^{c}_{\lmprime}}
        \bigg\}\\
        &=& 
        \frac{1}{4} L_{\ell\ell^\prime}\sum_{mm^\prime}\sum_\lmun\sum_\lmdeux \bigg\{ 
        C_{\ell_1}^{T_aT_c} C_{\ell_2}^{E_bE_d}W^{0,T_a}_{\lm\lmun}W^{0,T_c\star}_{\lmprime\lmun}W^{+,E_b\star}_{\lm\lmdeux}W^{+,E_d}_{\lmprime\lmdeux}
        +C_{\ell_1}^{T_aE_d}C_{\ell_2}^{E_bT_c}W^{0,T_a\star}_{\lm\lmun}W^{+,E_d}_{\lmprime\lmun}W^{+,E_b}_{\lmprime\lmdeux}W^{0,T_c\star}_{\lm\lmdeux}
        \bigg\} \nonumber
        \nonumber \\
        &\simeq&
        \frac{1}{2} L_{\ell\ell^\prime} C_{\ell\ell'}^{T_aT_c}C_{\ell\ell'}^{E_bE_d} \sum_{mm^\prime} W^{0,T_aT_c}_{\lm\lmprime} W^{+,E_bE_d\star}_{\lm\lmprime}
        + L_{\ell\ell^\prime} {C_{\ell\ell'}^{T_aE_d}C_{\ell\ell'}^{E_bT_c}} \sum_{mm^\prime} W^{0,T_aE_d}_{\lm\lmprime} W^{0,E_bT_c\star}_{\lm\lmprime}      
        \nonumber
\end{eqnarray}
\begin{equation}
\framebox{
$\displaystyle
        \VEV{ \Delta\tilde{C}^{T_aE_b}_\ell\Delta\tilde{C}^{T_cE_d\star}_{\ell^\prime}} 
        \simeq
        {C_{\ell\ell'}^{T_aT_c}C_{\ell\ell'}^{E_bE_d}} \tens{M}_{TP,TP}(\ell, \ell'; w^T_aw^T_c, w^P_bw^P_d) + 
        {C_{\ell\ell'}^{T_aE_d}C_{\ell\ell'}^{E_bT_c}} \tens{M}_{TT,TT}(\ell, \ell'; w^T_aw^P_d,w^P_bw^T_c)
$}
\end{equation}

\begin{eqnarray}
        \VEV{ \Delta\tilde{C}^{T_aT_b}_\ell\Delta\tilde{C}^{T_cE_d\star}_{\ell^\prime}} 
        &=& 
        L_{\ell\ell'} \sum_{mm'} \bigg\{
                \VEV{ T^{a}_{\lm}T^{c*}_{\lmprime}} \VEV{ T^{b*}_{\lm}E^{d}_{\lmprime}} + 
                \VEV{ T^{a}_{\lm}E^{d*}_{\lmprime}} \VEV{ T^{b*}_{\lm}T^{c}_{\lmprime}}
        \bigg\}\\
        &=& 
        \frac{1}{2} L_{\ell\ell^\prime}\sum_{mm^\prime}\sum_\lmun\sum_\lmdeux \bigg\{ 
        C_{\ell_1}^{T_aT_c} C_{\ell_2}^{T_bE_d}W^{0,T_a}_{\lm\lmun}W^{0,T_c\star}_{\lmprime\lmun}W^{0,T_b\star}_{\lm\lmdeux}W^{+,E_d}_{\lmprime\lmdeux}
        +C_{\ell_1}^{T_aE_d}C_{\ell_2}^{T_bT_c}W^{0,T_a\star}_{\lm\lmun}W^{+,E_d}_{\lmprime\lmun}W^{0,T_b}_{\lmprime\lmdeux}W^{0,T_c\star}_{\lm\lmdeux}
        \bigg\} \nonumber
        \nonumber \\
        &\simeq&
        L_{\ell\ell^\prime} C_{\ell\ell'}^{T_aT_c}C_{\ell\ell'}^{T_bE_d} \sum_{mm^\prime} W^{0,T_aT_c}_{\lm\lmprime} W^{0,T_bE_d\star}_{\lm\lmprime}
        + L_{\ell\ell^\prime} {C_{\ell\ell'}^{T_aE_d}C_{\ell\ell'}^{T_bT_c}} \sum_{mm^\prime} W^{0,T_aE_d}_{\lm\lmprime} W^{0,T_bT_c\star}_{\lm\lmprime}      
        \nonumber
\end{eqnarray}
\begin{equation}
\framebox{
$\displaystyle
        \VEV{ \Delta\tilde{C}^{T_aT_b}_\ell\Delta\tilde{C}^{T_cE_d\star}_{\ell^\prime}} 
        \simeq
        {C_{\ell\ell'}^{T_aT_c}C_{\ell\ell'}^{T_bE_d}} \tens{M}_{TT,TT}(\ell, \ell'; w^T_aw^T_c, w^T_bw^P_d) + 
        {C_{\ell\ell'}^{T_aE_d}C_{\ell\ell'}^{T_bT_c}} \tens{M}_{TT,TT}(\ell, \ell'; w^T_aw^P_d,w^T_bw^T_c)
$}
\end{equation}

\begin{eqnarray}
        \VEV{ \Delta\tilde{C}^{T_aT_b}_\ell\Delta\tilde{C}^{E_cE_d\star}_{\ell^\prime}} 
        &=& 
        L_{\ell\ell'} \sum_{mm'} \bigg\{
                \VEV{ T^{a}_{\lm}E^{c*}_{\lmprime}} \VEV{ T^{b*}_{\lm}E^{d}_{\lmprime}} + 
                \VEV{ T^{a}_{\lm}E^{d*}_{\lmprime}} \VEV{ T^{b*}_{\lm}E^{c}_{\lmprime}}
        \bigg\}\\
        &=& 
        \frac{1}{4} L_{\ell\ell^\prime}\sum_{mm^\prime}\sum_\lmun\sum_\lmdeux \bigg\{ 
        C_{\ell_1}^{T_aE_c} C_{\ell_2}^{T_bE_d}W^{0,T_a}_{\lm\lmun}W^{+,E_c\star}_{\lmprime\lmun}W^{0,T_b\star}_{\lm\lmdeux}W^{+,E_d}_{\lmprime\lmdeux}
        +C_{\ell_1}^{T_aE_d}C_{\ell_2}^{T_bE_c}W^{0,T_a\star}_{\lm\lmun}W^{+,E_d}_{\lmprime\lmun}W^{0,T_b}_{\lmprime\lmdeux}W^{+,E_c\star}_{\lm\lmdeux}
        \bigg\} \nonumber
        \nonumber \\
        &\simeq&
        L_{\ell\ell^\prime} C_{\ell\ell'}^{T_aE_c}C_{\ell\ell'}^{T_bE_d} \sum_{mm^\prime} W^{0,T_aE_c}_{\lm\lmprime} W^{0,T_bE_d\star}_{\lm\lmprime}
        + L_{\ell\ell^\prime} {C_{\ell\ell'}^{T_aE_d}C_{\ell\ell'}^{T_bE_c}} \sum_{mm^\prime} W^{0,T_aE_d}_{\lm\lmprime} W^{0,T_bE_c\star}_{\lm\lmprime}      
        \nonumber
\end{eqnarray}
\begin{equation}
\framebox{
$\displaystyle
        \VEV{ \Delta\tilde{C}^{T_aT_b}_\ell\Delta\tilde{C}^{E_cE_d\star}_{\ell^\prime}} 
        \simeq
        {C_{\ell\ell'}^{T_aE_c}C_{\ell\ell'}^{T_bE_d}} \tens{M}_{TT,TT}(\ell, \ell'; w^T_aw^P_c, w^T_bw^P_d) + 
        {C_{\ell\ell'}^{T_aE_d}C_{\ell\ell'}^{T_bE_c}} \tens{M}_{TT,TT}(\ell, \ell'; w^T_aw^P_d,w^T_bw^P_c)
$}
\end{equation}

\begin{eqnarray}
        \VEV{ \Delta\tilde{C}^{E_aE_b}_\ell\Delta\tilde{C}^{T_cE_d\star}_{\ell^\prime}} 
        & = &
        L_{\ell\ell'} \sum_{mm'} \bigg\{ 
        \VEV{ E^{a}_{\lm}T^{c*}_{\lmprime}} \VEV{ E^{b*}_{\lm}E^{d}_{\lmprime}} +
        \VEV{ E^{a}_{\lm}E^{d*}_{\lmprime}} \VEV{ E^{b*}_{\lm}T^{c}_{\lmprime}} 
        \bigg\}\\
        &=& 
        \frac{1}{       8} L_{\ell\ell^\prime}\sum_{mm^\prime}\sum_\lmun\sum_\lmdeux \bigg\{ 
        C_{\ell_1}^{E_aT_c} C_{\ell_2}^{E_bE_d}W^{+,E_a}_{\lm\lmun}W^{0,T_c\star}_{\lmprime\lmun}W^{+,E_b\star}_{\lm\lmdeux}W^{+,E_d}_{\lmprime\lmdeux}
        +C_{\ell_1}^{E_aE_d}C_{\ell_2}^{E_bT_c}W^{+,E_a}_{\lm\lmun}W^{+,E_d\star}_{\lmprime\lmun}W^{+,E_b\star}_{\lmprime\lmdeux}W^{0,T_c}_{\lm\lmdeux}
        \bigg\} \nonumber
        \nonumber \\
        &\simeq&
        \frac{1}{2} L_{\ell\ell^\prime} C_{\ell\ell'}^{E_aT_c}C_{\ell\ell'}^{E_bE_d} \sum_{mm^\prime} W^{0,E_aT_c}_{\lm\lmprime} W^{+,E_bE_d\star}_{\lm\lmprime}
        + \frac{1}{2} L_{\ell\ell^\prime} {C_{\ell\ell'}^{E_aE_d}C_{\ell\ell'}^{E_bT_c}} \sum_{mm^\prime} W^{+,E_aE_d}_{\lm\lmprime} W^{0,E_bT_c\star}_{\lm\lmprime}      
        \nonumber
\end{eqnarray}
\begin{equation}
\framebox{
$\displaystyle
        \VEV{ \Delta\tilde{C}^{E_aE_b}_\ell\Delta\tilde{C}^{T_cE_d\star}_{\ell^\prime}} 
        \simeq
        {C_{\ell\ell'}^{E_aT_c}C_{\ell\ell'}^{E_bE_d}} \tens{M}_{TP,TP}(\ell, \ell'; w^P_aw^T_c, w^P_bw^P_d) + 
        {C_{\ell\ell'}^{E_aE_d}C_{\ell\ell'}^{E_bT_c}} \tens{M}_{TP,TP}(\ell, \ell'; w^P_aw^P_d,w^P_bw^T_c)
$}
\end{equation}

\newpage
\section{\hillipop\ parameter list}
\label{hlp_params}

\begin{table}[!ht]
\begin{center}
\begin{tabular}{llc}
\hline
\hline
Name & Definition & Prior (if any) \\
\hline
\multicolumn{3}{c}{Instrumental}\\
\hline
$c_0$ & map calibration (100-A) &  $0.000 \pm 0.002$ \\  
$c_1$ & map calibration (100-B) &  $0.000 \pm 0.002$ \\  
$c_2$ & map calibration (143-A) &  fixed \\  
$c_3$ & map calibration (143-B) &  $0.000 \pm 0.002$ \\  
$c_4$ & map calibration (217-A) &  $0.002 \pm 0.002$ \\  
$c_5$ & map calibration (217-B) &  $0.002 \pm 0.002$ \\  
$A_{\rm pl}$ & absolute calibration & $1 \pm 0.0025$ \\
\hline
\multicolumn{3}{c}{Foreground modelling} \\
\hline
$A_{\rm PS}^{\rm radio}$        & scaling parameter for radio sources in TT                      &\\
$A_{\rm PS}^{\rm IR}$   & scaling parameter for IR sources in TT                                &\\
$A_{\rm SZ}$                    & scaling parameter for the tSZ in TT                           &\\
$A_{\rm CIB}$                   & scaling parameter for the CIB in TT                           & $1.00 \pm 0.20$\\
$A_{\rm dust}^{\rm TT}$         & scaling parameter for the dust in TT                          & $1.00 \pm 0.20$\\
$A_{\rm dust}^{\rm EE}$         & scaling parameter for the dust in EE                          & $1.00 \pm 0.20$\\
$A_{\rm dust}^{\rm TE}$         & scaling parameter for the dust in TE                          & $1.00 \pm 0.20$\\
$A_{\rm kSZ}$                   & scaling parameter for the kSZ effect                          &\\
$A_{\rm SZxCIB}$                & scaling parameter for cross correlation SZ and CIB      &\\
\hline
\end{tabular}
\caption{Nuisance parameters for the \hillipop\ likelihood}
\label{tab:hlp_nuisance}
\end{center}
\end{table}

\end{appendix}

\end{document}